\documentclass[11pt]{article}
\pdfoutput=1

\usepackage[DIV13]{typearea}
\usepackage{pdflscape}
\usepackage{amsfonts, amsmath, amssymb, textcomp}
\usepackage[a4paper,top=3.5cm,bottom=2.8cm,left=2.3cm,right=2.3cm,bindingoffset=5mm]{geometry}
\usepackage{graphicx}
\usepackage{cite}
\usepackage{bbold}
\usepackage{booktabs}

\usepackage{sidecap}
\usepackage{caption}
\usepackage{subcaption}
\usepackage{color}
\newcommand{\ba}{\begin{array}{c}}
\newcommand{\ea}{\end{array}}
\newcommand{\be}{\beta}

\def\be{\begin{equation}}
\def\ee{\end{equation}}
\def\beq{\begin{equation}}
\def\eeq{\end{equation}}
\def\bc{\begin{center}}
\def\ec{\end{center}}
\def\bea{\begin{eqnarray}}
\def\eea{\end{eqnarray}}

\definecolor{darkgreen}{rgb}{.1,0.4,0.3}

\begin{document}
\begin{titlepage}
\vspace*{-1cm}
\phantom{hep-ph/***}
\flushright
\hfil{FLAVOUR(267104)-ERC-76}
\hfil{TUM-HEP 951/14}
\hfil{RM3-TH/14-14}

\vskip 1.5cm
\begin{center}
\mathversion{bold}
{\LARGE\bf Lepton Mixing from $\Delta (3 \, n^2)$ and $\Delta (6 \, n^2)$ and CP}\\[3mm]
\mathversion{normal}
\vskip .3cm
\end{center}
\vskip 0.5  cm
\begin{center}
{\large Claudia Hagedorn}~$^{a)}$,
{\large Aurora Meroni}~$^{b), \, c)}$\\[2mm]
{\large and Emiliano Molinaro}~$^{d)}$
\\
\vskip .7cm
{\footnotesize
$^{a)}$~Excellence Cluster `Universe', Technische Universit\"{a}t M\"{u}nchen, Boltzmannstra\ss e 2, 85748 Garching, Germany
\vskip .1cm
$^{b)}$~Dipartimento di Matematica e Fisica, Universit\`a di Roma Tre, Via della Vasca Navale 84, 00146 Roma, Italy
\vskip .1cm
$^{c)}$~INFN, Laboratori Nazionali di Frascati, Via Enrico Fermi 40, 00044 Frascati, Italy
\vskip .1cm
$^{d)}$~Physik-Department T30d, Technische Universit\"{a}t M\"{u}nchen, James-Franck-Stra\ss e, 85748 Garching, Germany
\vskip .5cm
\begin{minipage}[l]{.9\textwidth}
\begin{center} 
\textit{E-mail:} 
\tt{claudia.hagedorn@ph.tum.de}, \tt{ameroni@fis.uniroma3.it}, \tt{emiliano.molinaro@tum.de}
\end{center}
\end{minipage}
}
\end{center}
\vskip 1cm
\begin{abstract}

We perform a detailed study of lepton mixing patterns arising from a scenario with three Majorana neutrinos in which a discrete flavor group
$G_f=\Delta (3\,n^{2})$ or $G_f=\Delta(6\, n^{2})$ and a CP symmetry are broken to residual symmetries $G_e=Z_3$
and $G_\nu=Z_2 \times CP$ in the charged lepton and neutrino sectors, respectively. While we consider all possible $Z_3$ and $Z_2$ generating elements,
we focus on a certain set of CP transformations. The resulting lepton mixing depends on group theoretical indices and one continuous parameter. 
 In order to study the mixing patterns comprehensively for all admitted $G_e$ and $G_\nu$, it is sufficient to discuss only three types of combinations. 
One of them requires as flavor group $\Delta (6 \, n^2)$.
Two types of combinations lead to mixing patterns with a trimaximal column, while the third one allows for a much richer structure.  
For the first type of combinations the Dirac as well as one Majorana phase
are trivial, whereas the other two ones predict in general all CP phases to be non-trivial and also non-maximal.
Already for small values of the index $n$ of the group, $n \leq 11$, experimental data on lepton mixing can be accommodated well for particular choices of the 
parameters of the theory. We also comment on the relation of the used CP transformations to the automorphisms of $\Delta (3 \, n^2)$ and $\Delta (6 \, n^2)$.

\end{abstract}
\end{titlepage}
\setcounter{footnote}{0}

\section{Introduction}
\label{intro}

Lepton mixing is encoded in the  Pontecorvo-Maki-Nakagawa-Sakata (PMNS) mixing matrix $U_{PMNS}$ that contains three mixing angles and 
up to three CP phases, one Dirac phase $\delta$ and two Majorana phases $\alpha$ and $\beta$. 
By now, all three lepton mixing angles have been measured in neutrino oscillation experiments \cite{nufit} (for other global fits reaching similar results see
\cite{otherglobal_1,otherglobal_2})
\begin{equation}
\label{anglesbfintro}
\sin^2 \theta_{13} = 0.0219^{+ 0.0010} _{-0.0011} \;\;\;  , \;\;\;  \sin^2 \theta_{12} = 0.304^{+ 0.012} _{-0.012} \;\;\; , \;\;\;
\sin^2 \theta_{23} = \left\{ \begin{array}{c} [0.451^{+ 0.06} _{-0.03}] \\[0.05in]
0.577^{+ 0.027} _{-0.035}
\end{array} \right. \; , 
\end{equation}
while there is only a weak indication for a preferred value of the Dirac phase $\delta$ \cite{nufit}
\begin{equation}
\label{deltabfintro}
\delta =4.38 ^{+1.17}_{-1.03} 
\end{equation}
and no measurement of the Majorana phases $\alpha$ and $\beta$.  An interesting approach is based on the idea that a flavor symmetry $G_f$
might be responsible for the peculiar mixing pattern, observed among leptons \cite{reviews,review_math}. This symmetry is 
usually chosen to be discrete, non-abelian and finite and is assumed to be broken to residual groups $G_e$ and $G_\nu$ 
in the charged lepton and neutrino sectors, respectively 
\cite{GLD3D4,A4first,Lam07,BHL07,dATFH}.  All mixing angles and the Dirac phase $\delta$ are then determined by $G_f$ and its breaking, if
the three generations of left-handed (LH) leptons form an irreducible
three-dimensional representation ${\bf 3}$ of $G_f$. The residual symmetry $G_e$
is taken as a (product of) cyclic group(s) with $G_e=Z_3$ being the simplest choice, while the group $G_\nu$ is fixed to be (a subgroup of)  a Klein group $Z_2 \times Z_2$ for Majorana neutrinos.\footnote{If the residual symmetry $G_\nu$ is only a subgroup of a Klein group, i.e. $G_\nu=Z_2$, then lepton mixing is not only determined by the symmetry breaking pattern of $G_f$, but a free parameter $\theta$ is present \cite{residualGnuZ2}. For neutrinos being Dirac particles $G_\nu$
can be any abelian subgroup of $G_f$ which allows the three generations of neutrinos to be distinguished like in the case of charged leptons, see e.g.
\cite{HDscan2,HMV}.} A drawback of this approach
is that Majorana phases cannot be constrained. In addition, surveys of mixing patterns which can be derived from flavor symmetries $G_f$ being subgroups of
$SU(3)$ or $U(3)$ have shown that the form of these mixing patterns is rather restricted, see \cite{D6n2mixing,D3n2mixing,HDscan,Lamscan,HMV} and, in particular, \cite{Grimus2014}, 
i.e. one of the columns of the PMNS mixing matrix turns out to be trimaximal \cite{TM_pheno} and the Dirac phase is trivial, if the pattern should be in reasonable agreement with the experimental data.

For this reason we follow here the approach \cite{S4CPgeneral} (see also \cite{GrimusRebelo,GfCPHD}) and consider a theory with a flavor and a CP symmetry which are broken to residual symmetries $G_e$ and $G_\nu$ in the charged lepton and neutrino sectors, respectively. The CP symmetry is represented by the CP transformation $X$ that acts on flavor space. Combining the latter
consistently with $G_f$ requires certain conditions to be fulfilled and thus constrains the choice of $X$ \cite{S4CPgeneral,GfCPHD}. 
The residual group $G_e$ is, like in the approach without a CP symmetry, taken to be 
an abelian subgroup of $G_f$ that allows the three generations of charged leptons to be distinguished. In contrast, the symmetry $G_\nu$ is assumed to be the direct product
of a $Z_2$ group contained in $G_f$ and the CP symmetry. All mixing angles and CP phases are then fixed in terms of a single free continuous parameter $\theta$, up to the possible
permutations of rows and columns of $U_{PMNS}$. These are admitted, since fermion masses are not constrained in this approach. 
All observables are strongly correlated and, in particular, predictions for Majorana phases are obtained.

In this paper we focus on the groups $\Delta (3 \, n^2)$ and $\Delta (6 \, n^2)$ as flavor symmetries $G_{f}$. Throughout our analysis we consider groups whose index $n$ is not
divisible by three and, if necessary, even.  We choose a class of CP transformations $X$ which fulfill all requirements
in order to be consistently combined with $G_f$.  As $G_e$ we consider the minimal possible symmetry, namely a $Z_3$ group, while for the residual $Z_2$ symmetry in the
neutrino sector we study all possible choices. We then find that the mixing arising from all combinations of such $G_e$ and $G_\nu$
can be comprehensively studied by considering only three types of combinations.

For the first type of combination, called case 1), the mixing angles only depend on the continuous parameter $\theta$ and are thus the same for all groups
$\Delta (3 \, n^2)$ and $\Delta (6 \, n^2)$. In addition, the Dirac phase as well as one of the Majorana phases are trivial, while the other Majorana phase depends on the chosen
CP transformation $X$. The mixing angles, the Dirac phase $\delta$ and the Majorana phase $\beta$ obtained for the second  
type of combination, called case 2), depend in general not only on the continuous parameter $\theta$, but also on an integer one whose value is determined
by the choice of the CP transformation $X$. The other Majorana phase $\alpha$ instead is 
also dependent on a third parameter
that is again related to the choice of the CP transformation $X$.
 One characteristic feature of the PMNS mixing matrix resulting from both these types of combinations is that its second column is trimaximal. 
 This originates from the choice of the generator of the residual $Z_2$
symmetry in the neutrino sector. The mixing arising from the third type of combination has a richer structure. In particular, we can
classify the mixing in two different cases, called case 3 a) and case 3 b.1). The reactor and the atmospheric mixing angles depend in case 3 a) only on the integer characterizing the residual $Z_2$ symmetry in the neutrino sector and the index $n$ of the flavor group. The expressions of the
solar mixing angle and the CP phases instead depend 
not only on these two parameters, but also on $\theta$ and the CP transformation $X$.
In case 3 b.1) all mixing angles and CP phases depend on these four parameters. 
Nevertheless, the requirement to accommodate the experimental data on the mixing angles well selects the residual $Z_2$ symmetry in the neutrino sector as well as 
requires particular values of the parameter $\theta$. Furthermore, it is interesting to note that in case 3 b.1) a particular choice of the residual $Z_2$ 
symmetry allows the PMNS
mixing matrix to have a first column whose elements have the same absolute values as those of the first column of the tribimaximal (TB) mixing matrix \cite{TB}.
We perform a numerical analysis in each of these cases and tabulate our results for the smallest (even and odd) values of the index $n$ that admit a reasonably good fit
to the experimental data on the mixing angles. We show that in most cases it is sufficient to consider groups with an index $n \leq 11$. 

Some particular cases of groups $\Delta (3 \, n^2)$ and $\Delta (6 \, n^2)$ combined with a CP symmetry have already been discussed in the literature:
 the groups with the smallest index $n=2$, $A_4$
and $S_4$ \cite{S4CPgeneral}, as well as the groups with $n=4$, $\Delta (48)$ \cite{Delta48CP} and $\Delta (96)$ \cite{Delta96CP}.\footnote{For a study 
of the group $\Delta (27)$ combined with a CP symmetry see \cite{Delta27CP}.} In \cite{D6n2CPZ2Z2} the groups $\Delta (6 \, n^2)$ for an arbitrary index $n$
are combined with a CP symmetry.
The fundamental difference between our approach and the one discussed there lies in the fact that the latter requires
  the residual symmetry in the neutrino sector to be a Klein group $Z_2\times Z_2$ and a CP symmetry (that do not necessarily form a direct product), while we only require 
 one $Z_2$ and a CP symmetry to be preserved. An immediate consequence is that the authors in \cite{D6n2CPZ2Z2} only discuss groups $\Delta (6 \, n^2)$
 with an even index $n$, whereas we also admit groups with an odd index, see case 3 a) and case 3 b.1). 
 The residual symmetry in the charged lepton sector, on the other hand, is in both approaches chosen as a $Z_3$ group.
 Since the symmetry preserved in the neutrino sector is larger in \cite{D6n2CPZ2Z2} than in our approach, 
 their results are more constrained, in particular all mixing angles are fixed, up to the possible permutations of rows and
 columns of the PMNS mixing matrix, one column of $U_{PMNS}$ is always trimaximal, 
 the Dirac phase is trivial as well as one of the Majorana phases, while the other one depends on the chosen CP transformation. 
 We can reproduce these results from ours for particular choices of the continuous parameter $\theta$, as we show explicitly in the discussion of case 1).
 
The paper is organized as follows: in section \ref{sec2} we recapitulate the essential ingredients of the approach with a flavor and a CP symmetry
and how lepton mixing is derived. Furthermore, we detail the relevant properties of the groups  $\Delta (3 \, n^2)$ and $\Delta (6 \, n^2)$ that we employ
as flavor symmetries. In section \ref{sec3} we list all possible elements of $\Delta (3 \, n^2)$ and $\Delta (6 \, n^2)$ that generate a $Z_3$ or a $Z_2$ group 
and thus can be used as generators of residual symmetries in the charged lepton and neutrino sectors, respectively.
As regards the CP transformations $X$, we focus on a certain set and show that these can
be consistently combined with the flavor groups under discussion and with the residual $Z_2$ group in the neutrino sector.
We also comment on the relation of these CP transformations to the automorphisms of $\Delta (3 \, n^2)$ and $\Delta (6 \, n^2)$ and study their 
properties, especially, the question whether they can be `class-inverting' or not \cite{ChenRatz}. 
The possibility to have  accidental CP symmetries in the theory is mentioned as well.
 Three types of different combinations of $Z_3$ and $Z_2$ generators and CP transformations $X$ turn out to be representative for all possible ones
 and for these lepton mixing is discussed in detail in section \ref{sec4}: we present analytic
formulae for mixing angles and CP invariants/phases, study constraints put on the parameters of the theory by the experimental data, discuss the possible presence of 
accidental CP symmetries, and analyze each mixing pattern numerically. In doing so, we first study the general dependences of mixing angles and CP phases
on the parameters of each combination and then perform a $\chi^2$ analysis in order to find the smallest values of the index $n$ that admit a good agreement with
experimental data. Our results are shown in various tables, see tables \ref{tab:caseu}-\ref{tab:caseushiftpminus2} and \ref{tab:case3an16}-\ref{tab:case3b1shift}. 
In section \ref{concl} we summarize
our main results and conclude. Our conventions for mixing angles, CP invariants and phases are found in appendix \ref{app1}
together with a summary of the global fit results \cite{nufit} and details of the $\chi^2$ analysis. Appendix \ref{app2} contains details about how to reduce the 
number of combinations of residual $Z_3$ and $Z_2$ symmetries and CP transformations $X$ to only three types that lead to distinct mixing patterns.

\section{Approach}
\label{sec2}

In this section we recapitulate the conditions which have to be fulfilled in order to consistently combine a flavor and a CP symmetry, represented by
the CP transformation $X$, and repeat the derivation of lepton mixing in a such a theory. Furthermore, we briefly summarize some relevant properties of the groups $\Delta (3 \, n^2)$ and $\Delta (6 \, n^2)$.

\subsection{Combination of flavor and CP symmetry and derivation of lepton mixing}
\label{sec21}

We consider in the following a theory that is invariant under a discrete, non-abelian and finite flavor symmetry $G_f$ and a CP symmetry
which in general also acts in a non-trivial way on the flavor space. Since we are interested in the description of lepton mixing and motivated by 
the existence of three generations, we focus on irreducible three-dimensional representations ${\bf 3}$ of $G_f$ to which
we will assign the three generations of LH leptons. The elements of the group $G_f$ can be represented by unitary three-by-three matrices
$g$ in ${\bf 3}$ and also the CP symmetry is represented by a three-by-three matrix $X$. This matrix has to be unitary and symmetric \cite{S4CPgeneral}
\begin{equation}
\label{Xcon}
X X^\dagger = X X^\star = \mathbb{1} \; .
\end{equation}
The latter constraint arises, because we only consider CP transformations that correspond to automorphisms of order two (involutions). 
As has been shown in \cite{S4CPgeneral}, constraints on the choice of $X$ arise from the requirement that the subsequent application of the CP transformation,
the flavor symmetry and the CP transformation can be represented by an element of the flavor group, i.e.
\begin{equation}
\label{gX}
(X^{-1} g X)^\star = g^\prime
\end{equation}
with $g$ and $g^\prime$ representing two elements of the flavor group $G_f$ which are in general not equal. In order to show that $X$ fulfills this condition it is sufficient
to check that it holds for a set of generators of $G_f$. The fact that
the residual symmetry of the neutrino sector should be a direct product of a $Z_2$ symmetry contained in $G_f$ and the CP symmetry imposes a further constraint
\begin{equation}
\label{XZ}
X Z^\star - Z X = 0 \;\;\; \mbox{and} \;\;\;  Z^2=\mathbb{1} 
\end{equation}
with $Z$ being the generator of this $Z_2$ symmetry in the representation ${\bf 3}$.
The lepton mixing is derived in this theory from the requirement that residual symmetries $G_e$ and $G_\nu=Z_2 \times CP$ are present in the charged lepton and
neutrino sectors, respectively. For $Q$ being the realization of the generator\footnote{Throughout the analysis we only discuss the case in which the group $G_e$ can be generated by a single generator. However, the generalization to the case in which $G_e$ is a (direct) product of cyclic symmetries is straightforward, see \cite{S4CPgeneral}.} of $G_e$ in ${\bf 3}$ we know that the combination $m_l^\dagger m_l$ ($m_{l}$ is written in the basis with right-handed (RH) charged leptons on the left-hand side and LH leptons on the right-hand side) fulfills
\begin{equation}
\label{Qml}
Q^\dagger m_l^\dagger m_l \, Q = m_l^\dagger m_l \; .
\end{equation}
For non-degenerate eigenvalues of $Q$ (i.e. we have the possibility to distinguish the three generations with the help of this symmetry) the unitary matrix $U_e$ which diagonalizes
$Q$ is determined, up to permutations of its columns and overall phases of each column, by the requirement that
\begin{equation}
\label{QUe}
U_e^\dagger \, Q \, U_e
\end{equation}
is diagonal. Given (\ref{Qml}) the matrix $U_e$ also diagonalizes $m_l^\dagger m_l$, i.e. also
\begin{equation}
\label{Ueml}
U_e^\dagger \, m_l^\dagger m_l \, U_e 
\end{equation}
is diagonal. The fact that lepton masses are not constrained in this approach is reflected by the possible permutations of the columns of $U_e$. Analogously, the neutrino sector and
thus the light neutrino mass matrix $m_\nu$ (for three Majorana neutrinos)  
is invariant under the residual symmetry $G_\nu=Z_2 \times CP$. Concretely, the matrix $m_\nu$ is
constrained by the conditions
\begin{equation}
\label{Gnumnu}
Z^T m_\nu \, Z = m_\nu \;\;\; \mbox{and} \;\;\; X m_\nu \, X = m_\nu^\star \; .
\end{equation}
Applying the basis transformation induced by the unitary matrix $\Omega$ that fulfills
\begin{equation}
\label{XZOmega}
X = \Omega \, \Omega^T \;\;\; \mbox{and} \;\;\; \Omega^\dagger \, Z \, \Omega = \left( \begin{array}{ccc} 
(-1)^{z_1} & 0 & 0\\
0 & (-1)^{z_2} & 0\\
0 & 0 & (-1)^{z_3}
\end{array}
\right)
\end{equation}
with $z_i=0,1$ and two $z_i$ being equal, we see that the combination $\Omega^T \, m_\nu \, \Omega$ is constrained to be block-diagonal and
real. Thus, this matrix is diagonalized  by a rotation $R_{ij} (\theta)$ through an angle $\theta$, $0 \leq \theta < \pi$, in the $(ij)$-plane.\footnote{\label{Rij} We define
the three different matrices $R_{ij} (\theta)$ as
\begin{equation}\nonumber
R_{12} (\theta) = \left( \begin{array}{ccc}
\cos \theta & \sin \theta & 0\\
-\sin \theta & \cos \theta & 0\\
0 & 0 & 1
\end{array} \right) \;\; , \;\;
R_{13} (\theta) = \left( \begin{array}{ccc}
\cos \theta & 0 &  \sin \theta\\
0 & 1 & 0\\
-\sin \theta & 0 & \cos \theta
\end{array} \right) \;\; , \;\;
R_{23} (\theta) = \left( \begin{array}{ccc}
1 & 0 &  0\\
0 & \cos \theta & \sin \theta\\
0 & -\sin \theta & \cos \theta
\end{array} \right)
\; .
\end{equation}}   
This plane is determined
by the $(ij)$-subspace of the matrix $\Omega^\dagger \, Z \, \Omega$ which has degenerate eigenvalues.
In addition, a diagonal matrix $K_\nu$ with elements equal to $\pm 1$ and $\pm i$ is necessary for making neutrino masses positive.
This matrix can be parametrized without loss of generality as 
\begin{equation}
\label{Knu}
K_\nu = \left( \begin{array}{ccc}
 1 & 0 & 0 \\
 0 & i^{k_1} & 0\\
 0 & 0 & i^{k_2}
\end{array}
\right) \; ,
\end{equation}
with $k_{1,2}=0,1,2,3$. So, the original matrix $m_\nu$ can be brought into
diagonal form with positive entries on its diagonal via the unitary matrix
\begin{equation}
\label{Unugeneral}
U_\nu= \Omega \, R_{ij} (\theta) \, K_\nu \; .
\end{equation}
Also the masses of the light neutrinos are not fixed and thus permutations of the columns of the matrix $U_\nu$ are admitted.
Altogether, we find that the lepton mixing matrix $U_{PMNS}$ is of the form
\begin{equation}
\label{PMNSgeneral}
U_{PMNS}= U_e^\dagger \, U_\nu= U^\dagger_e \, \Omega \, R_{ij} (\theta) \, K_\nu \; ,
\end{equation}
up to possible unphysical phases and permutations of rows and columns. Thus, in our analysis of the groups $\Delta (3 \, n^2)$ and $\Delta (6 \, n^2)$
we always consider 36 possible permutations of rows and columns for a given combination $(Q, Z, X)$.

Before concluding the discussion about the general approach let us mention that the formulae in (\ref{Xcon},\ref{XZ}) are co-variant under the basis
transformation with a unitary matrix $\tilde{\Omega}$, i.e.
\begin{equation}
\label{trafoOmegatilde1}
\tilde{Z}= \tilde{\Omega}^\dagger \, Z \, \tilde{\Omega} \;\;\; \mbox{and} \;\;\; \tilde{X}= \tilde{\Omega}^\dagger \, X \, \tilde{\Omega}^\star  
\end{equation}
do also fulfill the conditions in (\ref{Xcon},\ref{XZ}). If we also transform the generator $Q$ of $G_e$ in this way
\begin{equation}
\label{trafoOmegatilde2}
\tilde{Q}= \tilde{\Omega}^\dagger \, Q \, \tilde{\Omega} \,,
\end{equation}
we see that the PMNS mixing matrix in (\ref{PMNSgeneral}) does not change, since its result does not depend on the transformation $\tilde{\Omega}$.
Thus, both combinations $(Q, Z, X)$ and $(\tilde{Q}, \tilde{Z}, \tilde{X})$ related by $\tilde{\Omega}$ lead to the same results for lepton mixing.

\mathversion{bold}
\subsection{Group theory of $\Delta (3 \, n^2)$ and $\Delta (6 \, n^2)$}
\mathversion{normal}
\label{sec22}

The groups $\Delta (3 \, n^2)$ are isomorphic to the semi-direct product $(Z_n \times Z_n) \rtimes Z_3$ with the index $n$ being in general an integer.
Here we always assume that $n$ is not divisible by three, i.e. $ 3\nmid n$.
These groups can be defined  with the help of three generators $\tilde a$, $\tilde c$ and $\tilde d$ that fulfill the relations \cite{Delta3n2}\footnote{Some useful
relations that can be easily derived are:  $\tilde{c}^{-1} \tilde{a}^2 = \tilde{a}^2 \, \tilde{c} \, \tilde{d}$, $\tilde{d}^{-1} \tilde{a}^2 = 
\tilde{a}^2 \tilde{c}^{-1}$ and $\tilde{a} \, \tilde{c} \, \tilde{d} = \tilde{d}^{-1} \tilde{a}$.}
\begin{eqnarray}\nonumber
&& \tilde{a}^3=e \;\; , \;\;\; \tilde{c}^n=e \;\; , \;\;\; \tilde{d}^n=e \;\; , 
\\ \label{genD3n2}
&& \tilde{c} \, \tilde{d}=\tilde{d} \, \tilde{c} \;\; , \;\;\; \tilde{a} \, \tilde{c} \, \tilde{a}^{-1} = \tilde{c}^{-1} \tilde{d}^{-1} \;\; , \;\;\; \tilde{a} \, \tilde{d} \, \tilde{a}^{-1} = \tilde{c} 
\end{eqnarray}
with $e$ denoting the neutral element of the group $\Delta (3 \, n^2)$.
The explicit form of these generators in the irreducible three-dimensional representations can be chosen as 
\begin{eqnarray}
\label{acd3}
\tilde{a}=\left( \begin{array}{ccc}
 0&1&0\\
 0&0&1\\
 1&0&0
\end{array}
\right) \;\; , \;\; 
\tilde{c}=\left( \begin{array}{ccc}
 \eta^l&0&0\\
 0&\eta^k&0\\
 0&0&\eta^{-k-l}
\end{array}
\right)  \;\; , \;\; 
\tilde{d}=\left( \begin{array}{ccc}
 \eta^{-k-l}&0&0\\
 0&\eta^l&0\\
 0&0&\eta^k
\end{array}
\right) 
\end{eqnarray}
with $k, l= 0,1, ..., n-1$ and $\eta=e^{2 \pi i/n}$, i.e. $\eta^n=1$. The indices $k$ and $l$ label the three-dimensional representations (excluding the case $k=l=0$).
Since this labeling leads to an over-counting of representations, we find in general
 that this type of group has $\frac{n^2 -1}{3}$ inequivalent three-dimensional irreducible representations. In the following we choose $k=n-1$ and $l=1$, i.e.
\begin{equation}
\label{cdkn1l1}
\tilde{c}= \left( \begin{array}{ccc}
 \eta&0&0\\
 0&\eta^{-1}&0\\
 0&0&1
\end{array}
\right) \;\;\; \mbox{and} \;\;\;
\tilde{d}= \left( \begin{array}{ccc}
 1&0&0\\
 0&\eta&0\\
 0&0&\eta^{-1}
\end{array}
\right) \; ,
\end{equation}
that always give rise to a faithful representation of $\Delta (3 \, n^2)$,
i.e. each element of the abstract group is represented by a different matrix representative. Thus, in the following,
 for notational simplicity, we do not distinguish between the elements of the abstract
group $\Delta (3 \, n^2)$ and the representatives of these elements in the representation ${\bf 3}$ which we employ in our discussion of lepton mixing patterns.\footnote{One can
show that the usage of a faithful three-dimensional representation different from ${\bf 3}$ does not give rise to any new results for lepton mixing, see also \cite{D6n2mixing}.} 
It is convenient to change to a basis
in which $\tilde{a}$ becomes diagonal 
\begin{equation}
\label{forma}
a=U_a^\dagger \, \tilde{a} \, U_a=\left( \begin{array}{ccc}
 1&0&0\\
 0&\omega&0\\
 0&0&\omega^2
\end{array}
\right) \;\;\; \mbox{with} \;\;\; \omega=e^{2 \pi i/3}
\end{equation}
and the unitary matrix $U_a$ reads
\begin{equation} 
\label{Ua}
U_a = \frac{1}{\sqrt{3}} \left( \begin{array}{ccc}
 1&\omega&\omega^2\\
 1&\omega^2&\omega\\
 1&1&1
\end{array}
\right) \; .
\end{equation}
The generator $\tilde{c}$ reads in this basis
\begin{equation}
\label{formc}
c=U_a^\dagger \, \tilde{c} \, U_a = \frac 13 \left( \begin{array}{ccc}
 1+2 \cos \phi_n & 1-\cos \phi_n - \sqrt{3} \sin \phi_n & 1- \cos \phi_n + \sqrt{3} \sin \phi_n\\
 1-\cos \phi_n +\sqrt{3} \sin \phi_n & 1+ 2 \cos \phi_n & 1-\cos \phi_n - \sqrt{3} \sin \phi_n\\
 1-\cos \phi_n - \sqrt{3} \sin \phi_n & 1- \cos \phi_n +\sqrt{3} \sin \phi_n & 1+  2 \cos \phi_n
\end{array}
\right)
\end{equation}
where we introduced the abbreviation 
\begin{equation}
\label{phin}
\phi_n= \frac{2 \pi}{n} \; .
\end{equation}
The form of the remaining generator $\tilde{d}$ can also be easily computed in the new basis, e.g. by using the relation $d=a^2 c \, a$ in (\ref{genD3n2}).
It is important to note that all elements of the group can be written in the form
\begin{equation}
\label{gacd}
g= a^\alpha c^\gamma d^\delta \;\;\; \mbox{with} \;\;\; \alpha=0,1,2 \; , \;\; 0 \leq \gamma , \delta \leq n-1 \; . 
\end{equation}
In order to generate the groups $\Delta (6 \, n^2)$ that are isomorphic to $(Z_n \times Z_n) \rtimes S_3$ four generators $\tilde{a}$, $\tilde{b}$, $\tilde{c}$
and $\tilde{d}$ are necessary that fulfill the relations in (\ref{genD3n2}) and also  \cite{Delta6n2}
\begin{equation} \label{genD6n2}
\tilde{b}^2=e \;\; , \;\;  (\tilde{a} \, \tilde{b})^2=e \;\; , \;\;
\tilde{b} \, \tilde{c} \, \tilde{b}^{-1} = \tilde{d}^{-1} \;\; , \;\; \tilde{b} \, \tilde{d} \, \tilde{b}^{-1} = \tilde{c}^{-1} \; .
\end{equation}
Following \cite{Delta6n2} we define $\tilde{a}$ in the irreducible three-dimensional representations as in (\ref{acd3}), while $\tilde{c}$ and $\tilde{d}$
now depend on a single index $l$, $l=1, ... , n-1$,
\begin{equation}
\label{cdl}
\tilde{c}= \left( \begin{array}{ccc}
 \eta^l & 0 &0\\
 0& \eta^{-l} & 0\\
 0 & 0 & 1
\end{array} \right) \;\;\; \mbox{and} \;\;\;
\tilde{d}= \left( \begin{array}{ccc}
 1 & 0 &0\\
 0& \eta^l & 0\\
 0 & 0 & \eta^{-l}
\end{array} \right) \; ,
\end{equation}
and the additional generator $\tilde{b}$ is chosen to be of the form
\begin{equation} 
\label{genb}
\tilde{b}=\pm \left( \begin{array}{ccc}
0&0&1\\
0&1&0\\
1&0&0
\end{array} \right) \; .
\end{equation}
As can be checked, $2 \, (n-1)$ inequivalent irreducible three-dimensional representations are obtained. If we want to match $\tilde{c}$ and $\tilde{d}$ in (\ref{cdl}) to the ones already chosen for $\Delta (3 \, n^2)$ in (\ref{cdkn1l1})
we have to take $l=1$. We also apply the change of basis induced by $U_a$ in (\ref{Ua}) to $\tilde{b}$
\begin{equation}
\label{formb}
b=U_a^\dagger \, \tilde{b} \, U_a = \pm \, \left( \begin{array}{ccc}
 1 & 0 & 0\\
 0 & 0 & \omega^2\\
 0 & \omega & 0
 \end{array}
 \right) \; .
\end{equation}
Without loss of generality we can choose ``+" in (\ref{formb}).
Similar to (\ref{gacd}) all elements of the group $\Delta (6 \, n^2)$ can be uniquely written in the form
\begin{equation}
\label{gabcd}
g= a^\alpha b^\beta c^\gamma d^\delta \;\;\; \mbox{with} \;\;\; \alpha=0,1,2 \; , \;\; \beta=0,1 \; , \;\; 0 \leq \gamma , \delta \leq n-1 \; . 
\end{equation}
For reasons which we discuss below we not only assume  that $n$ is not divisible by three, but for $\Delta (3 \, n^2)$ also
always that $n$ is even. For $\Delta (6 \, n^2)$ the latter assumption is only made for case 1) and case 2).

{\mathversion{bold}
\section{Possible choices of $Q$, $Z$ and CP transformation $X$}
{\mathversion{normal}
\label{sec3}

In this section we detail our choices of the generator $Q$ of the residual symmetry $G_e$, the possible choices for $Z$, the generator of 
the $Z_2$ symmetry present in the neutrino sector, as well as our choice of the CP transformation $X$. We also comment on the properties
of the automorphisms corresponding to the presented $X$ as well as discuss the possible existence of accidental CP symmetries for certain
choices of combinations $(Q, Z, X)$.

{\mathversion{bold}
\subsection{Discussion of choices of $Q$}
{\mathversion{normal}
\label{sec31}

As regards the groups $\Delta (3 \, n^2)$, it is clear that the generator $Q$ of $G_e$ has to be of the form 
\begin{equation}
\label{Qacd}
Q= a \, c^\gamma d^\delta \;\;\; \mbox{or} \;\;\; Q=a^2 c^\gamma d^\delta \;\;\; \mbox{with} \;\;\; 0 \leq \gamma , \delta \leq n-1 \; ,
\end{equation}
since the remaining form $c^\gamma d^\delta$, see (\ref{gacd}), would lead to a generator $Q$ which commutes with
all the possible choices of $Z_2$ symmetry generating elements, see (\ref{Z2gen1}).  The admissible choices
of $Q$ for $\Delta (3 \, n^2)$ thus all generate a $Z_3$ symmetry ($(a \, c^\gamma d^\delta)^3=e$ and $(a^2 c^\gamma d^\delta)^3=e$
with $0 \leq \gamma , \delta \leq n-1$). Indeed, these are also all elements of the groups $\Delta (3 \, n^2)$ that can give rise to a $Z_3$ symmetry for
an index $n$ that is not divisible by three.\footnote{If the latter constraint did not hold, also elements of the form $c^\gamma d^\delta$ can give 
rise to a $Z_3$ symmetry, see also \cite{D6n2mixing}.} Thus, the choice of $Q$ in (\ref{Qacd}) is the most general one for the groups $\Delta (3 \, n^2)$.
In the case of the groups $\Delta (6 \, n^2)$ we still stick to the same choice for $Q$ and thus discuss in this case comprehensively only the
case $G_e=Z_3$ (again, additional $Z_3$ generating elements exist, if $n$ is divisible by three). 
As we show in Appendix \ref{app2} it is sufficient to consider the choice
\begin{equation}
\label{Qa}
Q=a
\end{equation}
in order to comprehensively study all cases $G_e=Z_3$.

{\mathversion{bold}
\subsection{Discussion of choices of $Z$}
{\mathversion{normal}
\label{sec32}

In the case of  $\Delta (3 \, n^2)$ the index $n$ has to be even in order for the group to have $Z_2$ generating elements.
These are 
\begin{equation}
\label{Z2gen1}
Z=c^{n/2} \;\; , \;\;\; Z=d^{n/2} \;\;\; \mbox{and} \;\;\; Z=(c \, d)^{n/2} \;\; .
\end{equation}
The number of $Z_2$ generating elements considerably increases, if we choose $\Delta (6 \, n^2)$, since also elements of the form
\begin{equation}
\label{Z2gen2}
Z=b \, c^m d^m \;\; , \;\;\;  Z=a \, b \, c^m \;\;\; \mbox{and} \;\;\; Z=a^2 b \, d^m \;\;\; \mbox{with} \;\;\; 0 \leq m \leq n-1 
\end{equation}
give rise to a $Z_2$ symmetry. Depending on whether $n$ is odd or even, we thus have $3 \, n$ or $3 \, (n+1)$ elements at our disposal
as generator of the residual $Z_2$ symmetry in the neutrino sector for $\Delta (6 \, n^2)$, see also \cite{D6n2mixing}.

{\mathversion{bold}
\subsection{Discussion of choices of CP transformation $X$}
{\mathversion{normal}
\label{sec33}

We do not attempt to perform a comprehensive study of all possible admissible CP transformations $X$. Rather we would like to focus on a
particular set. One representative of this set is the CP transformation $X_0$\footnote{Notice that $X_0$ in the ``un-rotated" basis is given by $\tilde X_0=U_a X_0 U_a^T= \mathbb{1}$.}
\begin{equation}
\label{X0}
X_0= \left( \begin{array}{ccc}
 1&0&0\\
 0&0&1\\
 0&1&0
\end{array}
\right) = P_{23} \; .
\end{equation}
A viable choice of $\Omega$ fulfilling $X_0=\Omega \, \Omega^T$ is 
\begin{equation}
\label{X0Omega}
\Omega= P_{123} \; U_a \;\;\; \mbox{with} \;\;\; P_{123}= \left( \begin{array}{ccc}
0&0&1\\
1&0&0\\
0&1&0
\end{array} \right) \; .
\end{equation}
As one can check $X$ fulfills (\ref{gX}) for the generators $a$, $c$ and $d$ as given in (\ref{forma}) and (\ref{formc})\footnote{Notice that 
$c^{-1}$ has the same form as $c$ in (\ref{formc}) with $\phi_n \; \rightarrow \; - \phi_n$. This is also the form of the matrix $c^T$ in this basis.}
\begin{equation}
\label{X0onacd}
X_0 \, a^\star \, X_0^\star=a \;\; , \;\; X_0 \, c^\star \, X_0^\star=c^{-1} \;\; , \;\; X_0 \, d^\star \, X_0^\star = d^{-1} \; .
\end{equation}
As shown in \cite{GfCPHD,ChenRatz}, CP transformations correspond to  automorphisms of the flavor symmetry $G_f$, here $\Delta (3 \, n^2)$. The action
of the automorphism corresponding to $X_0$ on the generators of the group is as follows
\begin{equation}
\label{X0autD3n2}
a \;\;\; \rightarrow \;\;\; a \;\;\; , \;\;\; c \;\;\; \rightarrow \;\;\;  c^{-1} \;\;\; \mbox{and} \;\;\; d \;\;\; \rightarrow \;\;\; d^{-1} \; .
\end{equation}
Since this transformation exchanges classes of $\Delta (3 \, n^2)$, e.g. the class  $\left\{ c, \,(c \, d)^{-1}, \, d \right\}$ is mapped into  
$\left\{ c^{-1}, \, c \, d, \, d^{-1} \right\}$
that is  different, if $n \neq 2$,  we conclude that this automorphism is an outer one (for a definition of outer automorphisms see \cite{GfCPHD}).
We also see that since $a$ is mapped into $a$ and $a$ is not in the same class as $a^{-1}=a^2$
that this automorphism cannot be `class-inverting'.\footnote{
An automorphism is called `class-inverting', if it maps each element into 
 an element that belongs to the same class as  the inverse of the former. 
For a discussion of the necessity of using CP transformations that correspond to `class-inverting' automorphisms see \cite{ChenRatz}.}

If we choose the CP transformation $X_{0}$ for $G_f=\Delta(6 \, n^2)$ we can additionally check that
\begin{equation} 
\label{X0onb}
X_0 \, b^\star \, X_0^\star=b 
\end{equation}
and thus the automorphism corresponding to this CP transformation maps
\begin{equation}
\label{X0autD6n2}
b \;\;\; \rightarrow \;\;\; b \; .
\end{equation}
We also note that in the case of $G_f=\Delta (6 \, n^2)$ $X_0$ can be written in terms of the matrices $a$ and $b$
\begin{equation}
\label{P23}
P_{23} = a \, b \; .
\end{equation}
 By studying only the classes containing the generators of the group
one could be tempted to claim this to be an inner automorphism (mainly because $c$, $d$ and $c^{-1}$, $d^{-1}$ are now in the same class). 
However, for example the class $\left\{ c \, d^{-1}, c^{-2} d^{-1}, c \, d^2 \right\}$ ($\rho=1$)  is mapped into $\left\{ c^{-1} d, c^{2} d, c^{-1} d^{-2} \right\}$ ($\rho=n-1$)
that is different (the general form of this type of classes is $\left\{ c^\rho \, d^{-\rho}, c^{-2 \rho} d^{-\rho}, c^\rho \, d^{2 \rho} \right\}$ with $\rho=1, ..., n-1$ \cite{Delta6n2}).
 Thus, again some classes are exchanged and the automorphism must be an outer one, unless we choose $n=2$ (the flavor symmetry is then $S_4$).
Considering the class structure of $\Delta (6 \, n^2)$ we see that the elements of the form $a \, c^z \, d^y$ and $a^2 c^{-y} d^{-z}$, $y, \, z=0, ..., n-1$, 
belong to the same class, i.e.
especially $a$ is now similar to $a^2=a^{-1}$. Furthermore, $b=b^{-1}$ because it has order two. Thus, we might guess that the automorphism is `class-inverting' with
respect to the group $\Delta(6\,n^{2})$.
This guess is confirmed by an explicit computation at the end of this subsection.

As is known, if $X_0$ is an admissible CP transformation also CP transformations of the form
\begin{equation}
\label{Xgeneral}
X= g X_0 = a^\alpha c^\gamma d^\delta P_{23} \;\;\; \mbox{and} \;\;\; X= g X_0 = a^\alpha b^\beta c^\gamma d^\delta P_{23} 
\end{equation}
with $\alpha=0,1,2$, $\beta=0,1$ and $0 \leq \gamma, \delta \leq n-1$ are admissible for $G_f=\Delta (3 \, n^2)$ and $\Delta (6 \, n^2)$, respectively,
as long as they lead to symmetric matrices in the representation ${\bf 3}$.
Applying this constraint  we find four types of CP transformations $X$
\begin{equation}
\label{possibleX}
X=c^s d^t P_{23} \;\; , \;\;\; X=b \, c^s d^{n-s} P_{23} \;\; , \;\;\; X=a \, b \, c^s d^{2 s} P_{23} \;\;\; \mbox{and} \;\;\; X=a^2 b \, c^{2 t} d^t P_{23}   
\end{equation}
with $0 \leq s, t \leq n-1$. 
In particular, we cannot find any $X$ of the form $a \, c^s d^t P_{23}$ or $a^2 c^s d^t P_{23}$ that corresponds to a symmetric matrix.
If we just count the number of admissible choices of $X$ that arise from $X_0$ and its conjugation with an element of the flavor group,
we arrive at $n^2$ such choices for $\Delta (3 \, n^2)$ and $n \, (n+3)$ possibilities for $G_f=\Delta (6 \, n^2)$. This, however, does not imply
that the last three CP transformations in (\ref{possibleX}) are in general not admitted, if $G_f=\Delta (3 \, n^2)$ is selected. It just implies
that such a CP transformation is not related to the automorphism corresponding to $X_0$ and we have to carefully check the properties of this
new automorphism. This indeed happens in case 1) see (\ref{QZXmin}).

\subsubsection*{Comment on `class-inverting' automorphisms}

A simple test to see whether an automorphism can be `class-inverting'
is related to the following observation: for an automorphism $\iota$ that is an involution and `class-inverting' 
the twisted Frobenius-Schur indicator $\epsilon_{\iota} (\mathrm{{\bf r}})$ equals $\pm 1$ for all irreducible representations {\bf r}. If 
 all $\epsilon_{\iota} (\mathrm{{\bf r}})=1$,  the automorphism $\iota$ is a Bickerstaff-Damhus automoprhism  \cite{ChenRatz}. The definition of
 $\epsilon_{\iota} (\mathrm{{\bf r}})$ is
\begin{equation}
\label{twistedFSindicator}
\epsilon_{\iota} ({\mathrm{{\bf r}}}) = \frac{1}{|G|} \, \sum_{g \in G} \chi_{\mathrm{{\bf r}}} (g \, {}^{\iota}g)
\end{equation}
for a group $G$, here $G_f$, $|G|$ being the number of elements of $G$, $\chi_{\mathrm{{\bf r}}} (h)$ the character of the element $h$ and ${}^{\iota}g$ being the image of the element $g$ under the automorphism $\iota$.
According to \cite{FSind_comp} for an automorphism $\iota$ being an involution and a finite group $G$ the following holds
\begin{equation}
\label{reltwistedFSgen}
\sum_{\mathrm{{\bf r}}} \chi_{\mathrm{{\bf r}}} (h) \epsilon_{\iota} (\mathrm{{\bf r}}) = \left| \left\{ g \in G \mid g \, {}^{\iota}g=h \right\} \right|
\end{equation}
for any element $h$ of $G$ and summing over all irreducible representations {\bf r} on the left-hand side.
In particular, it is true for $h$ being the neutral element of the group that  $\epsilon_\iota (\mathrm{{\bf r}})=1$ for every irreducible representation {\bf r} of $G$
if and only if
\begin{equation}
\label{reltwistedFSone}
\sum_{\mathrm{{\bf r}}} \chi_{\mathrm{{\bf r}}} (e) =  \left| \left\{ g \in G \mid {}^{\iota}g=g^{-1} \right\} \right|
\end{equation}
with $\chi_{\mathrm{{\bf r}}} (e)$ being the character of the neutral element in the representation {\bf r}, i.e. we sum over the dimensions of all
irreducible representations of $G$ on the left-hand side of (\ref{reltwistedFSone}).

So, we can check for all CP transformations $X$ mentioned in (\ref{possibleX}) whether the equality in (\ref{reltwistedFSone}) is fulfilled. If so,
the automorphism must be `class-inverting'. The explicit computation shows that the right-hand side of (\ref{reltwistedFSone})
turns out to be equal to $n \, (n+3)$ for all $X$ in the case of $G_f=\Delta (6 \, n^2)$, whereas it can be maximally $n^2$ for $X$ in (\ref{possibleX}), if $G_f=\Delta (3 \, n^2)$.\footnote{
The value $n^2$ is only obtained for the CP transformation $X=c^s d^t P_{23}$ and by evaluating (\ref{reltwistedFSgen}) for $h=a$ and $h=a^2$ we can show
that the twisted Frobenius-Schur indicator of the two non-trivial one-dimensional representations has to vanish, while $\epsilon_{\iota} (\mathrm{{\bf r}})=1$ holds for all
other representations ${\bf r}$.}
We can compare this result to the sum of the dimensions of the irreducible representations, i.e. the left-hand side of (\ref{reltwistedFSone}), 
and see that for $\Delta (3\, n^2)$ it is always equal to
$3 + n^2 -1=n^2 +2$ for $3 \nmid n$, (and $n^2+6$, if we considered $3 \mid n$), while for $\Delta (6 \, n^2)$ we 
always get $2 + 2 + 6 \, (n-1) + (n-1) \, (n-2) = n \, (n+3)$ for $3 \nmid n$ (and $n \, (n+3)+4$ for $3 \mid n$).
Thus, we find equality of left- and right-hand side of (\ref{reltwistedFSone}) for $\Delta (6 \,n^2)$, $3 \nmid n$, whereas in the other cases  
the value of the right-hand side is smaller than the one of the left-hand side. So, we know that the CP transformations $X$ for $\Delta (6 \, n^2)$, $3 \nmid n$,
correspond to `class-inverting' automorphisms that are of Bickerstaff-Damhus type. For $\Delta (3 \, n^2)$ instead this cannot be deduced and, indeed,
the arguments given above showed that the CP transformation $X_0$ is not `class-inverting'. As a consequence \cite{ChenRatz},
these groups cannot be consistently combined with any of the discussed CP transformations in general without enlarging the group.\footnote{Such an enlargement of the group
is always possible, but might not be desired.}
However, a consistent definition of CP is still possible for these groups, as long as we only consider representations fulfilling (\ref{Xcon}-\ref{XZ}).
This is the case for the representation ${\bf 3}$ under which the LH leptons transform. Since we only make explicit use of this representation in our
approach,  the discussion of lepton mixing is not affected. This is in accordance with the findings of \cite{ChenRatz} (see in particular subsection 3.1.4), since 
$\epsilon_{\iota} ({\mathrm{{\bf 3}}})=1$ for the automorphisms $\iota$ corresponding to the CP transformations presented in (\ref{possibleX}).

{\mathversion{bold}
\subsection{Accidental CP symmetries}
{\mathversion{normal}
\label{sec34}

Before summarizing all possible choices of combinations $(Q, Z, X)$ that we will study in the subsequent section we pay attention to the possibility
that an accidental CP symmetry can be present, different from the one corresponding to the CP transformation $X$ that we impose in our theory.
To remind the reader: a(n accidental) CP symmetry corresponding to a CP transformation $Y$ exists, if $Y$ fulfills
the conditions \cite{S4CPgeneral}
\begin{equation}
\label{defY}
Y^\star \, m_l^\dagger m_l \, Y = (m^\dagger_l m_l)^\star \;\;\; \mbox{and} \;\;\; Y m_\nu \, Y = m_\nu^\star \; .
\end{equation}
Clearly, then all CP phases $\delta$, $\alpha$ and $\beta$ have to be trivial
\begin{equation}
\label{CPtrivial}
\sin \delta =0 \;\; , \;\; \sin\alpha=0 \;\; \mbox{and} \;\; \sin\beta=0 \; .
\end{equation}
If $Y$ and $m_\nu$ only fulfill
\begin{equation}
\label{YdefD}
Y^\star \, m_\nu^\dagger m_\nu \, Y = (m^\dagger_\nu m_\nu)^\star \; ,
\end{equation}
the Majorana phases are in general non-trivial, while the Dirac phase $\delta$ has to be $0$ or $\pi$.
As has been shown in \cite{S4CPgeneral}, the first equality in (\ref{defY}) is fulfilled, if 
\begin{equation}
\label{QY}
Q \, Y -Y Q^T=0 \; ,
\end{equation}
while the fulfillment of the second equality implies 
\begin{equation}
\label{XZY}
Y Z^\star - Z \, Y = 0 \;\;\; \mbox{and} \;\;\; X \, Y^\star - Y X^\star = 0 
\end{equation}
as well as that the CP transformation $Y$ is diagonal and real in the neutrino mass basis, i.e. 
\begin{equation}
\label{tildeY}
\tilde{Y}= U_\nu^\dagger \, Y \, U_\nu^\star
\end{equation}
has to be diagonal and real. If only the equality in (\ref{YdefD}) should be fulfilled, it is sufficient that the first equation in (\ref{XZY}) is satisfied together with the 
condition that $\tilde Y$ has to be diagonal. In this case the (in general non-trivial) Majorana phases are determined by the differences of the phases of the diagonal
elements of $\tilde{Y}$, see \cite{S4CPgeneral}.

In particular, we see that the most general form of $Y$ compatible with a charged lepton mass matrix invariant under the residual symmetry $G_e$ generated
by $Q=a$ ($Q=a^2$) is 
\begin{equation}
\label{YQa}
Y= \left( \begin{array}{ccc}
 e^{i y_1} & 0 & 0 \\
 0 & e^{i y_2} & 0\\
 0 & 0 & e^{i y_3}
\end{array}
\right)
\end{equation}
with $0 \leq y_i < 2 \pi$.\footnote{If we had chosen $Q'=a \, c^\gamma d^\delta$ or $Q'=a^2 c^\gamma d^\delta$, $0 \leq \gamma , \delta \leq n-1$, as generator of $G_e$, 
the accidental CP transformation $Y'$ fulfilling $Q' Y' -Y' Q^{\prime \, T}=0$ would be of the form
\begin{equation}\nonumber 
Y'= g^\dagger Y g^\star
\end{equation}
with $g=c^x d^y$ being the similarity transformation relating $Q'=a \, c^\gamma d^\delta$ ($Q'=a^2 c^\gamma d^\delta$) 
to $Q=a$ ($Q=a^2$) via $Q'=g^\dagger Q \, g$ for certain values of $x$ and $y$.} 
As we will see in the following section such a CP transformation $Y$ can also be, for certain values of $y_i$, accidentally present
in the neutrino sector, if the latter is required to be invariant under $G_\nu=Z_2 \times CP$.

An accidental CP symmetry that is always present in the neutrino sector for given transformations $Z$ and $X$ is the one
represented by the CP transformation $Y=Z X$ that fulfills the constraints in (\ref{Xcon}-\ref{XZ}) and (\ref{XZY}). 

\begin{table}[t!]
\begin{center}
\begin{tabular}{|l|l|}
\hline
$Z=c^{n/2}$ & $X= c^s d^t P_{23}$, $X=a \, b \, c^s d^{2 s} P_{23}$\\
\hline
$Z=d^{n/2}$ & $X= c^s d^t P_{23}$, $X=a^2 b \, c^{2 t} d^t P_{23}$\\
\hline
$Z=(c \, d)^{n/2}$ & $X= c^s d^t P_{23}$, $X=b \, c^s d^{n-s} P_{23}$\\
\hline
$Z=b \, c^m d^m$ & $X= c^s d^t P_{23}$ with $t=n-2 \, m-s$, $X=b \, c^s d^{n-s} P_{23}$\\
\hline
$Z=a \, b \, c^m$ & $X= c^s d^t P_{23}$ with $t=2 \, (m +s)$, $X=a \, b \, c^s d^{2 s} P_{23}$\\
\hline
$Z=a^2 b \, d^m$ & $X= c^s d^t P_{23}$ with $s=2\, (m+t)$, $X=a^2 b \, c^{2 t} d^t P_{23}$\\
\hline
\end{tabular}
\end{center}
\begin{center}
\caption{\label{tab:ZXchoices}{\small Different types of $Z_2$ generators $Z$, contained in $G_f$, that can be combined with a CP transformation $X$ of the form 
$a^\alpha b^\beta c^\gamma d^\delta P_{23}$, when requesting the fulfillment of (\ref{XZ}). If not stated differently, $m$, $s$ and $t$ take integer
 values between $0$ and $n-1$. Obviously, for the first three choices of $Z$ the index $n$ of the flavor symmetry has to be even. Note furthermore that
 the last three types of $Z_2$ generators are only admitted for the groups $\Delta (6 \, n^2)$.}}
\end{center}
\end{table}
%

{\mathversion{bold}
\subsection{Summary of choices $(Q, Z, X)$}
{\mathversion{normal}
\label{sec35}

We take the residual symmetry $G_e$ in the charged lepton sector to be a $Z_3$ symmetry that is generated by
$Q=a \, c^\gamma d^\delta$ or $Q=a^2 c^\gamma d^\delta$, $0 \leq \gamma , \delta \leq n-1$. As generators of the $Z_2$ symmetry, we 
can use the ones mentioned in (\ref{Z2gen1}) and (\ref{Z2gen2}) and our possible choices of CP transformations $X$ are given in (\ref{possibleX}).
Since we require $G_\nu$ to be a direct product of the $Z_2$ symmetry generated by $Z$ and the CP symmetry corresponding to $X$, we additionally
have to ensure that the relation in (\ref{XZ}) is fulfilled. In doing so, we see that the six different types of $Z_2$ generating elements can
be each combined with two types of CP transformations $X$, that we list in table \ref{tab:ZXchoices}.
Thus, we should consider any generator $Q$ giving rise to a $Z_3$ symmetry to be combined
with any of the twelve possible combinations of $Z$ and $X$.
Instead of doing so, we can show, see appendix \ref{app2}, that it is sufficient to only analyze the following three types of choices of $(Q, Z, X)$
\begin{eqnarray}\nonumber
&&(Q= a, Z= c^{n/2}, X= a \, b \, c^s d^{2 s} P_{23}) \; ,
\\ \label{QZXmin}
&&(Q=a, Z= c^{n/2}, X= c^s d^t P_{23}) \; ,
\\ \nonumber
&&(Q=a, Z= b \, c^m d^m, X= b \, c^s d^{n-s} P_{23}) \,,
\end{eqnarray}
in order to comprehensively study the lepton mixing patterns.

\mathversion{bold}
\section{Mixing patterns derived from $(Q, Z, X)$}
\mathversion{normal}
\label{sec4}

In the following we discuss the mixing patterns arising from the choices of $(Q, Z, X)$ shown in (\ref{QZXmin}). We first present (one possible) form of 
 the matrix $\Omega$ and the PMNS mixing matrix. We then discuss the patterns originating from the 36 possible permutations of rows and columns
 of the latter matrix and detail analytical formulae for mixing angles and CP invariants $J_{CP}$, $I_1$ and $I_2$ for the permutations that allow the mixing angles to be 
 in accordance with the experimental data for particular values of the indices related to the choice of the flavor group, the residual symmetry
 in the neutrino sector as well as the continuous parameter $\theta$. Furthermore, we explain why and under which conditions (some) CP phases are trivial.
A numerical study shows the dependence of the mixing parameters on the quantities of the theory. The smallest values of the 
group index $n$ that admit a reasonably good fit to the experimental data are found with a $\chi^{2}$ analysis and are displayed in tables \ref{tab:caseu}-\ref{tab:caseushiftpminus2} and \ref{tab:case3an16}-\ref{tab:case3b1shift}.

\mathversion{bold}
\subsection{Case 1)  $(Q=a, Z= c^{n/2}, X= a \, b \, c^s d^{2 s} P_{23})$}
\mathversion{normal}
\label{sec41}

The first case for which we analyze the lepton mixing in detail can be represented by the following choice of the generator $Z$ of the $Z_2$ 
symmetry in the neutrino sector and of the CP transformation $X$
\begin{equation}
\label{ZXcase1}
Z=c^{n/2} \;\;\; \mbox{and} \;\;\; X=a \, b \, c^s d^{2 s} P_{23}  
\end{equation}
with $0 \leq s \leq n-1$. Since the CP transformation $X$ is a combination of the element $a \, b \, c^s d^{2 s}$ and the CP transformation $X_0$,
this case assumes as underlying flavor symmetry $\Delta (6 \, n^2)$. Nevertheless, it can also be realized for $G_f=\Delta (3 \, n^2)$. However, in the latter
case the automorphism corresponding to the CP transformation $X$ is different from the one related (via an inner automorphism) to $X_0$.\footnote{Using 
(\ref{P23}) we can rewrite $X$ as
$X=c^s d^{2 s}$ with $X_0$ now being $\mathbb{1}$ corresponding to the automorphism that maps $a \; \rightarrow \; a^2$, $c \; \rightarrow\; c$ and
$d \;\; \rightarrow \;\; c^{-1} d^{-1}$. Since $c$ and $c^{-1}$ do not belong to the same class in $\Delta (3 \, n^2)$ in general (only for $n=2$), see below (\ref{X0autD3n2}), this automorphism cannot be `class-inverting'. For $G_f=\Delta (6 \, n^2)$ the CP transformation $X_0=\mathbb{1}$ also maps $b \; \rightarrow \; a^2 b$ and, as discussed, the corresponding automorphism is `class-inverting' for these groups, see below (\ref{reltwistedFSone}).}
 The form of $Z$ is independent of $n$
 \begin{equation}
\label{Zcn2}
Z=c^{n/2} = \frac 13 \left( \begin{array}{ccc}
-1&2&2\\
2&-1&2\\
2&2&-1
\end{array}
\right)  \; .
\end{equation}
The non-degenerate eigenvalue of $Z$ is $+1$ and its corresponding eigenvector reads
\begin{equation}
\label{evcase1}
Z \, v_{+1} = + v_{+1} \;\;\; \mbox{with} \;\;\; v_{+1} \propto \left( \begin{array}{c} 1 \\ 1 \\ 1 \end{array} \right) \; .
\end{equation}
Thus, one of the columns of the resulting PMNS mixing matrix has to be trimaximal (up to phases). In order to achieve
compatibility with the experimental data on lepton mixing angles this column must be identified with the second one of $U_{PMNS}$.
As is well-known \cite{TM_pheno}, this implies a lower bound on the solar mixing angle
\begin{equation}
\label{th12TM}
\sin^2 \theta_{12} \gtrsim \frac 13 \; .
\end{equation}
 A choice of $\Omega$ which fulfills the conditions in (\ref{XZOmega}) for $X$ and $Z$ in (\ref{ZXcase1}) is
\begin{equation}
\label{Omegacase1}
\Omega_1 = e^{i \, \phi_s} \, U_{TB} \, \left( \begin{array}{ccc}
1 & 0 & 0\\
0 & e^{-3 \, i \, \phi_s} & 0\\
0 & 0 & -1
\end{array}
\right) 
\end{equation}
with $U_{TB}$ being the TB mixing matrix
\begin{equation}
\label{UTB}
U_{TB}=\left( \begin{array}{ccc}
\sqrt{\frac{2}{3}} & \frac{1}{\sqrt{3}} &0\\
-\frac{1}{\sqrt{6}} & \frac{1}{\sqrt{3}} & \frac{1}{\sqrt{2}}\\
-\frac{1}{\sqrt{6}} & \frac{1}{\sqrt{3}} & -\frac{1}{\sqrt{2}}
\end{array}
\right) 
\end{equation}
and
\begin{equation}
\label{phis}
\phi_s = \frac{\pi s}{n} \; .
\end{equation}
In particular, $Z$ reads after the basis transformation $\Omega_1$
\begin{equation}
\label{ZafterOmega1}
\Omega_1^\dagger \, Z \, \Omega_1= \left( \begin{array}{ccc}
-1 & 0 & 0\\
0 & 1 & 0\\
0 & 0 & -1
\end{array} \right)
\end{equation}
and thus the rotation $R_{ij} (\theta)$ has to be applied in the $(13)$-plane. So, the contribution to lepton mixing from the neutrino sector is of the form
\begin{equation}
\label{Unucase1}
U_{\nu , 1}= \Omega_1 \, R_{13} (\theta) \, K_\nu \; ,
\end{equation}
up to permutations of columns, with $K_\nu$ defined as in (\ref{Knu}). 
Given that the generator $Q=a$ of the residual symmetry in the charged lepton sector is diagonal in our chosen basis, it results $U_e=\mathbb{1}$, up to 
permutations of columns, and thus,
the PMNS mixing matrix is, up to possible permutations of rows and columns, of the form
\begin{equation}
\label{PMNS1}
U_{PMNS, 1}= \Omega_1 \, R_{13} (\theta) \, K_\nu \; .
\end{equation}

\mathversion{bold}
\subsubsection{Analytical results}
\mathversion{normal}
\label{sec411}

Out of the 36 possible permutations of rows and columns only twelve lead to a pattern compatible with data. As mentioned above, 
these are the ones with the second column being trimaximal
(the others either give rise to $\sin^2 \theta_{13}=1/3$  or to a relation between solar and reactor mixing angle which does not allow both to be fitted well simultaneously).
 Six of these twelve permutations lead to the same mixing pattern, if a possible shift in the continuous parameter $\theta$ and a possible re-labeling of $k_1$ and $k_2$
(including their sum or difference)\footnote{This can only affect the sign of the Majorana invariants $I_1$ and $I_2$.} are taken into account. Using the actual form of the 
 PMNS mixing matrix as quoted in (\ref{PMNS1}), we find
\begin{equation}
\label{anglescase1}
\sin^2 \theta_{13}= \frac 23 \, \sin^2 \theta \;\;\; , \;\;\;  \sin^2 \theta_{12} = \frac{1}{2+\cos 2 \theta}  \;\;\; , \;\;\;
\sin^2 \theta_{23} = \frac 12 \, \left( 1 + \frac{\sqrt{3} \sin 2 \theta}{2+\cos 2 \theta} \right) 
\end{equation}
and for the CP invariants we get
\begin{equation}
\label{CPinvcase1}
J_{CP}=0 \;\; , \;\;\; I_1 = \frac 29 \, (-1)^{k_1+1} \, \cos^2 \theta \, \sin 6 \, \phi_s  \;\; , \;\;\; I_2=0 \; .
\end{equation}
The remaining six permutations lead to very similar results with the only difference that the atmospheric mixing angle reads
 \begin{equation}
\label{theta23altcase1}
\sin^2 \theta_{23} = \frac 12 \, \left( 1 - \frac{\sqrt{3} \sin 2 \theta}{2+\cos 2 \theta} \right) \; ,
\end{equation}
i.e. the relative sign among the two terms in the expression of $\sin^2 \theta_{23}$ in (\ref{anglescase1}) changes. This pattern, for example, originates 
from the PMNS mixing matrix in (\ref{PMNS1}) with second and third rows exchanged. It is noteworthy that
the mixing angles only depend on the continuous parameter $\theta$ and so all groups $\Delta (3 \, n^2)$ and $\Delta (6 \, n^2)$ lead to the same results.
Thus, it is sufficient to consider the smallest such group, i.e. the case $n=2$. Indeed, this case has already been studied in the literature and our results
coincide with those, see case II in \cite{S4CPgeneral}.

The size of the parameter $\theta$ is mainly determined by the requirement to fit the reactor mixing angle well, i.e.
we expect $\theta$ to be either small ($0.17 \lesssim \theta \lesssim 0.2$) or close to $\pi$ ($2.94 \lesssim \theta \lesssim 2.97$). 
Since all mixing angles only depend on $\theta$, they fulfill certain (approximate) sum rules 
\begin{equation}
\label{sumrulescase1}
\sin^2 \theta_{12} =  \frac{1}{3 \cos^2 \theta_{13}} \approx \frac 13 \, \left( 1+\sin^2 \theta_{13} \right) \;\;\; \mbox{and} \;\;\; \sin^2 \theta_{23} \approx \frac 12 \left( 1 \pm \sqrt{2} \sin\theta_{13} \right)
\end{equation}
with ``+" for $\theta<\pi/2$ and ``-" for $\theta>\pi/2$. These have also been found in \cite{Delta48CP}.
For $(\sin^2 \theta_{13})^{\mathrm{bf}}=0.0219$ ($\theta \approx 0.18$ or $\theta \approx 2.96$) 
which is the best fit value from the latest global fit \cite{nufit} we find
\begin{equation}
\label{anglesbfcase1}
\sin^2 \theta_{12} \approx 0.341 \;\;\; \mbox{and} \;\;\; \sin^2 \theta_{23} \approx \left\{ \begin{array}{c} 0.605 \\  0.395 \end{array} \right. \; .
\end{equation}
As we see, the Dirac phase is trivial as well as one of the two Majorana phases, since $J_{CP}$ and $I_2$ both vanish. The vanishing of the former indicates an accidental
CP symmetry common to the charged lepton sector and to the combination $m_\nu^\dagger m_\nu$ of the neutrino mass matrix (see section 2.4 of \cite{S4CPgeneral}) that we explicitly confirm, see (\ref{tildeYcase1}).
The Majorana invariant $I_1$ is in general non-vanishing and can take different values. We can easily extract the
value of the Majorana phase $\alpha$ from $I_1$
\begin{equation}
\label{sinalcase1}
\sin \alpha = (-1)^{k_1+1} \, \sin 6 \, \phi_s  \; . 
\end{equation}
For the particular case $n=2$ which has been studied in the literature (see case II in \cite{S4CPgeneral}) $I_1$ vanishes, as the only allowed values of $s$ are $s=0$ and $s=1$ ($\phi_s=0$ or $\phi_s=\pi/2$).
For $n=4$ which has been presented in \cite{Delta48CP,Delta96CP} instead also non-vanishing $I_1$ can be achieved by the choice $s=1$ or $s=3$ (corresponding to $\phi_s=\pi/4$ or $\phi_s=3 \pi/4$).  
They both lead to a maximal Majorana phase $\alpha$. The behavior of $\sin\alpha$ for general values of $n$ and $s$ can be read off from the plot in the bottom-left panel
 of figure \ref{Fig:4} that belongs to case 2), $(Q=a, Z= c^{n/2}, X= c^s d^t P_{23})$, if we identify $6 \, s/n$ with $v/n$ (setting $k_1$ to zero in (\ref{sinalcase1})). 
 
We can understand the vanishing of the CP invariant $J_{CP}$ by recognizing that the accidental CP symmetry $Y$ of the charged lepton sector, see (\ref{YQa}), fulfills
the following conditions: the one involving $Z$ in (\ref{XZY}), if 
\begin{equation}
\label{YZcn2}
Y= e^{i y} \, \mathbb{1} \;\;\; \mbox{with} \;\;\; 0 \leq y < 2 \pi \; ,
\end{equation}
and it takes a diagonal form in the neutrino mass basis
\begin{equation}
\label{tildeYcase1}
\tilde{Y}=U^\dagger_{\nu , 1} Y U_{\nu , 1}^\star= e^{i  \, ( y-2 \, \phi_s)} \, \left(
\begin{array}{ccc}
 1 & 0 & 0\\
 0 & (-1)^{k_1} \, e^{6 \, i \, \phi_s} & 0\\
 0 & 0 & (-1)^{k_2}
\end{array}
\right) \; .
\end{equation}
As discussed in \cite{S4CPgeneral}, the fulfillment of these conditions tells us that the CP symmetry $Y$ of the charged lepton sector is in this case also
a CP symmetry of the combination $m_\nu^\dagger m_\nu$ of the neutrino mass matrix. Furthermore, the values of the Majorana phases $\alpha$ and $\beta$ can be read off
from $\tilde{Y}$ in (\ref{tildeYcase1})
\begin{equation}
\label{YMajoranaphasescase1}
|\sin \alpha| = \left| \sin 6 \, \phi_s \right| \;\;\; \mbox{and} \;\;\; \sin \beta=0
\end{equation}
that are consistent with our results for the CP invariants $I_1$ and $I_2$, see (\ref{CPinvcase1}) and (\ref{sinalcase1}). 
Only if $\tilde{Y}$ is also real, all CP violation vanishes, i.e. if $e^{6 \, i \, \phi_s}= \pm 1$ which is equivalent to $\sin 6 \, \phi_s =0$. This holds for 
$s=0$ and $s= \frac n2$. In these cases and for $y=0$ or $y=\pi$ the two CP symmetries $X$ and $Y$ also commute, see second equality in (\ref{XZY}).
The two values given for  $s$ are the only admissible ones, since $0 \leq s \leq n-1$ and three does not divide $n$.

At the end of this subsection, we would like to comment on the relations of the presented results to those found in the literature.
Our case 1) leads to results very similar to those obtained in \cite{D6n2CPZ2Z2} where an additional $Z_2$ symmetry is present in the neutrino sector. 
The CP invariants $J_{CP}$ and $I_2$ vanish in general like in \cite{D6n2CPZ2Z2}. Furthermore, the second column of the PMNS mixing matrix is 
also trimaximal.  If we identify the continuous parameter $\theta$ with $-\frac{\pi\gamma}{n}$ ($\gamma$ is related to one of the $Z_2$ symmetries, while $n$ is the index of the group
$\Delta (6 \, n^2)$),  we can achieve the same results for the mixing angles as found in \cite{D6n2CPZ2Z2}. In order to reproduce their result for the non-trivial
Majorana phase, $6 \, \phi_s$ in (\ref{CPinvcase1}) should be identified with $-(\varphi_1-\varphi_3)$ of \cite{D6n2CPZ2Z2}, since both 
parameter combinations depend on the choice of the CP transformation (Here we implicitly have set $k_1=0$.) Thus, $s=\gamma+x$ (see equation (3.40) in \cite{D6n2CPZ2Z2}).

\mathversion{bold}
\subsection{Case 2)  $(Q=a, Z= c^{n/2}, X= c^s d^t P_{23})$}
\mathversion{normal}
\label{sec42}

Also the choice $(Q=a, Z= c^{n/2}, X= c^s d^t P_{23})$ requires $n$ to be even.
The results of this choice have certain similarities with those found in case 1), but have
a richer structure, since now the mixing angles not only depend on the continuous parameter $\theta$, but also on the chosen CP transformation $X$,
i.e. on a certain combination of the exponents $s$ and $t$, see (\ref{defuv}) and (\ref{anglescase2}). In addition, all CP violating phases are in general non-trivial and depend on
$\theta$ as well as on $s$ and $t$ that characterize $X$.

Since also in this case $Z=c^{n/2}$, we know that the resulting PMNS mixing matrix will have a second column which is trimaximal. Consequently, the value
of $\sin^2 \theta_{12}$ is bounded from below, $\sin^2 \theta_{12} \gtrsim 1/3$, as is confirmed in the numerical analysis, see tables \ref{tab:caseu}-\ref{tab:caseushiftpminus2}.

It is useful to define the two parameters $u$ and $v$
\begin{equation}
\label{defuv}
 u= 2 \, s -t \;\;\;\;\; \mbox{and} \;\;\;\;\; v = 3 \, t 
\end{equation}
that take integer values in the intervals
\begin{equation}
\label{intervaluv}
- (n-1) \leq u \leq 2 \, (n-1) \;\;\; \mbox{and} \;\;\; 0 \leq v \leq 3 \, (n-1) \; ,
\end{equation}
since $s$ and $t$ are constrained to be $0 \leq s , t \leq n-1$. Furthermore, we also define, analogously to $\phi_s$ in (\ref{phis}) for case 1),
\begin{equation}
\phi_u = \frac{\pi u}{n} \;\;\; \mbox{and} \;\;\; \phi_v = \frac{\pi v}{n} \; .
\end{equation}
Then, the form of $\Omega_2$ which fulfills the conditions in (\ref{XZOmega}) for $Z=c^{n/2}$ and $X=c^s d^t P_{23}$ can be chosen as 
\begin{equation}
\label{Omegacase2}
\Omega_2= e^{i \, \phi_v/6} \, U_{TB} \, R_{13} \left( - \frac{\phi_u}{2} \right) \, \left( \begin{array}{ccc}
 1 & 0 & 0\\
0 & e^{-i \, \phi_v/2} & 0\\
0 & 0 & -i
\end{array} 
\right) \; .
\end{equation}
The $Z_2$ generator $Z$ is given as 
\begin{equation}
\label{ZafterOmega2}
\tilde{Z}= \Omega_2^\dagger \, Z \, \Omega_2 = \left( \begin{array}{ccc}
-1 & 0 & 0\\
0 & 1 & 0\\
0 & 0 & -1
\end{array}
\right) 
\end{equation}
in the basis transformed with $\Omega_2$ and thus also here the appropriate rotation $R_{ij} (\theta)$ is in the (13)-plane. The mixing matrix $U_{\nu , 2}$ in the 
neutrino sector hence reads, up
to permutations of its columns,
\begin{equation}
\label{Unucase2}
U_{\nu , 2}=\Omega_2 \, R_{13} (\theta) \, K_\nu \; 
\end{equation}
 and consequently, the PMNS mixing matrix, called $U_{PMNS, 2}$ in the following, is of the same form, up to permutations of rows and columns.

\mathversion{bold}
\subsubsection{Analytical results}
\mathversion{normal}
\label{sec421}

As in case 1), also in this case only twelve out of the 36 possible permutations can lead to a mixing pattern compatible with experimental data, namely those whose second column is trimaximal. 
Similarly to the above, also here all twelve permutations lead to the same type of results for the mixing angles and CP invariants.
Taking into account possible shifts in the parameter $\theta$ and a possible re-labeling of $k_1$ and $k_2$ two out of the twelve permutations of the PMNS mixing matrix in (\ref{Unucase2}) 
lead to 
 \begin{eqnarray}\nonumber
&&\sin^2 \theta_{13} = \frac 13 \left( 1 - \cos \phi_u \cos 2 \theta \right) \;\;\; , \;\;\;
 \sin^2 \theta_{12} = \frac{1}{2+\cos \phi_u \cos 2 \theta} \;\;\; , \;\;\; 
 \\ \label{anglescase2}
&&\sin^2 \theta_{23} = \frac 12 \, \left( 1 + \frac{\sqrt{3} \sin \phi_u \cos 2 \theta}{2+ \cos \phi_u\cos 2 \theta} \right)
\end{eqnarray}
 and  
 \begin{eqnarray}\nonumber
 && J_{CP}= -\frac{\sin 2 \theta}{6 \sqrt{3}} \;\; , \;\;  I_2 = \frac{1}{9} (-1)^{k_2} \,  \sin 2 \phi_u \sin 2 \theta \;\; ,
 \\ \label{CPinvcase2}
 && I_1 = \frac{1}{9} (-1)^{k_1+1} \, \left( \left[\cos\phi_u + \cos 2 \theta \right] \sin\phi_v-  \sin \phi_u \cos \phi_v \sin 2 \theta \right) \; .
  \end{eqnarray}
We easily see that the mixing angles fulfill the following sum rules
\begin{eqnarray}
\label{sumrulescase2}
&&\sin^2 \theta_{12} = \frac{1}{3 \, \cos^2 \theta_{13}} \approx \frac 13 \, \left( 1 + \sin^2 \theta_{13} \right)
\\ \nonumber
\mbox{and}\;\;&&6 \, \sin^2 \theta_{23} \, \cos^2 \theta_{13} = 3 + \sqrt{3} \, \tan \phi_u - 3 \, \left( 1+ \sqrt{3} \, \tan\phi_u\right) \,  \sin^2\theta_{13} \; .
\end{eqnarray}  
\begin{table}[t!]
\begin{center}
\begin{tabular}{|l|l|l|}
\hline
$\begin{array}{c}
u \; \rightarrow \;\; u+n\\
\left( \phi_u \; \rightarrow \;\; \phi_u+\pi \right)
\end{array}$ 
&$\theta \; \rightarrow \;\; \frac{\pi}{2}-\theta$&$\sin^2 \theta_{ij}$, $J_{CP}$, $I_2$ are invariant\\
&&$I_1$ changes sign\\
\hline
$\begin{array}{c}
u \; \rightarrow \;\; n-u\\
\left( \phi_u \; \rightarrow \;\; \pi-\phi_u \right)
\end{array}$ 
&$\theta \; \rightarrow \;\; \theta+\frac{\pi}{2}$&$\sin^2 \theta_{13}$, $\sin^2 \theta_{12}$, $I_2$ are invariant\\
&&$\sin^2\theta_{23}$ becomes $1-\sin^2\theta_{23}$; $J_{CP}$ and $I_1$ change sign\\
\hline
$\begin{array}{c}
u \; \rightarrow \;\; 2 \, n-u\\
\left( \phi_u \; \rightarrow \;\; 2 \, \pi-\phi_u \right)
\end{array}$ 
&$\theta \; \rightarrow \;\; \pi-\theta$&$\sin^2 \theta_{13}$, $\sin^2 \theta_{12}$, $I_1$ and $I_2$ are invariant\\
&&$\sin^2\theta_{23}$ becomes $1-\sin^2\theta_{23}$; $J_{CP}$  changes sign\\
\hline
\end{tabular}
\end{center}
\begin{center}
\caption{\label{tab:case2symmetries}{\small {\bf Case 2)}. Symmetry transformations of the formulae for mixing angles and CP invariants in (\ref{anglescase2}) and (\ref{CPinvcase2}).}}
\end{center}
\end{table}
Obviously, the first sum rule coincides with the one found in case 1).
Another two of the twelve permutations give rise to the same formulae, but $\sin^2\theta_{23}$ becomes $1-\sin^2 \theta_{23}$, i.e.
\begin{equation}
\label{sinth23sqaltcase2}
\sin^2 \theta_{23} = \frac 12 \, \left( 1 - \frac{\sqrt{3} \sin \phi_u \cos 2 \theta}{2+ \cos \phi_u\cos 2 \theta} \right) 
\end{equation}
and the corresponding sum rule reads
\begin{equation}
\label{sumrulessinth23sqaltcase2}
6 \, \sin^2 \theta_{23} \, \cos^2 \theta_{13} = 3 - \sqrt{3} \, \tan \phi_u - 3 \, \left( 1- \sqrt{3} \, \tan\phi_u\right) \,  \sin^2\theta_{13} \; .
\end{equation}
The results for the mixing angles and CP invariants calculated from the other eight permutations of the PMNS mixing matrix can be cast into the form of these formulae, if not only a shift in the continuous parameter $\theta$ and a re-labeling of $k_1$ and $k_2$ is taken into
account, but also a shift of $\pm \frac{n}{3}$ of the integer parameter $u$ (which means $\phi_u$ is shifted into $\phi_u \pm \frac{\pi}{3}$). It is important to mention that the latter shift does in general lead to physically
different results, since we consider a shift of an integer parameter through a non-integer number $\frac n3$ (remember three does not divide $n$). For this reason we discuss the numerical results of mixing angles and CP invariants
for $u$ and $u$ shifted into $u \pm \frac n3$ separately.
In particular, if we consider the PMNS mixing matrix in (\ref{Unucase2}) and multiply it from the left with the matrix 
\begin{equation}
\label{P1}
P_1=\left( \begin{array}{ccc}
0&1&0\\
0&0&1\\
1&0&0
\end{array}
\right) \; ,
\end{equation}
i.e. cyclicly permute the rows of this matrix, we can obtain the corresponding mixing angles and CP invariants from the formulae in (\ref{anglescase2}) and (\ref{CPinvcase2})
by simply shifting $u$, $\theta$ and by re-defining $k_1$
\begin{equation}
\label{shift1case2}
u \;\;\; \rightarrow \;\;\; u - \frac{n}{3} \;\; , \;\; \theta \;\;\; \rightarrow \;\;\; \frac{\pi}{2} - \theta \;\;\; \mbox{and} \;\;\; k_1 \;\;\; \rightarrow \;\;\; k_1 +1 \; ,
\end{equation}
while for a PMNS mixing matrix that is multiplied from the left by the matrix
\begin{equation}
\label{P2}
P_2=\left( \begin{array}{ccc}
0&0&1\\
1&0&0\\
0&1&0
\end{array}
\right)  = P_1^2 = P_1^T \; ,
\end{equation}
we get the formulae for mixing angles and CP invariants from (\ref{anglescase2}) and (\ref{CPinvcase2}), if we perform the transformations
\begin{equation}
\label{shift2case2}
u \;\;\; \rightarrow \;\;\; u + \frac{n}{3} \;\; , \;\; \theta \;\;\; \rightarrow \;\;\; \frac{\pi}{2} - \theta \;\;\; \mbox{and} \;\;\; k_1 \;\;\; \rightarrow \;\;\; k_1 +1 \; .
\end{equation}
From (\ref{anglescase2}) and (\ref{CPinvcase2}) we see that $J_{CP}$ only depends on $\theta$, while the mixing angles and $I_2$ depend on $\theta$ as well as on $u$. $I_1$ eventually is the only quantity which
also depends on $v$. 
The formulae in (\ref{anglescase2}) and (\ref{CPinvcase2}) have several symmetry properties which help us to understand the numerical results and which we summarize in table \ref{tab:case2symmetries}. The first of these symmetries is also valid, if we consider the formulae after applying the transformations in (\ref{shift1case2}) or (\ref{shift2case2}).
The other two instead relate results for $u-\frac{n}{3}$ to those for $u+\frac{n}{3}$, i.e. 
if the operations $u\; \rightarrow \; n-u$ and $\theta \; \rightarrow \; \theta + \frac{\pi}{2}$ are applied to the formulae (\ref{anglescase2}) and (\ref{CPinvcase2}) that are transformed with  (\ref{shift1case2}), we recover expressions that result from performing the transformations in (\ref{shift2case2}) on mixing angles and CP invariants in (\ref{anglescase2}) and (\ref{CPinvcase2}). Since the third symmetry in table \ref{tab:case2symmetries} is obtained from applying the other two ones subsequently (the ordering of the two transformations is irrelevant), also in this case we relate results for $u-\frac{n}{3}$ to those for $u+\frac{n}{3}$.
Furthermore, we note that the formulae in (\ref{anglescase2}) in the original version as well as if the transformations in (\ref{shift1case2}) or (\ref{shift2case2}) are applied, remain invariant, if we replace $\theta$ with $\pi-\theta$. Thus, we expect to find in our numerical analysis for each value $\theta=\theta_{\mathrm{bf}}$ that allows to accommodate the experimental data of the mixing angles well the same good fit for $\theta=\pi-\theta_{\mathrm{bf}}$. The CP invariants, on the other hand, do not remain invariant,
if $\theta$ is replaced by $\pi-\theta$, but instead $J_{CP}$ and $I_2$ change their sign, while $I_1$ does not transform in a definite way, since it contains terms that are even functions in $\theta$, but also one that is odd in $\theta$.

Note that for $u=0$ ($\phi_u=0$) the same results of the mixing angles are obtained for all $n$, i.e.
\begin{equation}
\label{anglesu0case2}
\sin^2 \theta_{13} = \frac 23 \, \sin^2 \theta \;\;\; , \;\;\; \sin^2 \theta_{12}= \frac{1}{2+\cos2 \theta} \;\;\; , \;\;\; \sin^2 \theta_{23}= \frac 12 \; .
\end{equation}
The formulae of reactor and solar mixing angles are the same as in case 1), see (\ref{anglescase1}), while atmospheric mixing turns out to be maximal.
Since the value of $\theta$ that gives rise to the best fit of the experimental data is (mainly) determined by $\theta_{13}$, also in this case the preferred values of $\theta$ are
$\theta \approx 0.18$ and $\theta \approx 2.96$. Furthermore, the Dirac phase extracted from $J_{CP}$ in (\ref{CPinvcase2}) and (\ref{anglesu0case2}) is maximal, $|\sin\delta|=1$ (for $\theta \neq 0, \pi/2, \pi$),
while the Majorana phase $\beta$ is trivial ($I_2$ vanishes independently of $\theta$) and from
\begin{equation}
\label{I1u0case2}
I_1 = \frac 29 \, (-1)^{k_1+1} \, \cos^2 \theta \, \sin \phi_v
\end{equation}
we can derive $|\sin\alpha|=|\sin\phi_v|$. Note that this formula for $I_1$  coincides with the one in case 1), see (\ref{CPinvcase1}), if we identify $\phi_v$ with $6 \, \phi_s$.
As one can check also for $u=n$ ($\phi_u=\pi$) the mixing angles become independent of $n$; a case that is clearly related to $u=0$ via a symmetry found in table \ref{tab:case2symmetries}.

If $u$, $v$ and $n$ are divisible by the same factor,  $\phi_u$ and $\phi_v$ do not change their values, e.g. a case with $u$ and $v$ even can always be reduced to a case with smaller $n'=\frac n2$ and $u'= \frac u2$ ($v'= \frac v2$), as long as $n$ is divisible by four, since also $n'$ has to be even. Thus, it frequently happens that the same results of mixing
angles and CP invariants are achieved with different groups $\Delta (3 \, n^2)$ (and $\Delta (6 \, n^2)$). In the case in which only $u$ and $n$ (but not $v$) have a common divisor $\rho$ larger than one, the mixing angles, the Dirac and the Majorana phase $\beta$ are the same, if computed for $u$ and $n$ as well as for $u/\rho$ and $n/\rho$, however, different values of the Majorana phase $\alpha$ can be obtained for the ``original" and the ``reduced" pair of $u$ and $n$. In the numerical analysis, in particular in tables 
 \ref{tab:caseu}-\ref{tab:caseushiftpminus2}, we only mention the smallest value of $n$ and $u$ that lead to a certain result for the mixing parameters.

The results for mixing angles and CP phases obtained in the present case reduce to the ones found in case 1), if 
\begin{equation}
\label{reducecase2}
\theta=0 \;\;\;  , \;\;\; \phi_u = 2 \, \theta_1 \;\;\; \mbox{and} \;\;\; v= 6 \, s_1
\end{equation}
with $\theta_1$ and $s_1$ being $\theta$ and $s$ as defined in case 1), see (\ref{anglescase1}) and (\ref{CPinvcase1}). Here we assume that $n$ is the same in both cases. Since we have to identify the discrete parameter $\phi_u$
with the continuous one $\theta_1$, it is clear that in general the results obtained are slightly different, see results for $n=8$ and $u=\mp 1$ in table \ref{tab:caseu}.
Using the symmetry transformations displayed in table \ref{tab:case2symmetries}  we see that a very similar identification can be made for $\theta=\pi/2$, i.e.
\begin{equation}
\label{reducecase22}
\theta=\frac \pi2 \;\;\;  , \;\;\; \phi_u = 2 \, \theta_1 +\pi \;\;\; , \;\;\; v= 6 \, s_1 \;\;\; \mbox{and} \;\;\; k_1= k_1^1 +1
\end{equation}
with $k_1^1$ denoting the parameter $k_1$ in case 1). Indeed, such cases are also found in the numerical analysis, see table \ref{tab:caseushiftpminus2}.
In these cases $J_{CP}$ and $I_2$ vanish, while the Majorana phase $\alpha$ fulfills $|\sin\alpha|=|\sin \phi_v|$.

Coming back to the general formulae in (\ref{anglescase2}) and (\ref{CPinvcase2}), the smallness of $\theta_{13}$ requires that
\begin{equation}
\label{conphiuth13case2}
\cos \phi_u \cos 2 \theta \approx 1 
\end{equation}
which is fulfilled for the following combinations of $(\phi_u, \theta)$
\begin{equation}\nonumber
(\phi_u, \theta) \approx (0,0) \, , \; (0,\pi) \, , \; (2 \, \pi,0) \, , \; (2 \, \pi, \pi) \;\;\; \mbox{or} \;\;\; (\phi_u, \theta) \approx (\pm\pi, \pi/2) \; .
\end{equation}
Taking into account the symmetries of the formulae in $(u, \theta)$ we see that it is sufficient to focus on the case $u \approx 0$ ($\phi_u \approx 0$) and $\cos 2 \theta \approx 1$.
Requiring $\theta_{13}$ to be within the experimentally preferred $3 \, \sigma$ interval we find an upper bound on $\phi_u$
\begin{equation}
\label{constrphiucase2}
|\phi_u| \lesssim 0.39 \;\;\; \mbox{corresponding to} \;\;\; |u/n| \lesssim  0.12 \; .
\end{equation}
Thus, for $n=20$ the maximum value of $u$ which can give rise to a good fit of the experimental data is $|u| \lesssim 0.12 \, n \approx 2.4$. This is confirmed in our numerical analysis and, indeed, one finds $u=\mp 1$ for $n=20$ in table \ref{tab:caseu} as well as the case $u=\mp 2$ and $n=20$ that can be ``reduced" to $u= \mp 1$ and $n=10$ -- a case
that is also mentioned in table \ref{tab:caseu}. Obviously, applying the symmetries listed in table \ref{tab:case2symmetries} further choices of $u$ can be found that lead to the 
same good accordance with experimental data. However, since these values are easily obtained using the table we refrain from listing them explicitly in the following. 

If we consider instead a pattern with $u$ shifted into $u \pm \frac n3$, we see that not $\phi_u$, but $\phi_u \pm \frac{\pi}{3}$ is constrained to lie in the interval $[-0.39, 0.39]$
in order to fit the 
reactor mixing angle well. Thus, the allowed range for $u/n$ we can derive is
\begin{equation}
\label{constrphiushiftcase2}
 0.66 \lesssim |\phi_u| \lesssim 1.44 \;\;\; \mbox{corresponding to} \;\;\;  0.21 \lesssim |u/n| \lesssim  0.46 \; .
\end{equation}
For $n=20$ we shall expect a good fit to the experimental data for
\begin{equation}
4.2 \lesssim |u| \lesssim 9.2 \; .
\end{equation}
Also this result can be compared with the findings of the numerical analysis and, indeed, the values $u=7$ and $u=9$ are mentioned in table \ref{tab:caseushiftpminus2}
for $n=20$ in the case of a PMNS mixing matrix that leads to (\ref{anglescase2}) and (\ref{CPinvcase2}) with replacements as in (\ref{shift1case2}). The other three values
that should lead to a good fit, $u=5$, $u=6$ and $u=8$, only appear implicitly, namely in table \ref{tab:caseushiftpminus}, since their common divisor with $n=20$ is larger than one 
($u=5$, $n=20$ is reduced to $u=1$, $n=4$ and $u=6$, $u=8$ and $n=20$ to $u=3$, $u=4$ and $n=10$).

In case 2) in general all CP violating phases are non-trivial. However, also here for particular choices of the parameters $\theta$, $u$ and $v$ some or all of these phases
can vanish. 
Consider again the CP symmetry $Y$ accidentally present in the charged lepton sector. Since $Z$ is the same as in case 1), also here
$Y$ is constrained to be of the form as in (\ref{YZcn2}). Its form is block-diagonal in the neutrino mass basis
\begin{equation}
\label{Ytcase2}
\tilde{Y}=U^\dagger_{\nu , 2} Y U_{\nu , 2}^\star= e^{i \left( y-\frac{\phi_v}{3} \right)} \, \left(
\begin{array}{ccc}
 \cos 2\theta & 0 & (-i)^{k_2} \sin 2\theta\\
 0 & (-1)^{k_1} e^{i \phi_v} & 0\\
 (-i)^{k_2} \sin 2\theta & 0 & (-1)^{k_2+1} \cos 2\theta
\end{array}
\right) \; .
\end{equation}
As one sees, $\sin 2\theta=0$ leads to a diagonal form of $\tilde{Y}$ and thus must imply the vanishing of $J_{CP}$. This is consistent
with the findings that $J_{CP}$ is proportional to $\sin 2\theta$. The Majorana phases can then be read off as
\begin{equation}
\label{YtMajoranacase2}
|\sin\alpha|= \left|  \sin \phi_v \right| \;\;\; \mbox{and} \;\;\; \sin\beta=0 \; . 
\end{equation}
This is again consistent with the form of the CP invariants $I_1$ and $I_2$ shown above. Note that for $v=0$ all CP violation vanishes. This 
is only fulfilled if $t=0$, since we only consider groups with an index $n$ that is not divisible by three. The CP transformation $\tilde Y$ in (\ref{Ytcase2}) becomes then
real (and diagonal) for the choice $y=k\, \pi$, $k=0,1$, and fulfills the second equation in (\ref{XZY}).

We note that expressions corresponding to those in (\ref{anglescase2}) and (\ref{CPinvcase2}) have been obtained in \cite{Delta48CP,Delta96CP} for the particular
choice $n=4$, i.e. for $\Delta (48)$ \cite{Delta48CP} and $\Delta (96)$ \cite{Delta96CP}. In particular, the sum rules in (\ref{sumrulescase2}) and (\ref{sumrulessinth23sqaltcase2}) with $\phi_u=\pi/12$ were also found in \cite{Delta96CP}.}}

\mathversion{bold}
\subsubsection{Numerical results}
\mathversion{normal}
\label{sec422}

%
\begin{figure}[t!]
\begin{center}
\begin{tabular}{cc}
\includegraphics[width=0.48\textwidth]{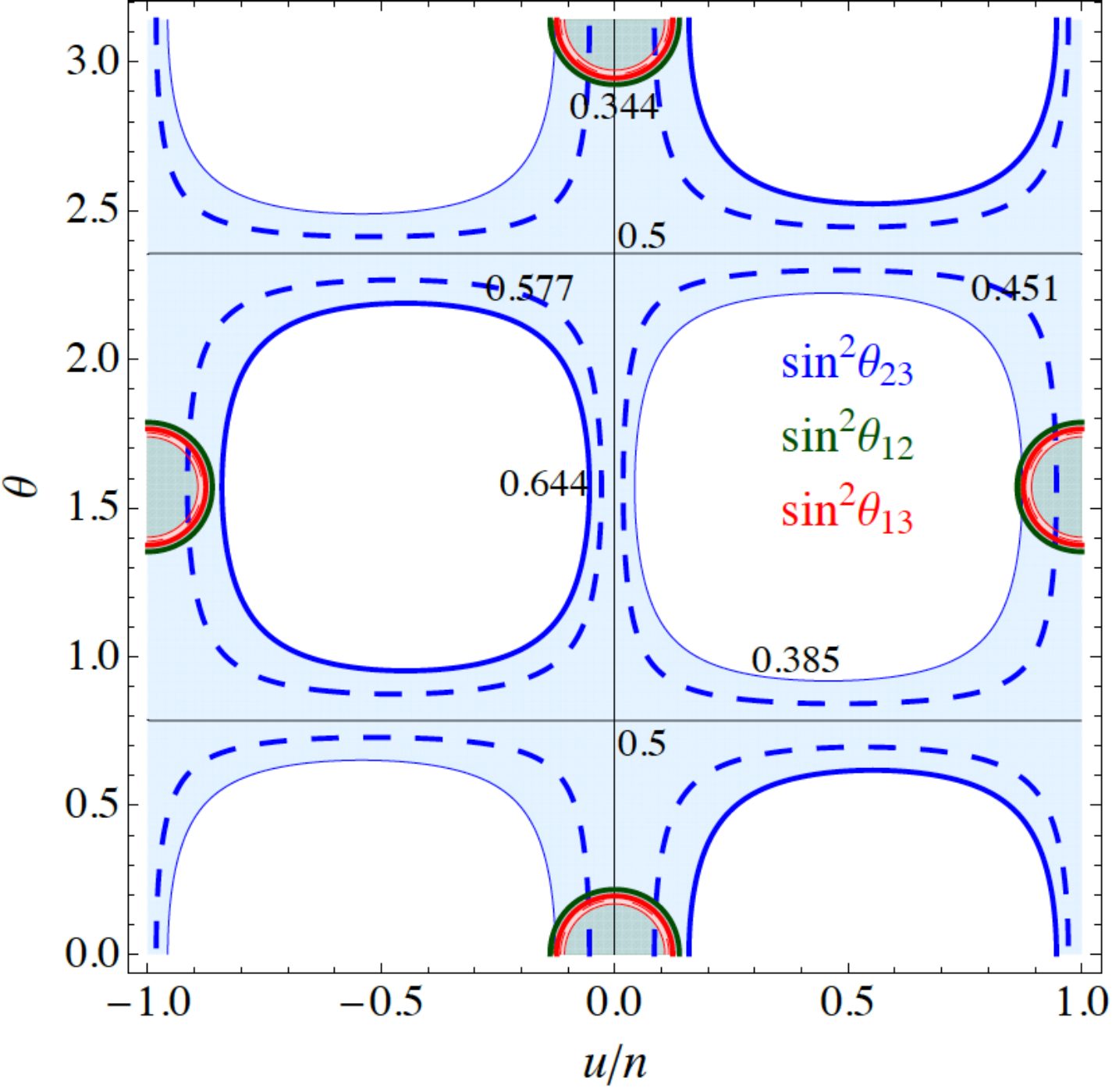} &
\includegraphics[width=0.48\textwidth]{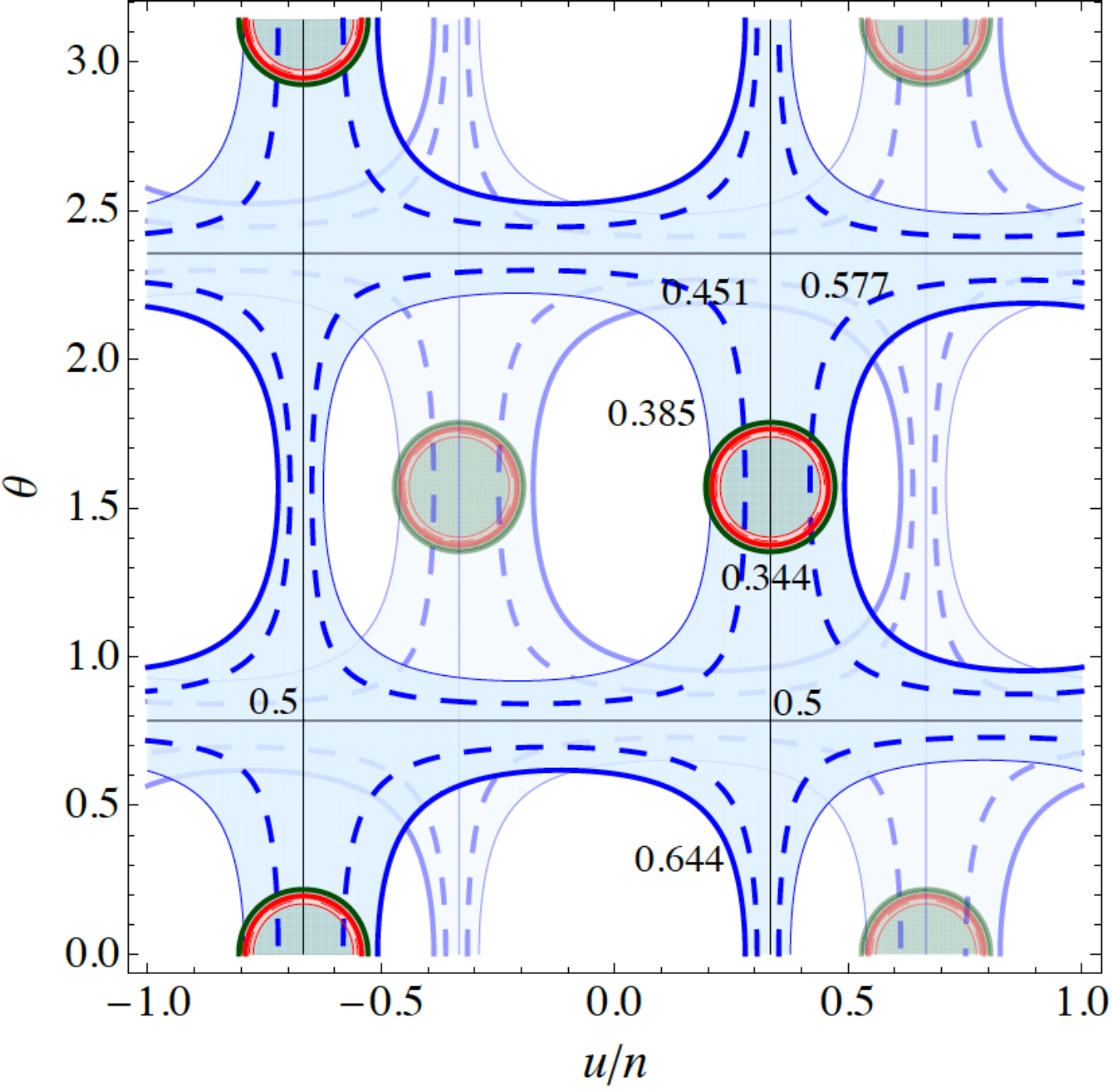}
\end{tabular}
\caption{\small{\textbf{Case 2)}. Contour plots of $\sin^{2}\theta_{ij}$ in the plane $\theta$ versus $u/n$. 
 The blue, green and red contour lines are associated with the atmospheric, solar and reactor mixing angles, respectively.
The thick (thin) plain lines represent the upper (lower) $3\,\sigma$ bounds of the lepton mixing angles, while the dashed lines refer to the corresponding best fit values. 
The $3 \, \sigma$ colored  regions in the left panel are computed from (\ref{anglescase2}). In the right panel, regions in the foreground and background follow from a different 
permutation of the PMNS mixing matrix  $U_{PMNS, 2}$ that leads to mixing angles  given by the formulae in (\ref{anglescase2}) with replacements (\ref{shift1case2}) and (\ref{shift2case2}), respectively. The plain black lines in both panels indicate maximal atmospheric mixing $\theta_{23}=\pi/4$.}}
\label{Fig:1}
\end{center}
\end{figure}

Here we study numerically mixing angles and CP phases that are obtained for the choice  $(Q=a, Z= c^{n/2}, X= c^s d^t P_{23})$. 
Since the mixing angles depend on $u/n$ ($\phi_u$) and the continuous parameter $\theta$
we first display the contour regions for the $3\,\sigma$ intervals of $\sin^2\theta_{ij}$ as well as the contour lines for their experimental best fit values 
in the plane $\theta$ versus $u/n$ in figure \ref{Fig:1}, using the data from the global fit analysis given in \cite{nufit} and summarized in appendix~\ref{app12}. 
In this figure we can restrict the discussion to the interval $-1<u/n\leq 1$
($-\pi < \phi_u \leq \pi$), since the mixing angles depend on $\cos \phi_u$ and $\sin\phi_{u}$, see (\ref{anglescase2}). As one can clearly see, the tightest constraint on the
parameters $u/n$ and $\theta$ arises from the requirement to accommodate the reactor mixing angle within the experimentally preferred $3\,\sigma$
interval (red ring-shaped areas in figure \ref{Fig:1}). If this is the case, also the value of the solar mixing angle is within its $3\,\sigma$ range (green disk).
This is also almost always true for the atmospheric mixing angle whose experimentally preferred regions in the $u/n$-$\theta$ plane are indicated in blue.
We note that for the solar mixing angle only the upper $3 \,\sigma$ bound, $\sin^2 \theta_{12}=0.344$, is visible in the figure, since 
the trimaximal column of the PMNS mixing matrix $U_{PMNS, 2}$ constrains $\theta_{12}$ to fulfill $\sin^2\theta_{12} \gtrsim 1/3$.
As discussed in the preceding section, there are three mixing patterns that can be distinguished for $(Q=a, Z= c^{n/2}, X= c^s d^t P_{23})$ corresponding
to three different permutations of $U_{PMNS, 2}$. If no permutation is applied, the formulae in (\ref{anglescase2}) are obtained
for the mixing angles and these are used in the left panel of figure \ref{Fig:1}. In this case we confirm the analytical estimates that $|u/n|$ has to be small
and $\theta$ close to $0$ or $\pi$, see (\ref{conphiuth13case2}) and (\ref{constrphiucase2}), or $u/n$ close to $\pm 1$ and $\theta\approx\pi/2$,  
if the symmetries in table \ref{tab:case2symmetries}
are taken into account. If we instead apply the permutation $P_1$ or $P_2$ to $U_{PMNS, 2}$, as described above, and thus obtain the formulae in (\ref{anglescase2})
with replacements (\ref{shift1case2}) or (\ref{shift2case2}), the corresponding figure of $\sin^2 \theta_{ij}$ in the plane $\theta$ versus $u/n$ is the one in the right panel of figure \ref{Fig:1}.
The shifts in the parameters $u/n$ and $\theta$ are clearly visible from figure \ref{Fig:1} and also the analytical estimates of $|u/n|$ that leads to a viable fit of the 
experimental data, see (\ref{constrphiushiftcase2}), are confirmed. The other regions in the $u/n-\theta$ plane that are indicated to accommodate the data well are, as expected,
related to the former region through symmetry transformations found in table \ref{tab:case2symmetries}. We would like to emphasize that the figures in the left and the right panel of
figure \ref{Fig:1} do lead in general to different results for mixing angles, simply because the shift $\pm \frac n3$ in the integer parameter $u$ that is necessary to relate these two figures
is not an integer for $3 \nmid n$. 

\begin{table}[t!]
\centering
\begin{tabular}{c}
$
\begin{array}{|l||c|c|c|c|c||c|}
\hline
 n& 8  & 10 &  14   & 16  & 20 & \text{even} \\
 \hline
 u & \mp1 &\mp1  & \mp1  &\mp1  & \mp1 & 0~(n)\\
\hline
 \chi^{2}_{\rm tot} & \ba26.4\\(23.9)\ea & \ba11.1\\(9.61)\ea & \ba9.60\\(9.55)\ea & \ba9.43\\(9.79)\ea & \ba9.40\\(10.3)\ea & 10.1 \\
\hline
  \theta _{\text{bf} } & 0 & 0.0932 & 0.144  & 0.154  & 0.165 & 2.96~(1.75)\\
\hline
 \sin^{2}\theta_{12} & 0.342 & 0.341 & 0.341  & 0.341  & 0.341 & 0.341 \\
\hline
 \sin^{2}\theta_{13} & 0.0254 & 0.0218  & 0.0218 & 0.0218 & 0.0218 & 0.0218 \\
\hline
 \sin^{2}\theta_{23} & \ba0.387\\(0.613)\ea & \ba0.410\\(0.590)\ea & \ba0.437\\(0.563)\ea & \ba0.445\\(0.555)\ea & \ba0.456\\(0.544)\ea & 1/2 \\
\hline
 J_{CP} & 0 & -0.0178  & -0.0274  & -0.0292 & -0.0311 & 0.0342 \\
\hline
 \sin\delta  & 0 & -0.529  & -0.807  & -0.858  & -0.913 & 1 \\
\hline
 I_2 & 0 & \mp 0.0121 & \mp 0.0137 & \mp 0.0129  & \mp 0.0111 & 0 \\
\hline
 \sin\beta  & 0 & \mp 0.861  & \mp 0.976 & \mp 0.917 & \mp 0.790 & 0 \\
\hline
\end{array}
$
\end{tabular}
\caption{\label{tab:caseu}{\small \textbf{Case 2)}. Results for fixed values of $n$ and $u$.
Expressions for $\sin^{2}\theta_{ij}$, $J_{CP}$ and $I_{2}$ are taken from (\ref{anglescase2}) and (\ref{CPinvcase2}) with $k_{1}=k_{2}=0$.
For all cases presented $\chi^2_{\rm tot} \lesssim 27$ and the mixing angles lie in their experimentally preferred $3\,\sigma$ intervals.
A second solution with the same $\chi^{2}_{\rm tot}$ is obtained in each case for $\theta=\pi-\theta_{\rm bf}$, however $J_{CP}$ and $I_{2}$ change sign.
Furthermore, different values of $u$ are obtained from the symmetry transformations in table~\ref{tab:case2symmetries}.
Notice that we do not display $I_1$ and the Majorana phase $\alpha$ since they depend also on $\phi_v$ ($v/n$), see (\ref{CPinvcase2}),
and thus several different values of $\sin\alpha$ can be achieved for a particular choice of $n$ and $u$, see (\ref{sinan8u1case2}) for example
and the plot in the bottom-left panel in figure \ref{Fig:4}. The trivial Dirac phase $\delta$ for $n=8$ and $u=\mp 1$ is related to an accidental
CP symmetry $\tilde{Y}$, see (\ref{Ytcase2}), that arises in this case since $\theta_{\rm bf}$ is zero. Additionally, $\sin\beta$ vanishes in this case,
see (\ref{YtMajoranacase2}). However, the other Majorana phase $\alpha$ is in general non-zero, see (\ref{sinan8u1case2}).
Here and in the following tables lower signs, if present, refer to the values given in parentheses.}}
\end{table}

With a $\chi^2$ analysis that includes the three mixing angles, uses the global fit results found in \cite{nufit}, see also appendix \ref{app12}, 
and that is described in detail in appendix \ref{app13}, we evaluate for even $n \leq 20$, $3 \nmid n$, and all corresponding values of $u$ whether  
the continuous parameter $\theta$ can take values such that a good fit to the experimental data ($\chi^2_{\rm tot} \lesssim 27$ and all mixing angles
within their $3\,\sigma$ intervals) can be achieved. Our results of such an analysis using the formulae in (\ref{anglescase2}) for the mixing angles 
are summarized in table \ref{tab:caseu} where we list for each case the values of $n$, $u$, the resulting $\chi^2_{\rm tot}$ obtained for the ``best fitting" value
$\theta=\theta_{\rm bf}$, the results for the mixing angles $\sin^2\theta_{ij}$ for this set of parameters as well as the values of the CP invariants $J_{CP}$
and $I_2$ and the corresponding CP phases $\sin\delta$ and $\sin\beta$. The results for $I_1$ and thus $\sin\alpha$ are not reported in this table, since
these quantities depend on an additional parameter $\phi_v$ ($v/n$), and are discussed in detail below, see (\ref{sinan8u1case2}) and the plot
in the bottom-left panel in figure \ref{Fig:4}. A second best fitting value for $\theta$ that leads to exactly the same results for the mixing angles is 
found at $\theta=\pi-\theta_{\rm bf}$, since the formulae of the mixing angles in (\ref{anglescase2}) remain invariant, if $\theta$ is replaced by $\pi-\theta$.

As can be seen from table \ref{tab:caseu}, $n$ has to be at least $8$ or $u$ has to be chosen as $u=0 \, (n)$, see discussion
around (\ref{anglesu0case2}) and below for the latter case. For the smallest value of $n$, $n=8$, the requirement to accommodate the mixing angles
well ($\chi^2_{\rm tot} \lesssim 27$) leads to $\theta_{\rm bf}=0$ that implies the presence of an accidental CP 
symmetry in the theory, see (\ref{Ytcase2}), such that the Dirac phase is trivial.
Then also one of the Majorana phases $\beta$ becomes trivial, see (\ref{YtMajoranacase2}). Nevertheless, the remaining Majorana phase $\alpha$ is in 
general non-trivial and for $n=8$ and $u=\mp 1$ it can take several values, since it also depends on $\phi_v$ ($v/n$), see (\ref{CPinvcase2}). Using the definition
of $u$ and $v$, see (\ref{defuv}), and the information that $0 \leq s, t \leq 7$ we find that 
$v$ can be 3, 9, 15 or 21 and the corresponding values of $\sin\alpha$ read (for $k_1=0$) 

\vspace{-0.15in}
\small
\begin{equation}
\label{sinan8u1case2}
\sin\alpha = -\sin \, (3 \pi/8) =-\sin\, (21\pi/8) \approx -0.924 \;, \; \sin\alpha = -\sin \, (9\,\pi/8)  = -\sin \, (15\,\pi/8) \approx 0.383 \; .
\end{equation}
\normalsize
%
\begin{figure}[t!]
\begin{center}
\begin{tabular}{cc}
\includegraphics[width=0.48\textwidth]{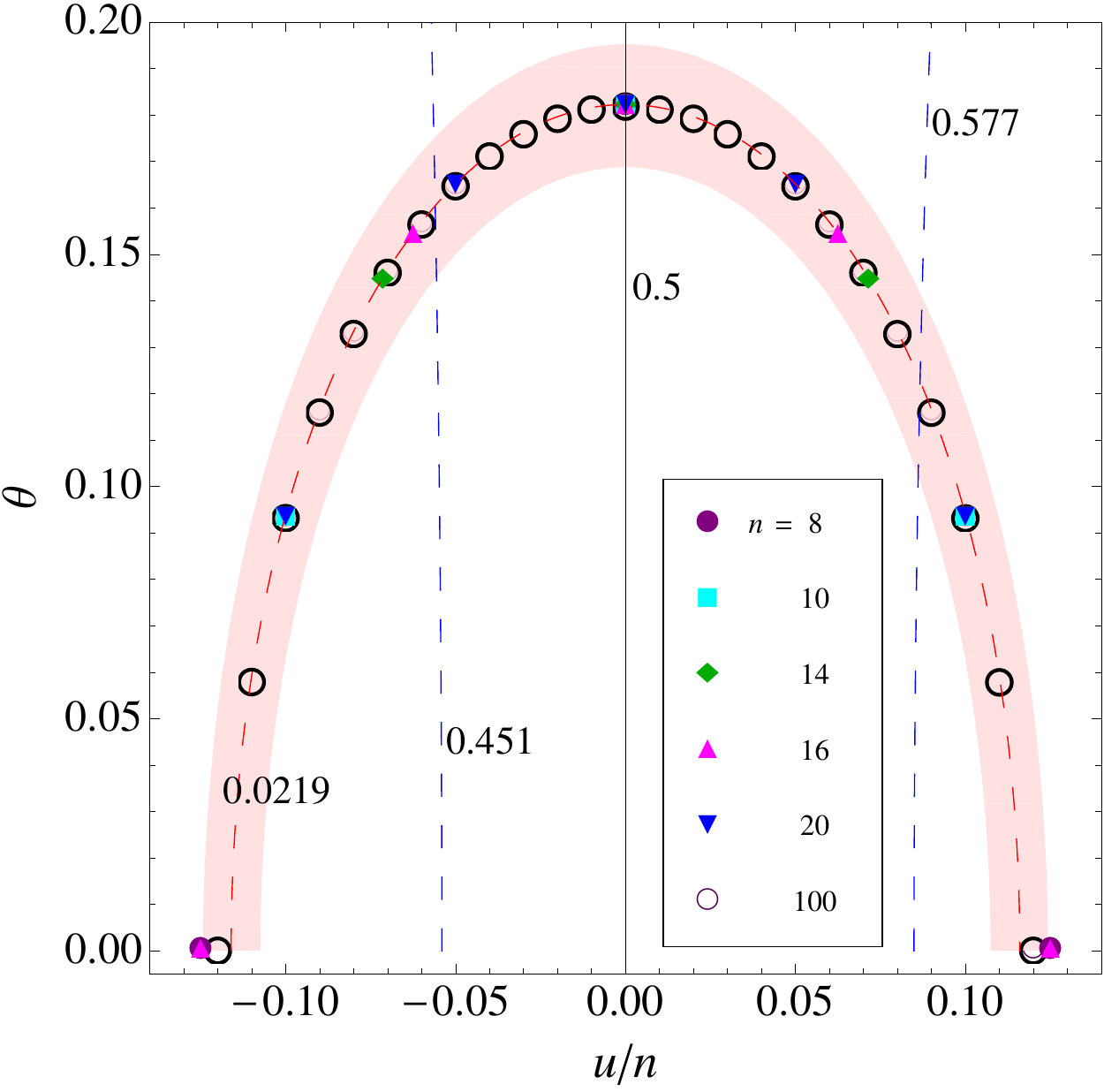} &
\includegraphics[width=0.48\textwidth]{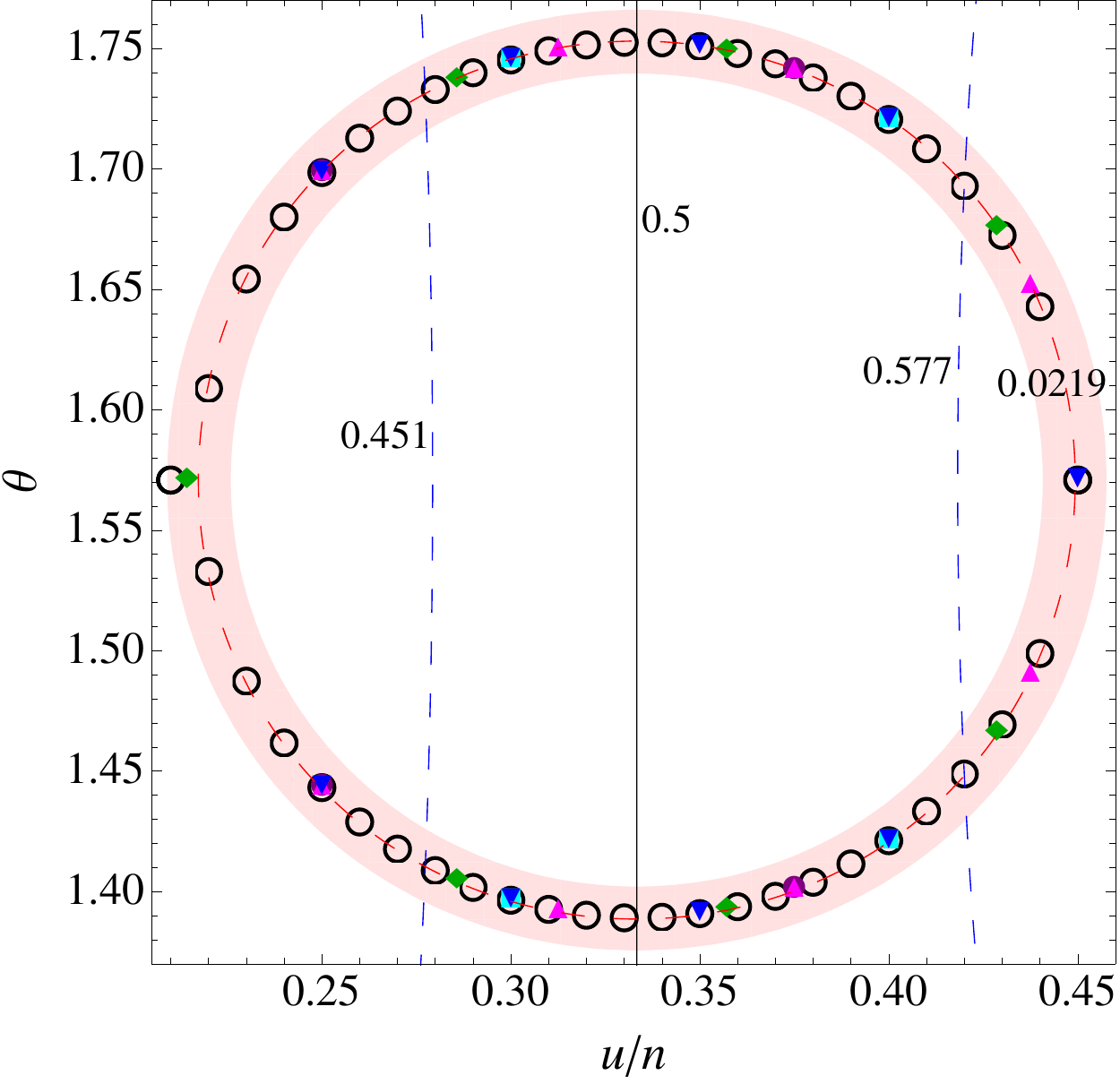}\\
\end{tabular}
\caption{\small{\textbf{Case 2)}. Pairs ($u/n$, $\theta_{\rm bf}$) that predict the lepton mixing angles in agreement with the 
experimental data, resulting from our $\chi^{2}$ analysis. The two plots correspond to two independent permutations of the PMNS mixing matrix.
 The red ring-shaped region and the contour lines in the left and right panels are extracted 
from figure~\ref{Fig:1}. The discrete points in the plane of the left (right) panel 
can be found  in table \ref{tab:caseu} (tables \ref{tab:caseushiftpminus} and \ref{tab:caseushiftpminus2}) for $n \leq 20$, 
taking into account the second solution with $\theta=\pi-\theta_{\rm bf}$ 
in the various cases.
Notice that  the red ring-shaped region is deformed in the plot on the left due to the  scales chosen for the axes.
}}
\label{Fig:2}
\end{center}
\end{figure}
%
As already mentioned above, cases in which $\theta$ vanishes reveal the same mixing pattern as the one obtained in case 1), if the identifications in (\ref{reducecase2})
are made. The case $n=8$ and $u=\mp1$ shows a characteristic feature common to the other cases, namely $u=-1$ entails $\theta_{23}$ smaller than $\pi/4$, while
values larger than $\pi/4$ are achieved for $u=1$, see table \ref{tab:caseu}. If smaller values of the index $n$ of the flavor group (and thus smaller groups) are 
desired, a possibility that, independently of the value of $n$, always admits a reasonable fit to the experimental data is to choose the parameter $u=0$ ($\phi_u=0$), i.e. require
a certain form of the CP transformation $X$, $X=c^s d^{2 s} P_{23}$, or, related by the symmetries in table \ref{tab:case2symmetries}, $u=n$ ($\phi_u=\pi$).
As shown
in (\ref{anglesu0case2}), this case always entails maximal atmospheric mixing, contributing $\chi^2_{23}\approx 0.69$ to $\chi^2_{\rm tot}$,
while the other two mixing angles can be accommodated equally well as in the other cases with $u\neq0$, see table \ref{tab:caseu}. Furthermore, the Dirac phase is fixed 
to its maximal value; for the value of $\theta_{\rm bf}$ displayed in table \ref{tab:caseu} $\sin\delta$ is positive, while $\sin\delta=-1$ is obtained for the choice
$\theta=\pi-\theta_{\rm bf} \approx 0.18 \, (1.39)$. One of the Majorana phases is trivial, $\sin\beta=0$, while the other one is determined by the parameter $v/n$:
$\sin\alpha=-\sin\phi_v=-\sin (\pi v/n)$ for $k_{1}=0$, see above. In particular, for the smallest value of the index $n$, $n=2$, also this phase is trivial, since the only possible value of 
$v$ is $v=0$. This feature has already been observed in \cite{S4CPgeneral}. For the next smallest choice $n=4$, also already known in the
literature \cite{Delta48CP,Delta96CP}, the Majorana phase $\alpha$ is either trivial (for $v=0$) or maximal ($\sin\alpha=-1$ for $v=6$; taking $k_1=1, \, 3$ also
$\sin\alpha=1$ can be achieved). Clearly, for larger values of $n$ also other values of $\alpha$ can be achieved that all lie on the curve displayed in the bottom-left panel in figure \ref{Fig:4}.
Larger values of $n$, $n>8$, all allow for $u=\mp1$ a reasonable fit to the experimental data of the lepton mixing angles and, at the same time, in general predict
non-trivial CP phases. Results corresponding to different choices of the parameter $u$ can be derived by applying the symmetry transformations reported in table ~\ref{tab:case2symmetries}. These all lead to the same results for the mixing parameters that are shown in table \ref{tab:caseu}.\footnote{We note that the symmetry 
transformations in table \ref{tab:case2symmetries} are exact as regards the analytic formulae shown in (\ref{anglescase2})
and (\ref{CPinvcase2}). However, when performing the $\chi^2$ analysis very minor differences in the results might be obtained, in particular if a symmetry transformation entails
that $\sin^2 \theta_{23}$ becomes $1-\sin^2\theta_{23}$. This happens because the best fit value as well as the $1 \, \sigma$ and $3 \, \sigma$ errors of $\sin^2\theta_{23}$ are not
(completely) symmetric with respect to $\sin^2\theta_{23}=1/2$ \cite{nufit}. Similar statements hold also for the numerical analysis of case 3 a) as well as of case 3 b.1).}
%
\begin{table}[t!]
\centering
\begin{tabular}{c}
$
\begin{array}{|l||c||c||c|c|}
\hline
n  & 4 &   8 &  \multicolumn{2}{c|}{10} \\
\hline
 u & 1 & 3 & 3 & 4   \\
\hline
 \chi^{2}_{\rm tot} & 10.0~[9.4] & 10.7~[9.44] & 9.51~[11.3] & 9.67~[9.49]  \\
\hline
  \theta _{\rm bf } & 1.70 & 1.40 & 1.40 & 1.72  \\
\hline
 \sin^{2}\theta_{12} & 0.341 & 0.341 & 0.341 & 0.341   \\
\hline
 \sin^{2}\theta_{13} & 0.0218 & 0.0218 & 0.0218 & 0.0218 \\
\hline
 \sin^{2}\theta_{23} & 0.426~[0.574] & 0.536~[0.464] & 0.471~[0.529] & 0.559~[0.441]  \\
\hline
 J_{CP} & \pm0.0243 & \mp0.0321 & \mp0.0329 & \pm0.0284   \\
\hline
 \sin\delta  & \pm0.718 & \mp0.941 & \mp0.963 & \pm0.835  \\
\hline
 I_2 & 0.014 & 0.0096 & -0.0079 & -0.0133   \\
\hline
 \sin\beta & 0.998 & 0.683 & -0.562& -0.949   \\
\hline
\end{array}
$
\end{tabular}
\caption{\label{tab:caseushiftpminus}{\small \textbf{Case 2)}. Results  for $n=4$, 8 and 10 obtained with the PMNS mixing matrix $U_{PMNS,2}$ whose
rows are cyclicly permuted with $P_1$, see (\ref{P1}), i.e. $u$ is replaced by $u-\frac n3$ in (\ref{anglescase2}) and (\ref{CPinvcase2}).
Again, we set $k_{1}=k_{2}=0$. The values in the square brackets, as well as the opposite sign of $J_{CP}$ ($\sin\delta$), 
are valid for the mixing pattern resulting from an additional permutation of the second and third rows of the PMNS mixing matrix. 
The mixing angles are accommodated to the same values as reported, if $\theta=\pi-\theta_{\rm bf}$ is used instead; clearly, the same value
of $\chi^2_{\rm tot}$ is achieved. Only $J_{CP}$ and $I_{2}$ in the table change sign.
Applying the symmetry transformations in table \ref{tab:case2symmetries} and taking into consideration the comments above additional values of $u$
are found that lead to the same fits.
 Numerical values of $\sin\alpha$ are displayed in figure~\ref{Fig:4} (bottom-right panel).} }
\end{table}

The results presented in table \ref{tab:caseu} together with the results for $n=100$ (empty circles) are displayed in the plane $\theta$ versus $u/n$ in figure \ref{Fig:2}, restricting the range of $u/n$ to $|u/n| \lesssim 0.12$, as estimated in (\ref{constrphiucase2}). These are superimposed with the red ring-shaped area indicating the $3\,\sigma$ interval of the reactor
mixing angle. The results for $n=100$ are shown in figure \ref{Fig:2} in order to improve the figure and to indicate the limit of large $n$. 
The Dirac phase $\sin\delta$ and the Majorana phase $\sin\beta$ are shown as functions of $u/n$ for $-1 < u/n \leq 1$ in the left panels of figures \ref{Fig:3} and \ref{Fig:4}.
They are computed for all the pairs ($u/n$, $\theta_{\rm bf}$) shown in figure \ref{Fig:2}. The possible values of the other Majorana phase $\sin\alpha$
are shown in the lower panel of figure~\ref{Fig:4}. We show the predictions of $\sin\alpha$ obtained for the cases reported in table~\ref{tab:caseu}, i.e. the value 
of $u$ is chosen as $\mp1$ and $\theta=\theta_{\rm bf}$, as well as for $n=100$, $u=\mp1$ and the corresponding $\theta_{\rm bf}$
 so that $v/n$ remains as variable, see (\ref{CPinvcase2}). As its fundamental interval we consider $0\leq v/n\leq 2$, since $I_1$ is a periodic function in $\phi_v=\pi v/n$
 with periodicity $2 \, \pi$. However, notice that for each $n$ some of the allowed values of $\sin\alpha$ are actually obtained for 
values of $v$ in the interval $2<v/n<3$, as it happens for example for $n=8$, $u=\mp1$ in (\ref{sinan8u1case2}).

In exactly the same manner we can discuss the results for the mixing originating from the permutation of the PMNS mixing matrix $U_{PMNS, 2}$
that leads to the $3\,\sigma$ allowed regions displayed in the foreground in the right panel of figure \ref{Fig:1}, i.e. the mixing angles and CP
invariants obtained from (\ref{anglescase2}) and (\ref{CPinvcase2}) with $u$ shifted into $u-\frac{n}{3}$.
The outcome of our analysis for $n \leq 20$ is collected in tables \ref{tab:caseushiftpminus} and \ref{tab:caseushiftpminus2}. The numbers mentioned in square brackets are obtained, if the second and third rows of the PMNS mixing matrix are exchanged, and represent a solution with the atmospheric mixing angle in the other octant (and 
$J_{CP}$ changes its sign). Additionally,
as in the case above, we find a further best fitting value $\theta$ at $\theta=\pi-\theta_{\rm bf}$ in each case that leads to the same mixing angles. 
%
\begin{figure}[t!]
\begin{center}
\begin{tabular}{cc}
\includegraphics[width=0.48\textwidth]{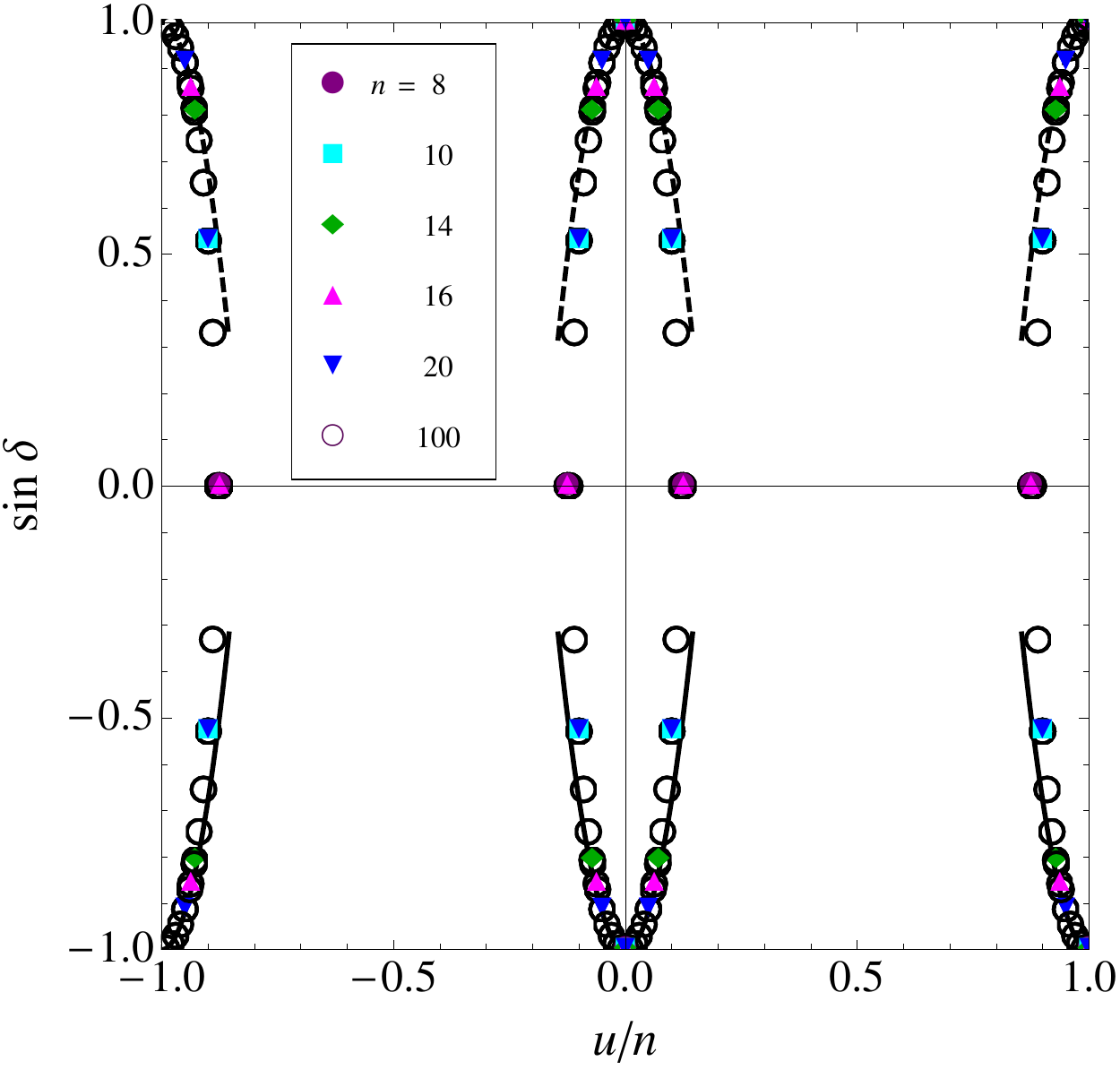} &
\includegraphics[width=0.48\textwidth]{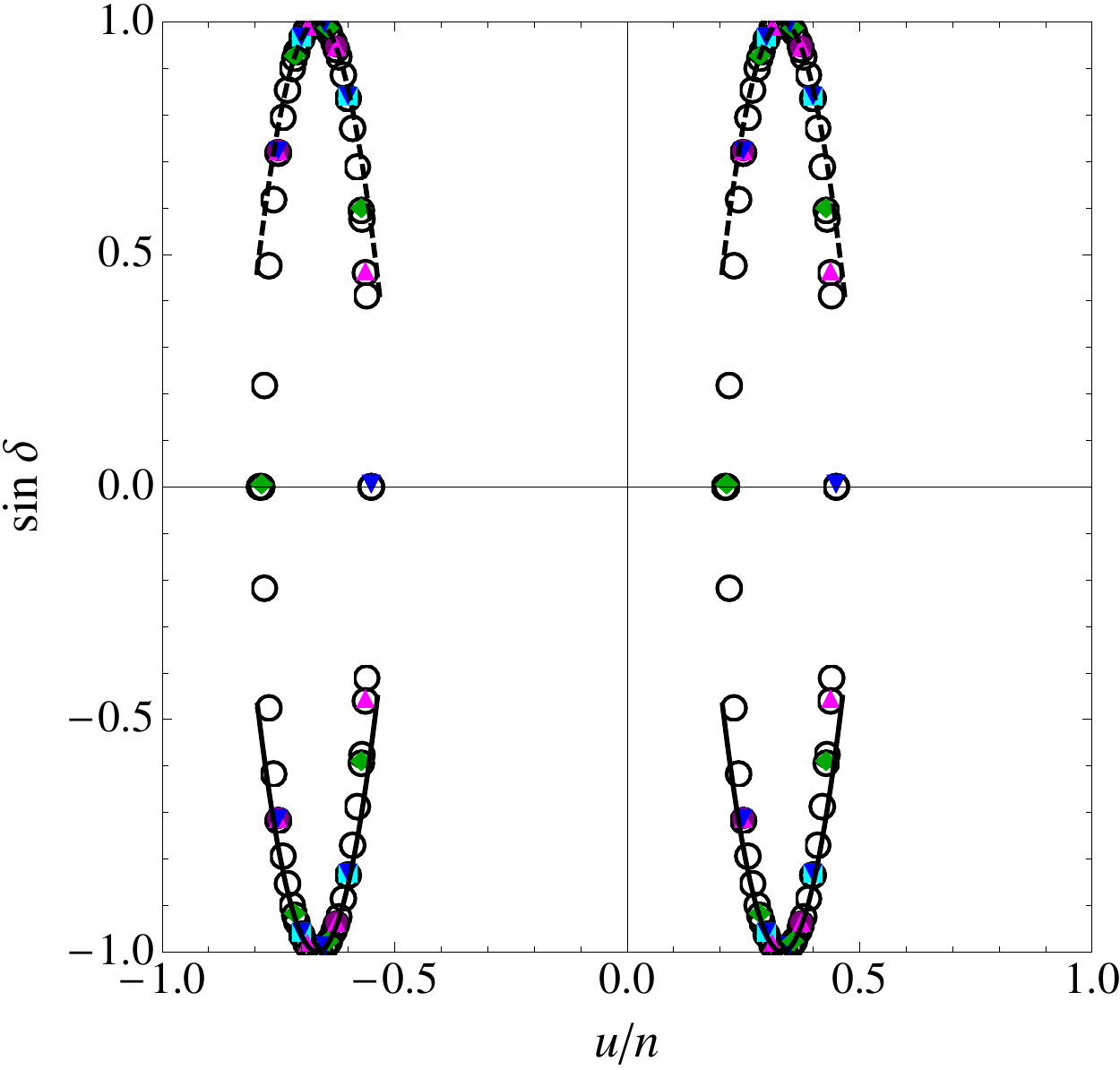}
\end{tabular}
\caption{\small{\textbf{Case 2)}. Predictions for the Dirac phase corresponding to different choices of $n$. 
The values of $\sin\delta$ are computed from the pairs ($u/n$, $\theta_{\rm bf}$) shown in figure~\ref{Fig:2},
which are obtained from our $\chi^{2}$ analysis.  The plots on the left and right panels are related to two independent permutations of the PMNS
mixing matrix in (\ref{Unucase2}). The analytic approximations represented by the dashed and continuous lines are found in (\ref{sindbapprox}) and refer to $\theta_{\rm bf} \gtrless \pi/2$.}}
\label{Fig:3}
\end{center}
\end{figure}
%
The smallest value of $n$
that allows a reasonable fit to the experimental data for this type of mixing pattern is $n=4$ and $u=1$. All CP phases  are non-trivial in this case.
In particular, the Majorana phase $\alpha$ reads for the values $v=3 \, t$, see (\ref{defuv}), admitted by the 
constraint $u=1$,
\begin{equation}
\label{sinan4u1shiftcase2}
\sin\alpha \approx 0.731 \;\; (v=3 \, , \, t=1) \;\;\; \mbox{and} \;\;\; \sin\alpha \approx 0.683 \;\; (v=9 \, , \, t=3) \; .
\end{equation}
These results are in agreement with those found in \cite{Delta48CP,Delta96CP}. Let us focus on the two particular cases $n=14$, $u=3$ and $n=20$, $u=9$
in table \ref{tab:caseushiftpminus2} that both lead to predictions $\sin\delta=0$ and $\sin\beta=0$. This result is obtained, since the best fitting value $\theta_{\rm bf}$
is in both cases $\pi/2$. For this value, as discussed above, the accidental CP symmetry of the charged lepton sector is also a CP symmetry of the neutrino mass matrix
combination $m_\nu^\dagger m_\nu$. This explains $\sin\delta=0$. The fact that also the Majorana phase $\beta$ is trivial is due to the special form of CP transformation
$\tilde{Y}$ in the neutrino mass basis, see (\ref{Ytcase2}) and (\ref{YtMajoranacase2}). Instead the Majorana phase $\alpha$ takes in both cases only non-trivial
values $|\sin\alpha|=|\sin \phi_v|=|\sin \pi v/n|$. In particular, for $n=14$ and $u=3$ we find (for $k_1=0$)
\begin{equation}
\label{sinan14u3case2shift}
\sin\alpha \approx 0.623 \;\; , \;\; \sin\alpha \approx 0.901 \;\; , \;\; \sin\alpha \approx -0.223 \;\;\; \mbox{and} \;\;\; \sin\alpha=-1
\end{equation}
valid for $v=3, \, 39$, $v=9, \, 33$, $v=15, \, 27$ and the maximal value of the Majorana phase $\alpha$ is attained for $v=21$. Notice that $\sin\alpha=1$ cannot
be achieved, simply because the parameter $v$ is always constrained to be divisible by three, see its definition in (\ref{defuv}).
Likewise, we find for $n=20$ and $u=9$ also always a non-vanishing value for the CP phase $\alpha$. Again, we set $k_1=0$
and achieve
\begin{equation}
\label{sinan20u9case2shift}
\sin\alpha \approx 0.454 \;\; , \;\; \sin\alpha \approx 0.988 \;\; , \;\; \sin\alpha \approx 0.707 \;\; , \;\; \sin\alpha \approx -0.156 \;\; \mbox{and} \;\; \sin\alpha \approx -0.891
\end{equation}
valid for $v=3, \, 57$, $v=9, \, 51$, $v=15, \, 45$, $v=21, \, 39$ and $v=27, \, 33$.
Numerical values of $\sin\alpha$  for all choices of $n$ and $u$ reported in tables~\ref{tab:caseushiftpminus} and \ref{tab:caseushiftpminus2} are displayed in figure~\ref{Fig:4} (bottom-right panel). Similarly to the mixing pattern derived from $U_{PMNS, 2}$ in (\ref{Unucase2}) also in this case additional values of $u$ that lead to the 
same results for the mixing angles are found, if the symmetry transformations in table  \ref{tab:case2symmetries} are applied. However, note that the second and third
transformations now relate the pattern with $u$ shifted into $u-\frac n3$ to the one with $u$ shifted into $u+\frac n3$. Again, we show in figure \ref{Fig:2} all pairs ($u/n$, $\theta_{\rm bf}$), this
time in the right panel of the figure, that reproduce the experimental data on the lepton mixing angles well, for $8\leq n \leq 20$ and $n=100$. As in the left panel of figure 
\ref{Fig:2} we restrict the interval of $u/n$ to the one estimated above, $0.21 \lesssim u/n \lesssim 0.46$, since this embraces all solutions $u/n \approx 1/3$
that allow a reasonable fit to the experimental data. The (upper) plot on the right in figure \ref{Fig:3} and \ref{Fig:4} is obtained in the analogous way as those on the left for the other permutation of the PMNS mixing matrix. Instead for the figure of $\sin\alpha$ on the bottom-right in figure \ref{Fig:4} the values of $u$ used for $n=100$ are now three, namely
$u=23$, 29 and 31, each of them leading to a set of fifty different values of $\sin\alpha$. 
Notice again that some of them lie in the interval $2 < v/n < 3$ that we do not report in the figure.

\begin{figure}[t!]
\begin{center}
\begin{tabular}{cc}
\includegraphics[width=0.48\textwidth]{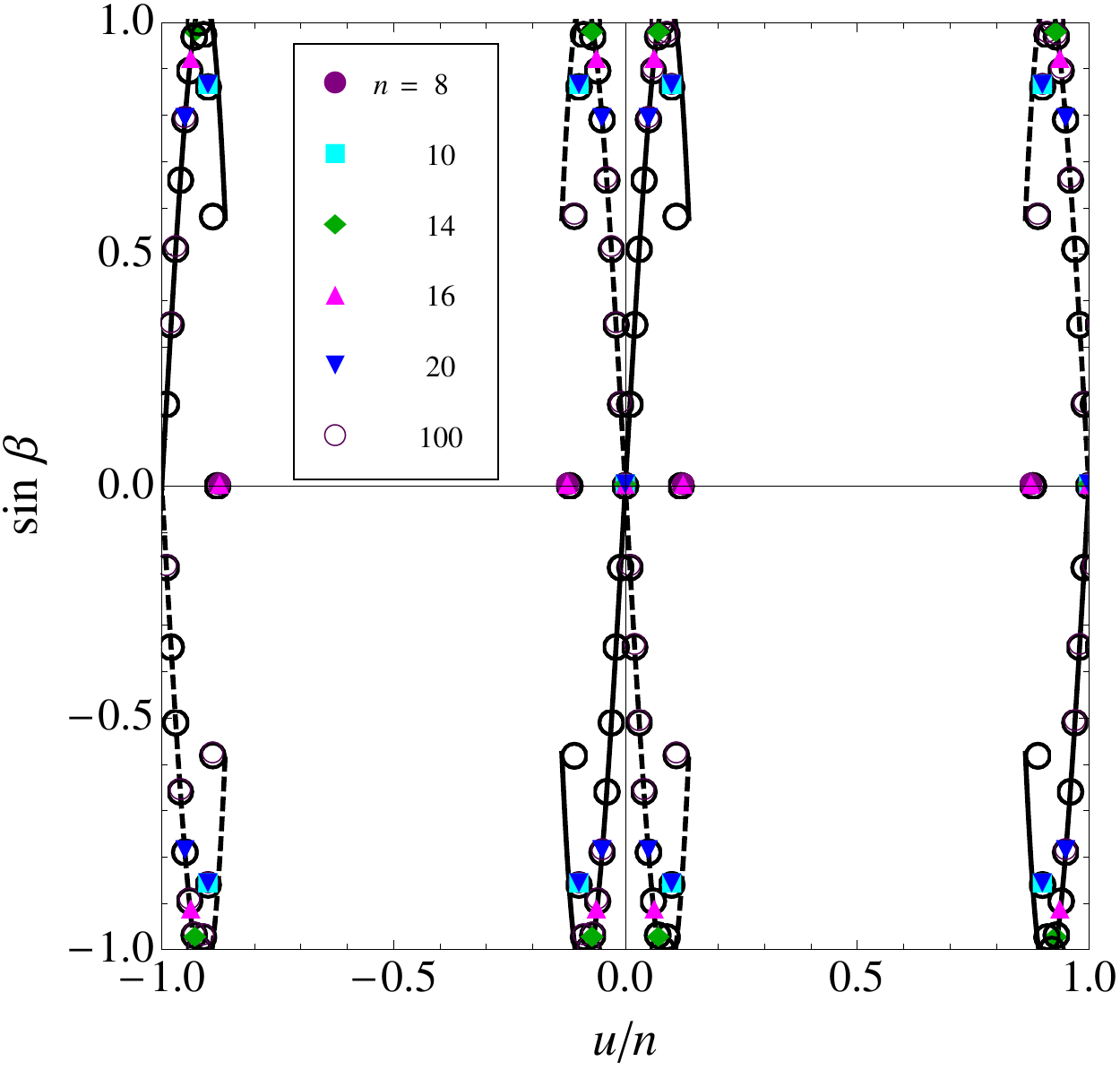} &
\includegraphics[width=0.48\textwidth]{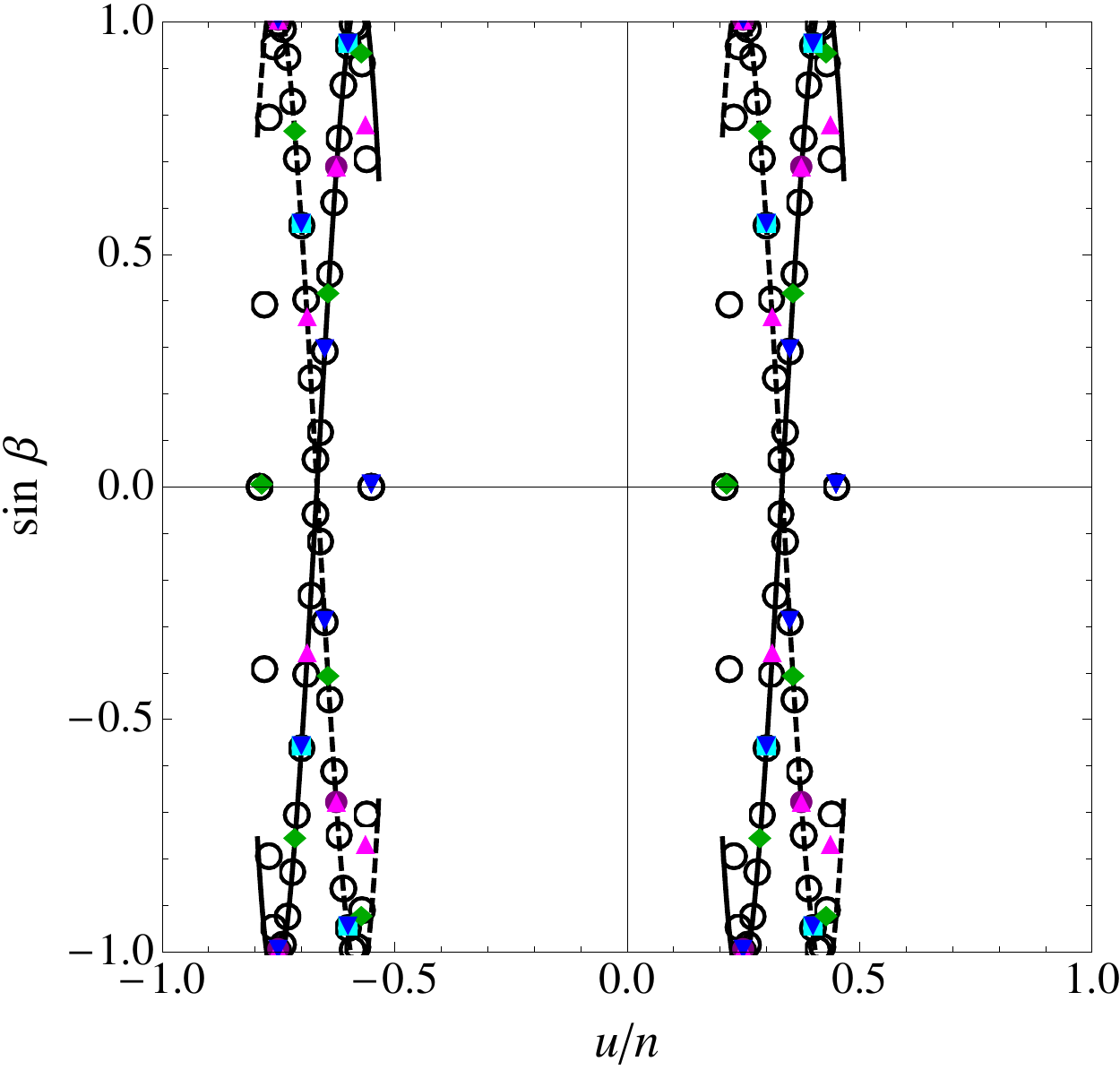}\\
\includegraphics[width=0.48\textwidth]{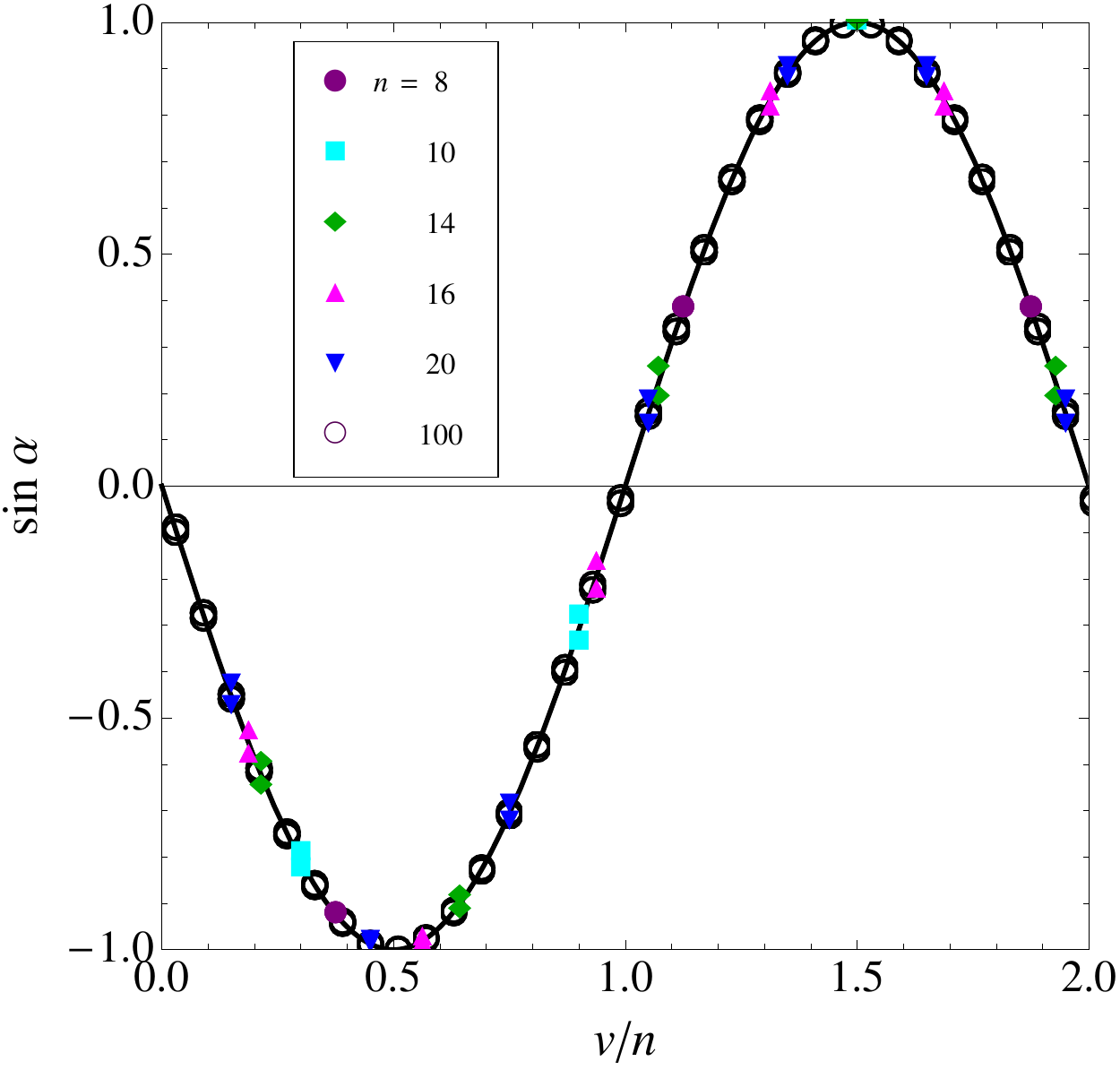} &
\includegraphics[width=0.48\textwidth]{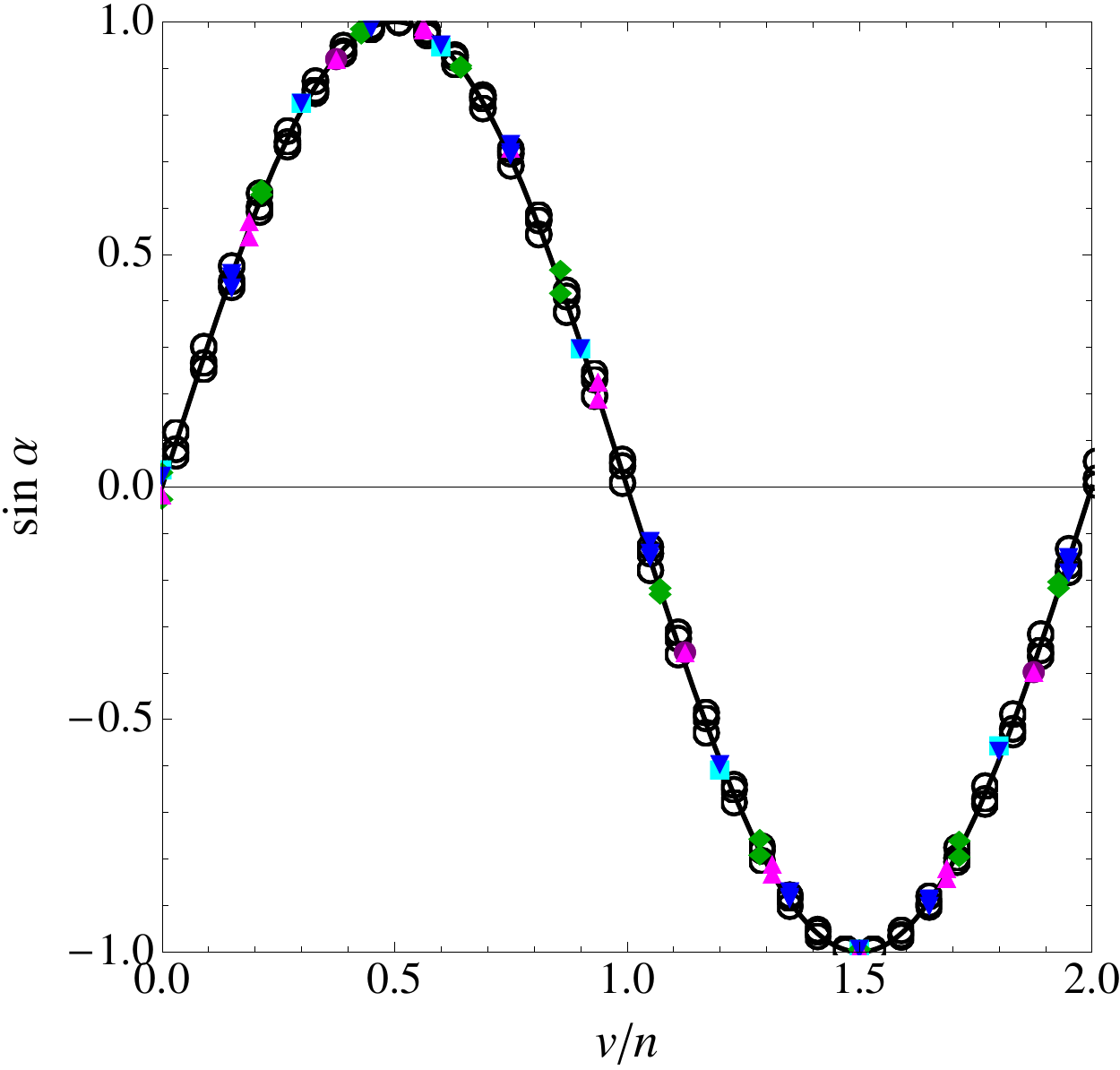}
\end{tabular}
\caption{\small{\textbf{Case 2)}. Predictions for the Majorana phases $\beta$ and $\alpha$ for different choices of $n$. 
The plots in the left and right panels are related to two independent permutations of the PMNS mixing matrix. 
In the upper panels $\sin\beta$ is displayed against $u/n$ for the pairs ($u/n$, $\theta_{\rm bf}$) shown in figure~\ref{Fig:2},
that result from our $\chi^{2}$ analysis. Similarly,
in the lower panels  $\sin\alpha$ against $v/n$ is presented with $\sin\alpha$ computed for the values of $\theta_{\rm bf}$ and $u$ reported 
in table~\ref{tab:caseu} (left lower panel) and in tables~\ref{tab:caseushiftpminus} and \ref{tab:caseushiftpminus2} (right lower panel).
In the case $n=100$, we set $u=\pm1$ ($u=23$, 29, 31) for the plot in the left (right) lower panel. 
The analytic approximations represented by the dashed and continuous lines in $\sin\beta$ are given in (\ref{sindbapprox}) and are valid for $\theta_{\rm bf} \gtrless \pi/2$.
For $\sin\alpha$ an excellent analytic approximation is found, see (\ref{sinalfaapprox}), that is indicated
by the continuous line in $\sin\alpha$.}}
\label{Fig:4}
\end{center}
\end{figure}


\begin{landscape}

\begin{table}[h!]
\centering
\begin{tabular}{c}
$
\begin{array}{|l||c|c|c|c||c|c||c|c|}
\hline
n  &  \multicolumn{4}{c||}{14} &  \multicolumn{2}{c||}{16} &  \multicolumn{2}{c|}{20}\\
\hline
 u & 3 & 4 & 5 & 6 & 5 & 7 & 7 & 9  \\
\hline
 \chi^{2}_{\rm tot} & 14.5~[12.2] & 9.41~[10.4] & 12.0~[9.62] & 9.48~[10.7] & 9.67~[12.2] & 9.77~[11.5] & 12.6~[9.74] & 10.6~[12.9]   \\
\hline
  \theta _{\rm bf} & \pi/2 & 1.40 & 1.75 & 1.47 & 1.39 & 1.49 & 1.39 & \pi/2 \\
\hline
 \sin^{2}\theta_{12} & 0.341 & 0.341 & 0.341 & 0.341 & 0.341 & 0.341 & 0.341 & 0.341   \\
\hline
 \sin^{2}\theta_{13} & 0.0230 & 0.0218 & 0.0218 & 0.0218 & 0.0218 & 0.0218 & 0.0218 & 0.0221 \\
\hline
 \sin^{2}\theta_{23}  & 0.392~[0.608] & 0.458~[0.542] & 0.521~[0.479] & 0.585~[0.415] & 0.482~[0.518] & 0.594~[0.406] & 0.514~[0.486] & 0.606~[0.394]   \\
\hline
 J_{CP} & 0 & \mp0.0314 & \pm0.0336 & \mp0.0200 & \mp0.0337 & \mp0.0155 & \mp0.0339 &  0  \\
\hline
 \sin\delta & 0 & \mp0.921 & \pm0.981 & \mp0.594 & \mp0.986 & \mp0.460 & \mp0.991 & 0 \\
\hline
 I_2  & 0 & -0.0107 & -0.0058 & 0.0130 & -0.0051 & 0.0109 & 0.0041 & 0  \\
\hline
 \sin\beta  & 0 & -0.761 & -0.411 & 0.928  & -0.362 & 0.774 & 0.291 & 0  \\
 \hline
\end{array}
$
\end{tabular}
\caption{\label{tab:caseushiftpminus2}{\small  \textbf{Case 2)}. Results for $n= 14$, 16 and 20 using the same permutation of the PMNS mixing matrix as in table~\ref{tab:caseushiftpminus}. The values in the square brackets as well as the opposite sign of $J_{CP}$ ($\sin\delta$)
are related to the mixing pattern resulting from an additional permutation of the second and third rows of the PMNS mixing matrix. 
Due to the properties of the formulae (\ref{anglescase2}) and (\ref{CPinvcase2})
the same good fit, i.e. the same $\chi^{2}_{\rm tot}$, is obtained for $\theta=\pi-\theta_{\rm bf}$, while $J_{CP}$ and $I_{2}$ change sign.
Additional values of $u$ giving rise to the same results for the mixing angles are found applying the symmetry transformations mentioned  
in table~\ref{tab:case2symmetries} and by taking into account the comments made above. 
For $n=14$ and $u=3$, the Dirac phase $\delta$ is trivial, since the value $\theta_{\rm bf}=\pi/2$ leads to an accidental CP symmetry, see (\ref{Ytcase2}).
Additionally, the Majorana phase $\beta$ is trivial. However, the other Majorana phase is in general non-vanishing, see (\ref{sinan14u3case2shift}).
Similarly, an accidental CP symmetry guarantees that $\sin\delta=0$ for $n=20$ and $u=9$ and additionally $\sin\beta=0$. Also in this case, the remaining CP phase
$\alpha$ is in general non-trivial, see (\ref{sinan20u9case2shift}).
Numerical values of $\sin\alpha$ for the choices of $n$ and $u$ in the table are displayed in figure~\ref{Fig:4} (bottom-right panel).}}
\end{table}

\end{landscape}

We end the discussion of case 2) by deriving approximate expressions for the sines of the CP phases that 
help to understand the distribution of the points in figures~\ref{Fig:3} and ~\ref{Fig:4}.
In fact, we can always express $\theta$ as a function of $\phi_{u}$ ($u/n$) and the best fit value of $\sin^{2}\theta_{13}$ determined 
from the global fit analysis, namely $(\sin^{2}\theta_{13})^{\rm bf}=0.0219$ \cite{nufit}, using (\ref{anglescase2}) (either for $u$ or for $u-\frac n3$).
Thus, we can write  $\sin\delta$ and $\sin\beta$, see (\ref{CPinvcase2}), in terms of $u/n$ only and can expand in the parameter $\phi_{u}$ (always setting $k_1=k_2=0$)
around $\bar{\phi}$ 
\begin{eqnarray}\label{sindbapprox}
	\sin\delta &\approx & \pm\,1\,\mp\,3.3\left(\phi_{u}\,-\,\bar{\phi}\right)^{2}\,,  \\
	\nonumber
	\sin\beta & \approx & \mp \,5.6\left(\phi_{u}\,-\,\bar{\phi}\right) \,\pm\,23\left(\phi_{u}\,-\,\bar{\phi}\right)^{3}\,,
\end{eqnarray}
with $\bar{\phi}=0,\,\pm\pi~(-2\pi/3,\,\pi/3)$. The different values of $\bar{\phi}$ 
correspond to the left (right) panel in figures~\ref{Fig:3} and \ref{Fig:4}. This approximation is reasonably good for $|\phi_{u}-\bar{\phi}|\lesssim 0.3$ in both cases.
The different signs in (\ref{sindbapprox}) refer to the different possible values of $\theta$, namely
the upper ones are valid for $\theta > \pi/2$ and the lower ones for $\theta< \pi/2$. These different solutions are represented with 
dashed and continuous lines, respectively, in figures \ref{Fig:3} and \ref{Fig:4}. The approximation of $\sin\delta$ nicely shows that a large
 CP phase can be achieved for small values of $u/n$ and for $u=0$ it becomes maximal, see also table \ref{tab:caseu}.
Likewise, we obtain for the sine of the Majorana phase $\alpha$ at leading order in $|\phi_{u}-\bar{\phi}|$
\begin{equation}\label{sinalfaapprox}
	 \sin\alpha \;\approx\; - \sin\phi_{v} \;\;\; \mbox{for} \;\;\; \bar{\phi}=0 \;\;\; \mbox{and} \;\;\; 
	 \sin\alpha \;\approx\;  \sin\phi_{v} \;\;\; \mbox{for} \;\;\; \bar{\phi}=\pi/3  \; .
\end{equation}
This approximation is shown as continuous line in the lower plots in figure~\ref{Fig:4} and, indeed, agrees very well with the 
numerical solutions in the whole range of $v/n$. The next term in the expansion contributing to $\sin\alpha$ reads
 $\pm\,0.18\,\cos\phi_{v}\,(\phi_{u}-\bar{\phi})$ for $\theta<\pi/2$, with the upper (lower) sign valid for $\bar{\phi}=0\,(\pi/3)$. In the case $\theta>\pi/2$ the sign should be reversed.
The form of the approximation in (\ref{sinalfaapprox}) coincides with the exact result for the Majorana phase $\alpha$ derived in case 1), see (\ref{sinalcase1}).
This approximation as well as figure \ref{Fig:4} show that large values of the Majorana phase $\alpha$ are achieved for $v/n\approx1/2$ and $v/n\approx3/2$,
while the choice $v/n\approx 0,\, 1, ...$ leads to small values of $\sin\alpha$.

\mathversion{bold}
\subsection{Case 3)  $(Q=a, Z= b \, c^m d^m, X= b \, c^s d^{n-s} P_{23})$}
\mathversion{normal}
\label{sec43}

The last case  can be represented by the choice
\begin{equation}
\label{ZXcase3}
Z=b \, c^m d^m \;\;\; , \;\;\; X=b \, c^s d^{n-s} P_{23}
\end{equation}
with $0 \leq m, s \leq n-1$. Since the $Z_2$ generator contains the element $b$, this case can only be realized, if the flavor symmetry is $\Delta (6 \, n^2)$.
First, we note that in the case $Z=b \, c^m d^m$ the eigenvector belonging to the non-degenerate eigenvalue is proportional to
\begin{equation}
\label{evcase3}
\frac{1}{\sqrt{6}}
\left( \begin{array}{c}
 -1+e^{2 \, i \, \phi_m} \\   -\omega^2+e^{2 \, i \, \phi_m} \\  -\omega+e^{2 \, i \, \phi_m} 
\end{array} \right)
\end{equation}
with
\begin{equation}
\label{phim}
\phi_m = \frac{\pi m}{n} \; .
\end{equation}
If this eigenvector is identified with the third column of the PMNS mixing matrix, the reactor as well as the atmospheric mixing angle are only determined
 by the ratio $m/n$. This case we call case 3 a) in the following and twelve possible permutations of the PMNS mixing matrix represent this situation.
 As we see, the smallness of the reactor mixing angle can be explained by small $m/n$ (or small $1-m/n$). However, the particular choice $m=0$ is excluded.
 If we consider instead $m=\frac n2$, the eigenvector in (\ref{evcase3}) takes  the special form 
\begin{equation}
\label{evcase3mn2}
\frac{1}{\sqrt{6}}
\left( \begin{array}{c}
 -2 \\   \omega \\  \omega^2 
\end{array} \right)
\end{equation}
whose components have the same absolute value as the ones of the first column of the TB mixing matrix, see (\ref{UTB}). Thus, such a vector can be identified with the 
first column of the PMNS mixing matrix. This is a particular choice in our case 3 b.1). As regards the mixing, we know that in this case the solar mixing angle is bounded
from above \cite{TB_first_column}
\begin{equation} 
\label{sinth12sqTM1}
\sin^2 \theta_{12} \lesssim \frac 13 \; .
\end{equation}
A possible choice of the matrix $\Omega$ that satisfies both equalities in (\ref{XZOmega}) for $Z$ and $X$ in (\ref{ZXcase3}) is
\begin{equation}
\label{Omegacase3}
\Omega_3 =\left(
\begin{array}{ccc}
1&0&0\\
0&\omega&0\\
0&0&\omega^2
\end{array}
\right) \, \Omega_1 \, R_{13} (\phi_m)
\end{equation}
with $\Omega_1$ as in (\ref{Omegacase1}). Note that $\Omega_1$ contains as parameter $s$ ($\phi_s$) that is constrained to be 
$0 \leq s \leq n-1$. Applying $\Omega_{3}$ to $Z$ we find
\begin{equation}
\label{ZafterOmega3}
\Omega_3^\dagger \, Z \, \Omega_3 = \left( \begin{array}{ccc}
1&0&0\\
0&1&0\\
0&0&-1
\end{array}
\right) \; .
\end{equation}
Thus, the appropriate rotation $R_{ij} (\theta)$ is in the (12)-plane. The neutrino mixing matrix takes the form
\begin{equation}
\label{Unucase3}
U_{\nu, 3} = \Omega_3 \, R_{12} (\theta) \, K_\nu \; ,
\end{equation}
as usual, up to permutations of its columns. Again, since $U_e=\mathbb{1}$, up to permutations of columns, the PMNS mixing matrix $U_{PMNS, 3}$ is of the same
form as $U_{\nu ,3}$, up to permutations of rows and columns. In this case, none of the 36 possible permutations can be obviously excluded and thus we study all of them
in the following. We can distinguish two types of mixing
\begin{itemize}
\item[a)] twelve permutations that lead to $\sin^2 \theta_{13}$ and $\sin^2 \theta_{23}$ depending on $n$ and $m$, but not on $s$ and $\theta$; 
for these permutations the third
column of the PMNS mixing matrix is identified with the eigenvector of $Z$ mentioned in (\ref{evcase3}) [up to permutations of its components]
 \item[b.1)] twelve permutations that lead to $\sin^2 \theta_{13}$ depending on $n$, $m$, $s$ as well as $\theta$; here the first column of $U_{PMNS, 3}$ is identified with
 the eigenvector in (\ref{evcase3})
 \item[b.2)] eventually, twelve permutations with the second column of the PMNS mixing matrix corresponding to (\ref{evcase3}).
 \end{itemize}
As we will show below,  case 3 b.2) cannot accommodate the experimental data on the mixing angles well. 

Before discussing the lepton mixing patterns analytically and also numerically we first comment on the possible 
presence of the accidental CP symmetry
$Y$ of the charged lepton sector, see (\ref{YQa}), in the neutrino sector: in general, there are two possible structures of $Y$ that fulfill the first equality in (\ref{XZY}) for $Z$ in (\ref{ZXcase3})
either
\begin{equation}
\label{Y1case3}
Y_1= e^{i y} \, \left(
\begin{array}{ccc}
 \omega^2 & 0 & 0\\
 0 & \omega & 0\\
 0 & 0 & 1
\end{array}
\right) \;\;\; \mbox{with} \;\;\; 0 \leq y < 2 \pi \; 
\end{equation}
or, if we take $m=0$, 
\begin{equation}
\label{Y2case3}
Y_2= \left(
\begin{array}{ccc}
 e^{i y_1} & 0 & 0\\
 0 & \omega \, e^{i y_2} & 0\\
 0 & 0 & e^{i y_2}
\end{array}
\right) \;\;\; \mbox{with} \;\;\; 0 \leq y_i < 2 \pi \; .
\end{equation}
However, the latter case is only of theoretical use, since $m=0$ cannot be chosen, if we want to accommodate the experimental data of all mixing angles well.

\mathversion{bold}
\subsubsection{Case 3 a)}
\mathversion{normal}
\label{sec431}

\mathversion{bold}
\subsubsection*{Analytical results}
\mathversion{normal}
\label{sec4311}

We first discuss the case in which the rows and columns of the PMNS mixing matrix in (\ref{Unucase3}) are not permuted. The mixing angles are found to be of the form
\begin{eqnarray}
\label{anglescase3a}
&& \sin^2 \theta_{13}= \frac 23 \, \sin^2 \phi_m  \;\;\; , \;\;\; \sin^2 \theta_{23}= \frac 12 \, \left( 1+ \frac{\sqrt{3} \sin 2 \, \phi_m}{2+\cos 2 \, \phi_m} \right) \;\; ,
\\ \nonumber
&& \sin^2 \theta_{12}=\frac{1+\cos 2 \, \phi_m \sin^2 \theta+\sqrt{2} \cos \, \phi_m \cos 3 \, \phi_s \sin 2 \theta}{2+\cos 2 \, \phi_m}
\end{eqnarray}
and for the CP invariants we find
\begin{eqnarray}
\label{CPinvcase3a}
&& J_{CP}= -\frac{1}{6 \sqrt{6}} \, \sin 3 \, \phi_m \sin 3 \, \phi_s \sin 2 \theta \;\; ,
\\ \nonumber
&& I_1= \frac{1}{9} \, (-1)^{k_1+1} \, \cos \phi_m \, \sin 3 \, \phi_s \, \left( 4 \cos \phi_m \cos 3 \, \phi_s \cos 2 \theta + \sqrt{2}  \, \cos 2 \, \phi_m \, \sin 2 \theta \right) \;\; ,
\\ \nonumber
&& I_2=\frac 49 \, (-1)^{k_2} \, \sin^2 \phi_m \sin 3 \, \phi_s \sin\theta \, \left( \cos 3 \, \phi_s \sin \theta - \sqrt{2} \cos \phi_m \cos \theta \right) \; .
\end{eqnarray}
All other permutations of the PMNS mixing matrix that leave the third column in its place
also give rise to this mixing pattern, if shifts in the continuous parameter $\theta$,
re-labeling of $k_1$ and $k_2$ as well as shifts in the integer parameter $m$ are taken into account. In particular, $m$ has to be shifted into $n-m$ and/or
$m \pm \frac n3$. Since $n$ is not divisible by three, the latter type of shifts in $m$ does lead in general to different results for the mixing angles and is thus treated
separately in our numerical analysis.\footnote{\label{foot15}Examples of permutations of the PMNS mixing matrix in (\ref{Unucase3}) which are related to the original
pattern by such shifts in $m$ are: the PMNS mixing matrix that is multiplied from the left with the matrix $P_1$ in (\ref{P1}) leads to mixing angles and CP invariants 
as in (\ref{anglescase3a}) and (\ref{CPinvcase3a}), if we replace in these formulae $m$ with $m+\frac n3$ and $\theta$ with $\pi-\theta$, while the PMNS mixing matrix
 in (\ref{Unucase3}) multiplied from the left with $P_2$ in (\ref{P2}) gives rise to mixing angles and CP invariants whose dependence from the parameters
$m$, $n$, $s$ and $\theta$ is obtained, if $m$ is replaced by $m-\frac n3$ and $\theta$ by $\pi-\theta$ in (\ref{anglescase3a}) and (\ref{CPinvcase3a}).} 
The shift of $m$ into $n-m$ also embraces the case in which the relative sign in the bracket of $\sin^2 \theta_{23}$ is changed\footnote{
This mixing pattern can be easily achieved by exchanging the second and third rows of the PMNS mixing matrix in (\ref{Unucase3}) and all mixing angles
and CP invariants can be obtained from (\ref{anglescase3a}) and (\ref{CPinvcase3a}) by replacing $m$ with $n-m$ and $\theta$ with $\pi-\theta$ in these formulae.}
 \begin{equation}
\label{sinth23sqaltcase3a}
\sin^2 \theta_{23}= \frac 12 \, \left( 1- \frac{\sqrt{3} \sin 2 \, \phi_m}{2+\cos 2 \, \phi_m} \right) \; .
\end{equation}
It is interesting to note that the formulae in (\ref{anglescase3a}) and (\ref{CPinvcase3a}) reveal certain symmetries that are collected in table \ref{tab:case3symmetries}.
Furthermore, we note that the solar mixing angle and CP invariants become even or odd functions in $\theta$ for $s=\frac{n}{2}$ (assuming $n$ is even), since 
terms with $\cos 3 \, \phi_s$ vanish. Mixing angles remain then the same, while the CP invariants change sign, if $\theta$ is replaced by $\pi-\theta$. 
If we apply one of the transformations that changes $m$ by $\pm \frac{n}{3}$ to (\ref{anglescase3a}) and (\ref{CPinvcase3a}), only the third symmetry in table \ref{tab:case3symmetries} remains intact, while the other two ones now relate results of mixing angles and CP invariants for $m-\frac{n}{3}$ to those obtained for 
$m+\frac{n}{3}$. This is very similar to what happens in case 2).

\begin{table}[t!]
\begin{center}
\begin{tabular}{|l|l|l|}
\hline
$\begin{array}{c}
m \; \rightarrow \;\; n-m\\
\left( \phi_m \; \rightarrow \;\; \pi-\phi_m \right)
\end{array}$ 
&$\;\;\;\;\;\theta \;\rightarrow \;\;\pi-\theta$&$\sin^2\theta_{13}$, $\sin^2 \theta_{12}$, $I_1$, $I_2$ are invariant\\
&&$\sin^2\theta_{23}$ becomes $1-\sin^2\theta_{23}$; $J_{CP}$  changes sign\\
\hline
$\begin{array}{c}
m \; \rightarrow \;\; n-m\\
\left( \phi_m \; \rightarrow \;\; \pi-\phi_m \right)
\end{array}$
&$\begin{array}{c}
s \; \rightarrow \;\; n-s\\
\left( \phi_s \; \rightarrow \;\; \pi-\phi_s \right)
\end{array}$
&$\sin^2 \theta_{13}$, $\sin^2 \theta_{12}$, $J_{CP}$ are invariant\\
&&$\sin^2\theta_{23}$ becomes $1-\sin^2\theta_{23}$; $I_1$ and $I_2$ change sign\\
\hline
$\begin{array}{c}
s \; \rightarrow \;\; n-s\\
\left( \phi_s \; \rightarrow \;\; \pi-\phi_s \right)
\end{array}$
&$\;\;\;\;\;\theta \; \rightarrow \;\; \pi-\theta$&$\sin^2 \theta_{ij}$ are invariant\\
&&$J_{CP}$, $I_1$ and $I_2$ change sign\\
\hline
\end{tabular}
\end{center}
\begin{center}
\caption{\label{tab:case3symmetries}{\small {\bf Case 3)}. Symmetry transformations of the formulae for mixing angles and CP invariants in (\ref{anglescase3a}) and (\ref{CPinvcase3a}).}}
\end{center}
\end{table}
Since $\sin^2 \theta_{13}$ only depends on $m/n$, the latter is fixed by the experimentally measured value of $\theta_{13}$ and has to be small or close to one. 
Thus, we already know that for a certain group $\Delta (6 \, n^2)$ only very few choices of the $Z_2$ symmetry generator $Z$ are admissible, since the value of the parameter
$m$ characterizes this symmetry, see (\ref{ZXcase3}).
For $0.0188 \lesssim \sin^2 \theta_{13} \lesssim 0.0251$ \cite{nufit} we find as allowed range of $m/n$
\begin{equation}
\label{mnallowedcase3a}
0.054 \lesssim m/n \lesssim 0.062 \;\;\; \mbox{or} \;\;\; 0.938 \lesssim m/n \lesssim 0.946 \; ,
\end{equation}
since $0 \leq m \leq n-1$. Also $\sin^2 \theta_{23}$ only depends on $m/n$ and thus we can express it in terms of $\sin^2 \theta_{13}$ 
\begin{equation}
\label{sinth23sqinsinth13sqcase3a}
\sin^2 \theta_{23} = \frac 12 \, \left( 1 \pm  \sin \theta_{13} \, \frac{\sqrt{2-3 \sin^2 \theta_{13}}}{1-\sin^2 \theta_{13}} \right)
\approx  \frac 12 \, \left( 1 \pm \sqrt{2} \, \sin \theta_{13} \right)
\end{equation}
with ``+" being valid for $m/n$ small and ``-" for $1-m/n$ small and it varies in the interval
\begin{equation}\nonumber
0.387 \lesssim \sin^2 \theta_{23} \lesssim 0.403 \;\;\; \mbox{for} \;\; 1-m/n \;\; \mbox{small and} \;\;\; 0.597 \lesssim \sin^2 \theta_{23} \lesssim 0.613 \;\;\; \mbox{for} \;\; m/n \;\; \mbox{small} \; .
\end{equation}
We can also approximate the result of the solar mixing angle for $\cos\phi_m \approx \pm 1$ (and thus $\cos 2 \, \phi_m \approx 1$) by
\begin{equation}
\label{sinth12sqapproxcase3a}
\sin^2 \theta_{12} \approx \frac 13 \, \left( 1 + \sin^2 \theta \pm \sqrt{2} \cos 3 \, \phi_s \sin 2 \theta \right)
\end{equation}
which is close to $1/3$, if 
\begin{equation}
\label{sinth12sqcondcase3a}
\sin \theta  \left( \sin \theta \pm 2 \, \sqrt{2} \cos 3 \, \phi_s \cos \theta \right) \approx 0 \; .
\end{equation}
This allows two types of solutions
\begin{equation}
\label{thetacase3a}
\theta \approx 0, \, \pi \;\;\; \mbox{or} \;\;\; \tan \theta \approx \mp \, 2 \, \sqrt{2} \, \cos 3 \, \phi_s
\end{equation}
with ``-" holding for $m/n$ small and ``+" being relevant for $1-m/n$ small.
These solutions are also found in the numerical analysis, see figure \ref{Fig:5} (in particular, the black lines represent the second type of solution for small $m/n$).
We note that for $s=n/2$ only the solution $\theta \approx 0, \, \pi$ remains and the solar mixing angle is bounded from below,
$\sin^2 \theta_{12} \gtrsim 1/3$, since 
\begin{equation}
\label{sinsqth12sn2case3a}
 \sin^2 \theta_{12} \approx  \frac 13 \, (1\,+\sin^{2}\theta)\,.
\end{equation}
 Thus, the experimental best fit value of the solar mixing angle can be accommodated
best for $\theta_{\rm bf} = 0$. This entails together with $s=n/2$ that all CP phases vanish, see table \ref{tab:case3an16}.

Using the fact that $m/n$ or $1-m/n$ is small and $\theta$ is constrained to fulfill (\ref{thetacase3a}) we can also derive approximations for the sines of the CP phases
from the expressions in (\ref{CPinvcase3a}). For the Dirac phase $\delta$ we get
\begin{equation}
\label{sindeltasolcase3a}
\sin\delta \left( \theta \approx 0, \pi \right) \approx 0 \;\;\; \mbox{and} \;\;\; |\sin\delta \left(\tan\theta \approx \mp \, 2 \, \sqrt{2} \, \cos 3 \, \phi_s \right)| \approx \left| \frac{3 \sin 6 \, \phi_s}{5+4 \cos 6 \, \phi_s} \right|
\end{equation}
showing that we can achieve a maximal Dirac phase in the latter case e.g. for $s/n \approx 0.13$ and $s/n \approx 0.2$. For the Majorana phase $\alpha$ we find analogously
\begin{equation}
\label{sinasolcase3a}
|\sin\alpha| \approx |\sin 6 \, \phi_s|
\end{equation}
for all possible values of $\theta$ in (\ref{thetacase3a}). The second Majorana phase $\beta$ instead behaves similar to the Dirac phase, i.e.
\begin{equation}
\label{sinbsolcase3a}
\sin\beta \left( \theta \approx 0, \pi \right) \approx 0 \;\;\; \mbox{and} \;\;\; |\sin\beta \left(\tan\theta \approx \mp \, 2 \, \sqrt{2} \, \cos 3 \, \phi_s\right)| \approx 2 \, |\sin 6 \, \phi_s| \, \left| \frac{2 + \cos 6 \, \phi_s}{5+4 \cos 6 \, \phi_s} \right|\,.
\end{equation}
We see that this phase cannot be maximal and its maximally achieved value is $|\sin\beta|=\sqrt{3}/2\approx 0.866$ for e.g. $s/n \approx 0.11$ and $s/n \approx 0.22$. For $\theta \not\approx 0, \, \pi$ it becomes very small for $s/n$ close to $k/6$, $k=0, ..., 5$. All statements made are consistent with our numerical results, see figure \ref{Fig:6}.

The CP transformation $Y_1$ in (\ref{Y1case3}) reads in the neutrino mass basis as follows
\begin{eqnarray}
\label{Y1tcase3a}
&&\tilde{Y}_1 = U^\dagger_{\nu, 3} Y_1 U^\star_{\nu, 3} 
\\ \nonumber
&&\phantom{\tilde{Y}_1} = \omega^2 \, e^{i \left(y -2 \, \phi_s \right)} \, \left(
\begin{array}{ccc}
  \cos^2 \theta + e^{6 \, i \, \phi_s} \sin^2 \theta & (-i)^{k_1+1} e^{3 \, i \, \phi_s} \sin 3 \, \phi_s \sin 2 \theta & 0\\
 (-i)^{k_1+1} e^{3 \, i \, \phi_s} \sin 3 \, \phi_s \sin 2 \theta &   (-1)^{k_1} \left( e^{6 \, i \, \phi_s} \cos^2 \theta + \sin^2 \theta \right) & 0 \\
0 & 0 & (-1)^{k_2} 
\end{array}
\right) \; .
\end{eqnarray}
This matrix becomes diagonal for $\sin 2\theta=0$ or $s=0$ (which is the only solution, since $0 \leq s \leq n-1$ and three does not divide $n$). In the first case,
$\sin 2\theta=0$, we find for the remaining diagonal matrix
\begin{eqnarray}
\label{Y1tthetafixedcase3a}
&&\tilde{Y}_1 \left(\theta=0\right) =  \omega^2 \, e^{i \left(y -2 \, \phi_s \right)} \, \left(
\begin{array}{ccc}
 1 & 0 & 0\\
 0 &   (-1)^{k_1} e^{6 \, i \, \phi_s}  & 0 \\
0 & 0 & (-1)^{k_2} 
\end{array}
\right)
\\
\text{and}&&\tilde{Y}_1 \left(\theta=\frac{\pi}{2}\right) =  \omega^2 \, e^{i \left(y - 2 \, \phi_s \right)} \, \left(
\begin{array}{ccc}
  e^{6 \, i \, \phi_s}  & 0 & 0\\
 0 &   (-1)^{k_1} & 0 \\
0 & 0 & (-1)^{k_2} 
\end{array}
\right)
\end{eqnarray}
and thus for the Majorana phases
\begin{eqnarray}\label{sinabY1ttheta0}
&& |\sin\alpha| = \left| \sin 6 \, \phi_s \right| \;\;\; \mbox{and} \;\;\; \sin\beta=0
\\
\mbox{and}\;\;\;&&  |\sin\alpha| = |\sin\beta| = \left| \sin 6 \, \phi_s \right|  \; ,
\end{eqnarray}
respectively, while for $s=0$ we find
\begin{equation}
\label{Y1ts0case3a}
\tilde{Y}_1 \left(s=0\right) = \omega^2 \, e^{i y} \, \left( \begin{array}{ccc}
1 & 0 & 0\\
0 & (-1)^{k_1} & 0\\
0 & 0 & (-1)^{k_2} 
\end{array}
\right)
\end{equation}
and thus all CP phases are trivial. This observation is consistent with the fact that $J_{CP}$, $I_1$ and $I_2$ are all proportional to $\sin 3 \, \phi_s$
and is, indeed, confirmed by the numerical analysis, see tables \ref{tab:case3an16}-\ref{tab:case3an11shift}. The CP transformation $\tilde{Y}_{1} (s=0)$
in (\ref{Y1ts0case3a}) fulfills the second equation in (\ref{XZY}), if we choose $y= \frac{2 \, \pi}{3} + k \, \pi$, $k=0, 1$.

\begin{figure}[t!]
\begin{center}
\begin{tabular}{cc}
\includegraphics[width=0.48\textwidth]{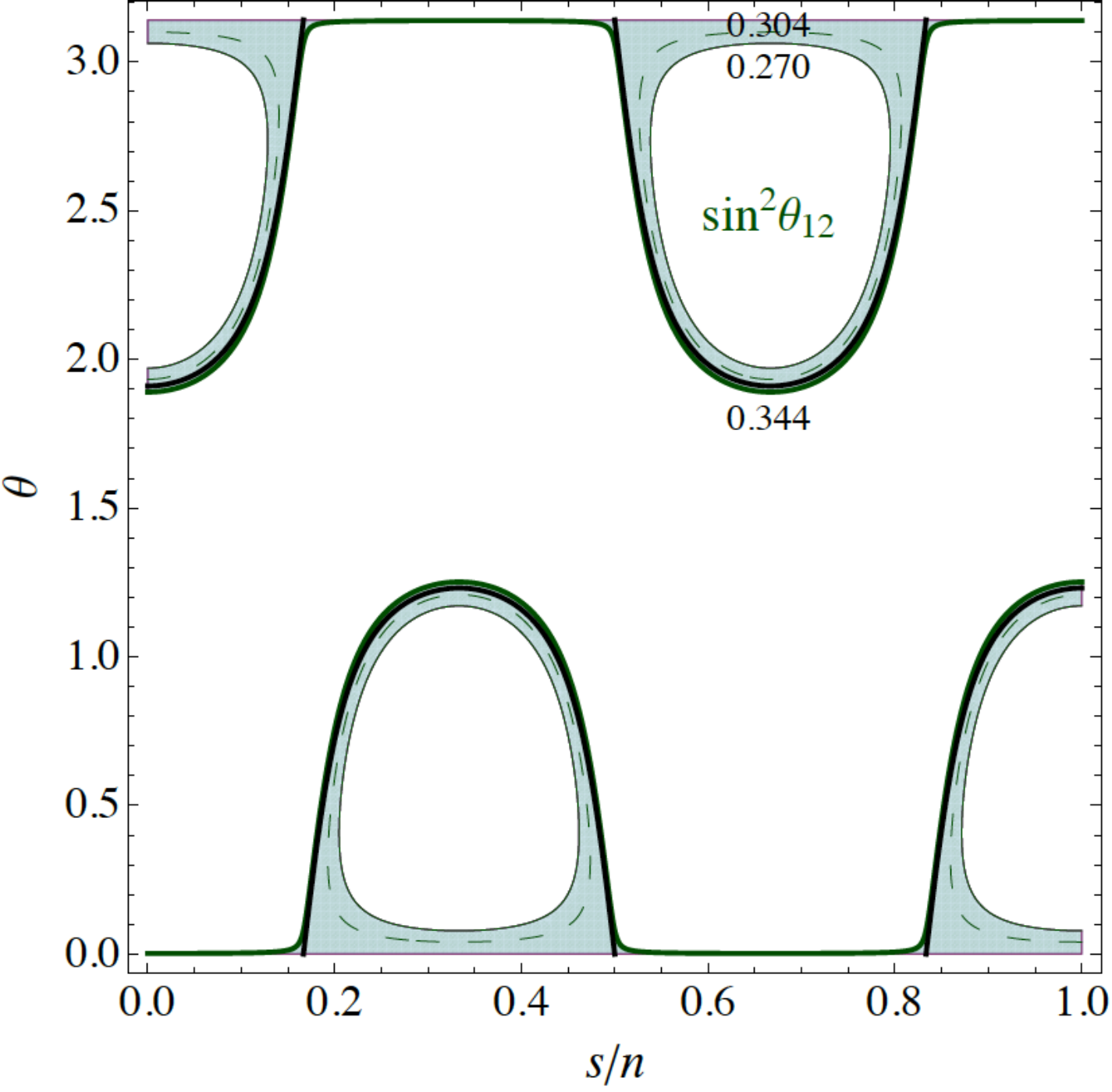}
\end{tabular}
\caption{\small{\textbf{Case 3 a)}. Contour region of $\sin^{2}\theta_{12}$ defined in (\ref{anglescase3a}), in the plane $\theta$ versus $s/n$ for $m/n=1/16$. 
The colored region in the plane is realized by taking the $3\,\sigma$ limits of $\sin^{2}\theta_{12}$ (continuous green lines) and thus all values in this area
lead to $\chi^2_{12} \lesssim 9$.
The dashed lines indicate the best fit value, $(\sin^{2}\theta_{12})^{\rm bf}=0.304$. The black plain curve
is an analytic approximation, assuming $\sin^2\theta_{12}\approx 1/3$ and is given in the second equation in (\ref{thetacase3a})
with ``-", since $m/n\ll 1$.
For this choice of $m/n$ the other two mixing angles are fixed to $\sin^{2}\theta_{13}\approx0.0254$ and $\sin^{2}\theta_{23}\approx0.613$.
In order to obtain the corresponding figure for $m/n=15/16$ we have to reflect the $3\,\sigma$ contour region 
of $\sin^{2}\theta_{12}$ in the line defined by $\theta=\pi/2$, as can be seen using the first transformation in table~\ref{tab:case3symmetries}.
In this case the atmospheric mixing angle is given by  $\sin^{2}\theta_{23}\approx0.387$.
}}
\label{Fig:5}
\end{center}
\end{figure}

%
\mathversion{bold}
\subsubsection*{Numerical results}
\mathversion{normal}
\label{sec4311}

\begin{figure}[t!]
    \begin{subfigure}[b]{0.48\textwidth}
      \includegraphics[width=\linewidth]{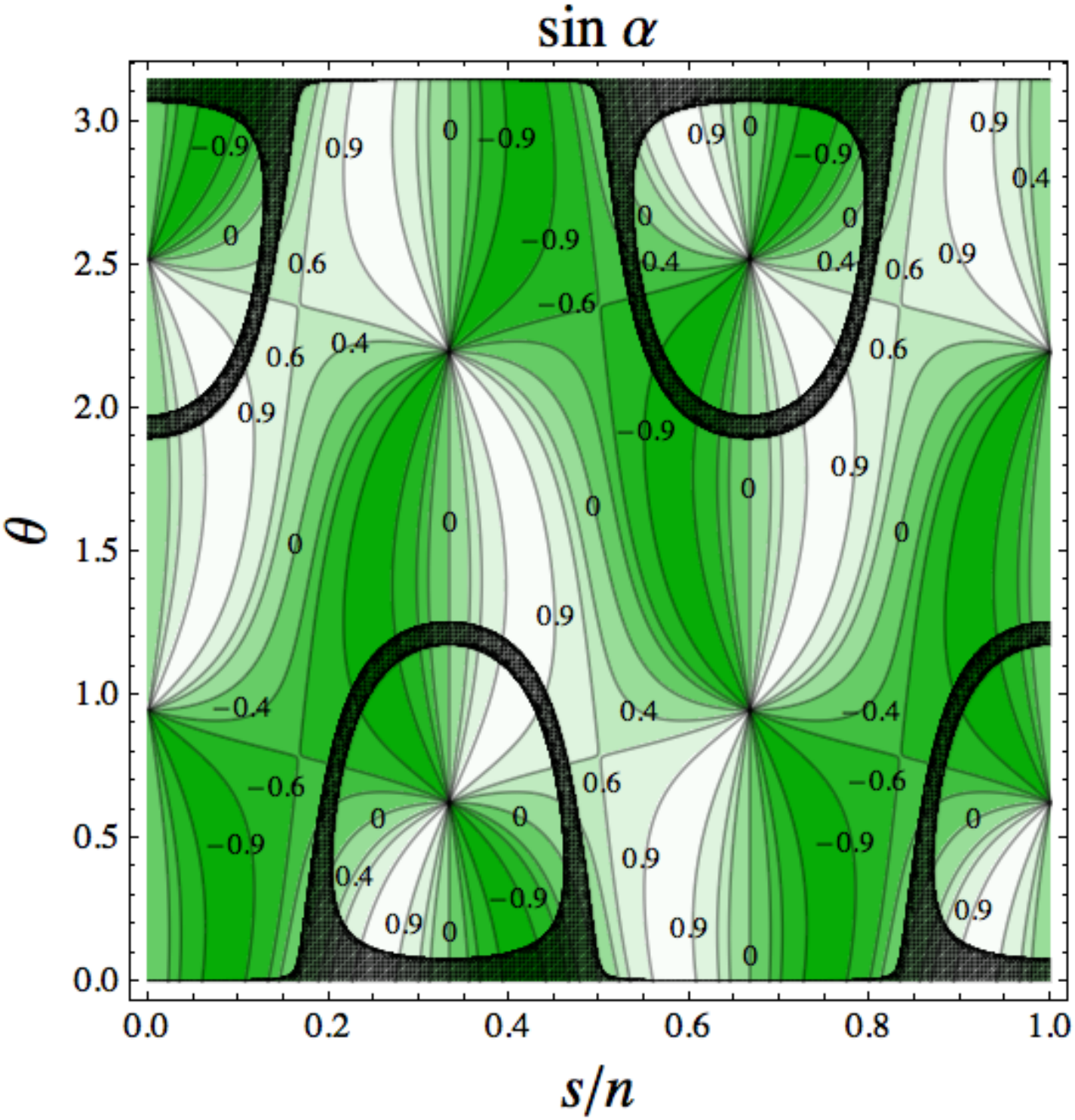}
      \caption*{}
    \end{subfigure}
    \begin{subfigure}[b]{0.48\textwidth}
     \includegraphics[width=\linewidth]{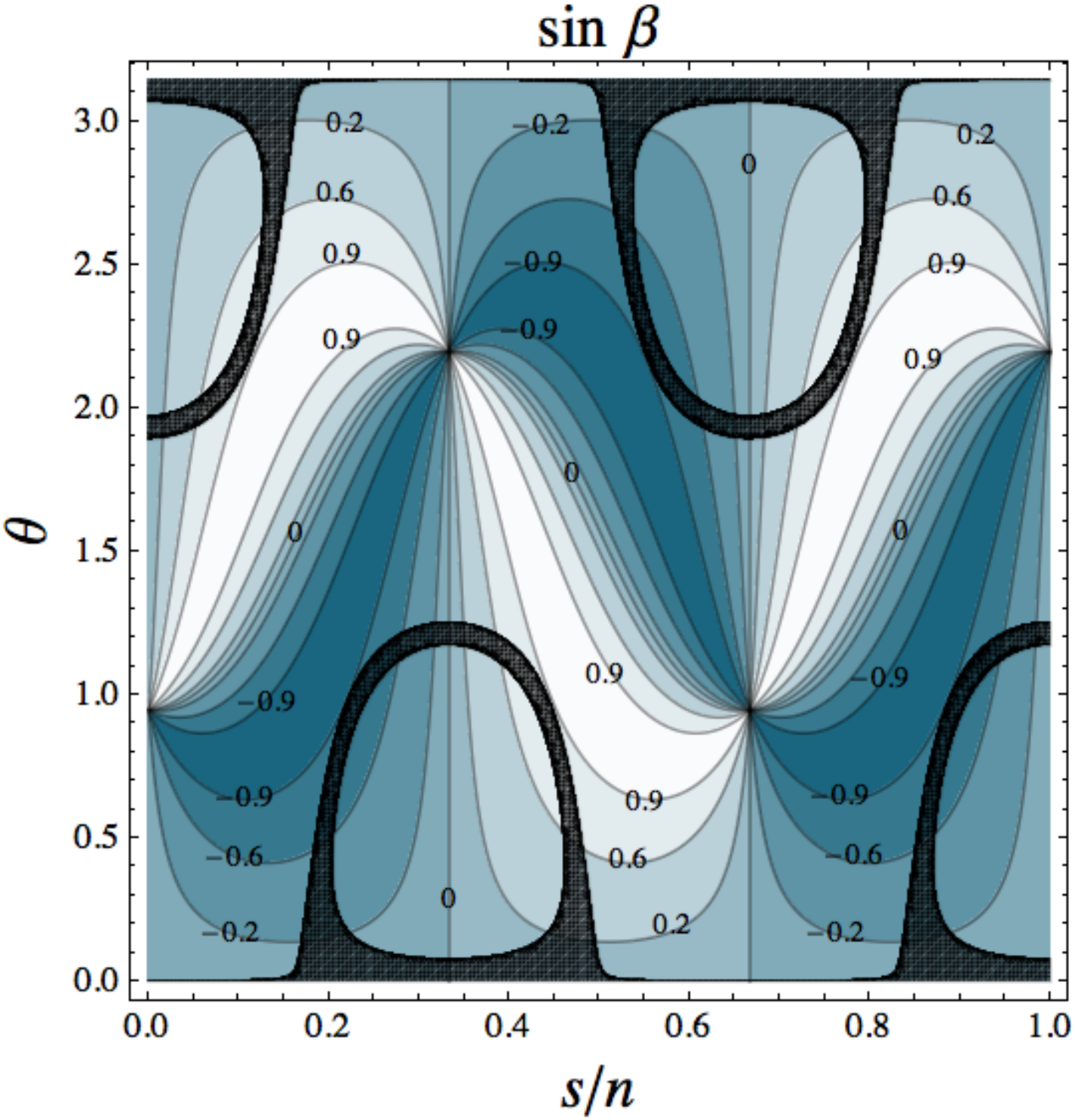}
      \caption*{}
    \end{subfigure}
    \begin{subfigure}[b]{0.48\textwidth}
      \includegraphics[width=\linewidth]{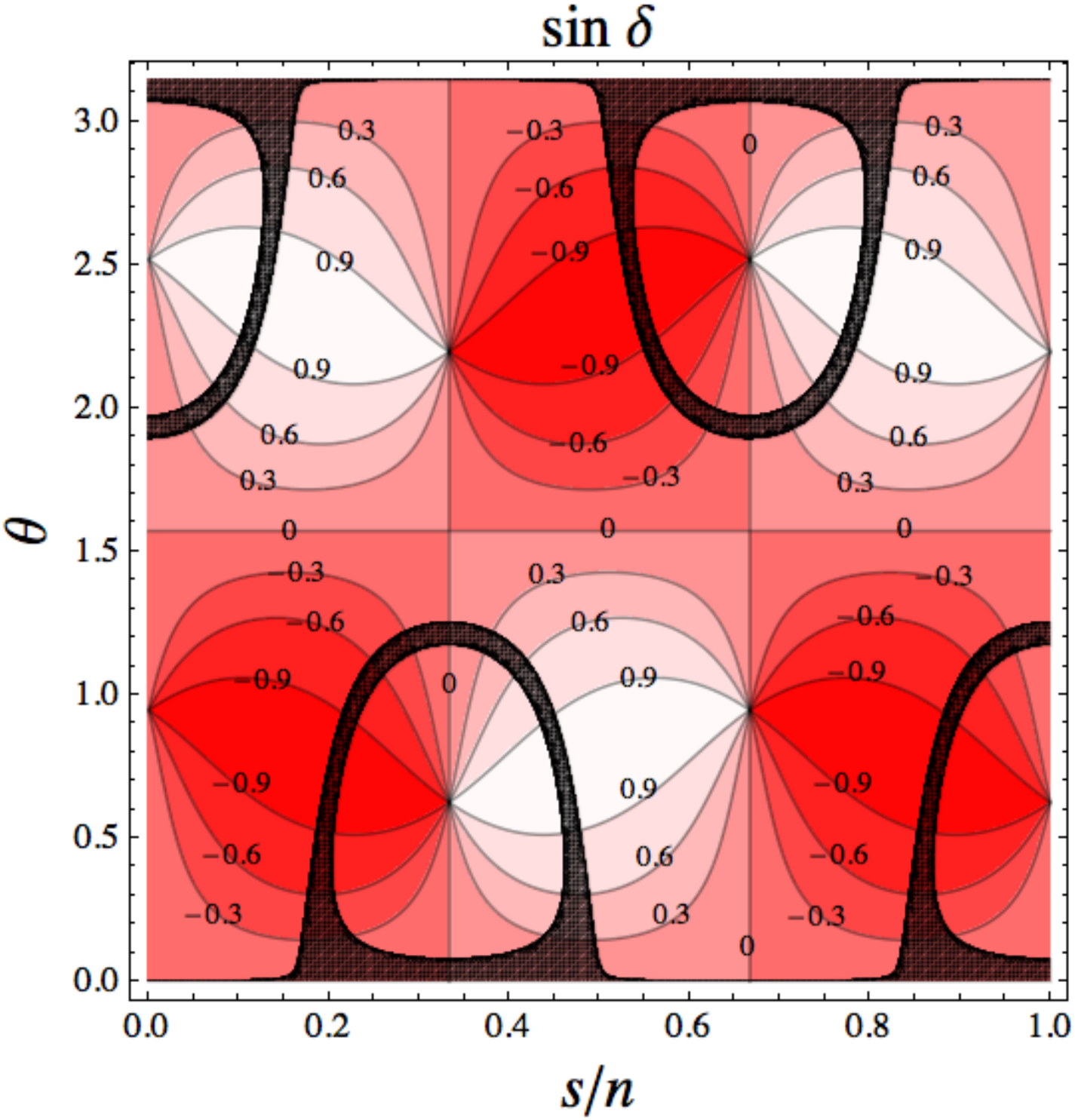}
      \caption*{}
    \end{subfigure}
        \hfill
    \begin{minipage}[b]{0.48\textwidth}
      \caption{{\small \textbf{Case 3 a)}. Predictions for the Majorana and Dirac phases obtained from (\ref{anglescase3a}) and (\ref{CPinvcase3a}) for $m/n=1/16$. 
      The shaded areas in the three figures correspond to
      the $3\,\sigma$ regions of $\sin^{2}\theta_{12}$, see figure~\ref{Fig:5} for details. The parameters $k_{1}$ and $k_{2}$ are set to zero. 
    For the choice $m/n=15/16$ the corresponding contour plots of the CP phases are obtained by performing the transformations 
      $\theta\to\pi/2-\theta$ and $\theta\to\pi-\theta$ for $\sin\delta$ and $\sin\alpha$, $\sin\beta$, respectively, see (\ref{phaserel3a}). 
      For numerical values obtained from a $\chi^{2}$ analysis see table~\ref{tab:case3an16}. \vspace{1.9cm}}}
      \label{Fig:6}
    \end{minipage}
  \end{figure}

As in case 2), we first discuss our numerical results for the mixing parameters derived from $U_{PMNS, 3}$ in (\ref{Unucase3}) and found in (\ref{anglescase3a})
and (\ref{CPinvcase3a}). Since in this case groups $\Delta (6 \, n^2)$ with an even as well as those with an odd index $n$ are admitted, we present results for both types of
choices. The ratio $m/n$ ($1-m/n$) is practically fixed by the requirement to accommodate the reactor and the atmospheric mixing angles well, see (\ref{mnallowedcase3a}), and we find as smallest indices $n$ and $m$ that allow for a good fit $n=16$ and $m=1$ ($m=15$) and $n=17$ and $m=1$ ($m=16$).
For these cases we study the dependence of the solar mixing angle and the CP phases on the continuous parameter $\theta$
as well as on $s$ that characterizes the chosen CP transformation $X$, see (\ref{ZXcase3}).

For $n=16$ and $m=1$ ($m=15$) the values of the reactor and the atmospheric mixing angle read $\sin^{2}\theta_{13}\approx0.0254$ and $\sin^{2}\theta_{23}\approx0.613$
($\sin^2\theta_{23}\approx0.387$). 
Note that $\theta_{23}$ is in agreement  with the $3\,\sigma$ range given in the current global fit analysis\cite{nufit}, whereas $\theta_{13}$
is marginally too large. The contribution to $\chi^2_{\rm tot}$ from each of these quantities is $\chi^{2}_{13}\approx12.1$ and $\chi^{2}_{23}\approx1.81$ ($\chi^2_{23}\approx4.31$), respectively.
In figure~\ref{Fig:5} we display the $3\,\sigma$ contour region for $\sin^{2}\theta_{12}$ in the plane $\theta$ versus $s/n$, see (\ref{anglescase3a}),
 for the choice $m/n=1/16$. The thick and thin plain lines in green correspond to the 
experimental upper and lower bounds of $\sin^{2}\theta_{12}$, as reported in appendix~\ref{app12}, while the dashed curve represents its
best fit value, $(\sin^{2}\theta_{12})^{\rm bf}=0.304$. 
As discussed above, 
there are two possible ways to accommodate the solar mixing angle within its $3\,\sigma$ interval: either $\theta \approx 0, \, \pi$ for $s/n$ arbitrary,
then $\sin^2 \theta_{12} \approx 1/3$, see (\ref{anglescase3a}), or the value of $\theta$ depends on $s/n$ and in general values of $\sin^2\theta_{12}$ smaller
than $1/3$ can be achieved. An analytic approximation of the latter is given in (\ref{thetacase3a}) (with the choice ``-", since $m/n$ is small) and 
 is indicated by the thick black lines in figure \ref{Fig:5}. For the choice $m/n=15/16$ the corresponding figure of $\sin^2\theta_{12}$ in the $s/n$-$\theta$ plane 
 is obtained from figure \ref{Fig:5} by performing a reflection of the $3\,\sigma$ regions 
in the line defined by $\theta=\pi/2$, compare to first symmetry transformation in table \ref{tab:case3symmetries}. 
Even though we have chosen the value $m/n=1/16$ in figure \ref{Fig:5}, the latter will practically be the same for other 
choices of $m/n$, provided these fall into the range given in (\ref{mnallowedcase3a}). In particular, the figure obtained for $m/n=1/17$
will be very similar and is thus not separately reported here.

We continue with the discussion of the CP phases and show results for them in form of contour plots in the $s/n$-$\theta$ plane
in figure~\ref{Fig:6}. We compute the CP phases from (\ref{CPinvcase3a}) for $k_1=k_2=0$. Again, we choose $m/n=1/16$.
The black areas in the figures represent the $3\,\sigma$ allowed regions of $\sin^{2}\theta_{12}$ in the same plane, also displayed in figure \ref{Fig:5}.
As can be seen, the CP phases $\delta$ and $\alpha$ can assume maximal values in these areas. This observation has also been made using the
analytic approximations in (\ref{sindeltasolcase3a}) and (\ref{sinasolcase3a}). For example, we can read off from figure \ref{Fig:6}

\vspace{-0.15in}
\small
\begin{equation}\nonumber
\label{sindmn116case3a}
\sin\delta\gtrsim0.9~(\lesssim-0.9) \;\;\; \mbox{for} \;\;\; 0.11\lesssim s/n \lesssim 0.14 \;\; (0.19\lesssim s/n \lesssim0.22) \;\;\; \mbox{and} \;\;\; 2.2\lesssim \theta\lesssim2.6 
\;\; (0.5\lesssim \theta\lesssim0.9)
\end{equation}
\normalsize
and

\vspace{-0.15in}
\small
\begin{equation}
\label{sinamn116case3a}\nonumber
\sin\alpha\gtrsim0.9~(\lesssim-0.9) \;\;\; \mbox{for} \;\;\;0.055\lesssim s/n \lesssim 0.11 \;\; (0.23\lesssim s/n \lesssim0.28) \;\;\; \mbox{and} \;\;\; 2\lesssim \theta\lesssim2.2
\;\; (1\lesssim \theta\lesssim1.2) \; .
\end{equation}
\normalsize
On the other hand, as also remarked in the analytical study, the absolute value of the Majorana phase $\beta$ has a non-trivial upper limit $|\sin\beta|\lesssim0.87$
and large values are obtained e.g.

\small
 \begin{equation}\nonumber
\label{sinbmn116case3a}
0.6\lesssim\sin\beta\lesssim0.87 \;\;\; \mbox{for} \;\;\; 0.05\lesssim s/n\lesssim0.15 \;\;\; \mbox{and} \;\;\; 1.9\lesssim \theta\lesssim2.7
\end{equation}
 \normalsize
 and
 
 \vspace{-0.15in}
\small
 \begin{equation}\nonumber
\label{sinbmn116case3a}
-0.87\lesssim\sin\beta\lesssim-0.6 \;\;\; \mbox{for} \;\;\; 0.18\lesssim s/n\lesssim0.28 \;\;\; \mbox{and} \;\;\; 0.44\lesssim \theta\lesssim1.2 \; .
\end{equation}
\normalsize
The various points in the plots in figure \ref{Fig:6} in which all contour lines converge correspond to points at which the CP phase(s)
are not physical, because some of the mixing angles vanish or become $\pi/2$. In particular, the points found in the figures of all three
CP phases indicate $\cos\theta_{12}=0$ ($\theta_{12}=\pi/2$), while those present only in the figures of $\sin\delta$ and $\sin\alpha$ correspond
to points with $\sin\theta_{12}=0$. Since these points are far away from the regions in which the solar mixing angle is accommodated well, these
have no impact on our results. 
In the case $m/n=15/16$ the contour lines for the CP phases can be obtained from those shown in the plots in figure \ref{Fig:6} 
by applying the following identities
 \begin{eqnarray}\label{phaserel3a}
 	&&\sin\delta\,(n-m,\,\theta)  =  \sin\delta\,(m,\,\pi/2-\theta)\,,\\
	&&\sin\alpha\,(n-m,\,\theta)  =  \sin\alpha\,(m,\,\pi-\theta) \;\;\; , \;\;\;
	\sin\beta\,(n-m,\,\theta)  =  \sin\beta\,(m,\,\pi-\theta)\nonumber\,.
 \end{eqnarray}
 Note that the appearance of $\pi/2-\theta$ as argument on the right-hand side of the first equality in (\ref{phaserel3a}) takes into account that $J_{CP}$ changes sign, if 
 the first transformation in table \ref{tab:case3symmetries} is applied. 

In table \ref{tab:case3an16} the results of our $\chi^2$ analysis for $n=16$ and $m=1$ are shown (again, always setting $k_1=k_2=0$). 
Since the value of the reactor and the atmospheric mixing angles
are fixed by the choice of $n$ and $m$, we display in the table only the solar mixing angle for each value of the parameter $s$ for which a value of the parameter $\theta$
is found that permits a reasonably good fit to the experimental data. As can be seen from the table and also from figure \ref{Fig:5}, this, indeed, happens for all value of $s$.
We only show those that fulfill $s \leq n/2=8$, since the results for the others are easily obtained by exploiting the symmetry transformations found in table \ref{tab:case3symmetries}.
As already mentioned when discussing figures \ref{Fig:5} and \ref{Fig:6} also the results for the choice $n=16$ and $m=15$ can be derived 
 by making use of the symmetries shown in table \ref{tab:case3symmetries}. For this choice, obviously, the value of $\chi^2_{\rm tot}$ is slightly different, since
 the atmospheric mixing angle then reads $\sin^2\theta_{23}\approx0.387$, see also caption of table \ref{tab:case3an16}.
 As regards the solar mixing angle, it is interesting to note that in most cases two values of ``best fitting" $\theta$ are obtained and that $\sin^2\theta_{12}$ 
 at these points coincides
 with the experimental best fit value $(\sin^2\theta_{12})^{\rm bf}=0.304$. If only one value of $\theta_{\rm bf}$ appears in the table, the solar mixing angle is not 
 accommodated so well (still within its $3 \,\sigma$ range). We comment on this observation in more detail at the end of this subsection. 
 Furthermore, also notice that the CP invariant $I_1$
and the Majorana phase $\alpha$ evaluated at the two different best fitting points $\theta_{\rm bf}$ have opposite signs and vanish, if only one value $\theta_{\rm bf}$ exists, while the other two CP phases take in general different values at the different $\theta_{\rm bf}$ and are still non-vanishing in the case with only one $\theta_{\rm bf}$, see $s=3$.
Also this behavior can be understood, as is shown at the end of this subsection. 
The cases $s=0$ and $s=8$ are peculiar, since in both cases all CP phases are trivial. Thus, an accidental CP symmetry
must be present in the theory. This, indeed, happens, since
$s=0$ always entails an accidental CP symmetry, see (\ref{Y1ts0case3a}), while for $s=8$ no CP violation is observed, because the best fitting value of $\theta$ is $\theta_{\rm bf}=0$, see (\ref{Y1tthetafixedcase3a}), and, in addition, $\sin 6 \, \phi_s$ vanishes, see (\ref{sinabY1ttheta0}). 
 
As expected, the results for $n=17$ and $m=1$, corresponding to the smallest value of an odd index $n$ for which the reactor and the atmospheric mixing angles can
be accommodated well, are pretty similar to those obtained for $n=16$ and $m=1$. Due to the slightly smaller value of $m/n$ both mixing angles agree slightly better with 
the data in this case: $\sin^{2}\theta_{13}\approx0.0225$ and $\sin^{2}\theta_{23}\approx0.607$ leading to contributions to $\chi^2_{\rm tot}$ of $\chi^{2}_{13}\approx0.371$ and $\chi^{2}_{23}\approx1.21$, respectively. An atmospheric mixing angle in the first octant is obtained for the choice $n=17$ and $m=16$, $\sin^{2}\theta_{23}\approx0.393$, as expected. Its contribution to $\chi^2_{\rm tot}$ is  $\chi^{2}_{23}\approx3.46$. Detailed numerical results for this case are found in table \ref{tab:case3an17}.
As already mentioned for $n=16$, $m=1$, the solar mixing angle at $\theta_{\rm bf}$ is for most values of the parameter $s$ equal to the experimental best fit value.
If so, also here two different values of $\theta_{\rm bf}$ are found. If not, see $s=3$ in table \ref{tab:case3an17}, only one value of $\theta_{\rm bf}$ is found. 
All statements made
above concerning the CP phases are also valid in this case. In particular, the statements referring to the values of $I_1$ and, consequently, $\sin\alpha$ as well as the observation that for $s=0$ an accidental CP symmetry is present are true. For values $s>n/2=17/2$ and for $n=17$ and $m=16$
numerical results are easily deduced from table \ref{tab:case3an17}, simply by applying the symmetry transformations in table \ref{tab:case3symmetries}.

Up to now, we have focussed on the mixing angles and CP invariants, shown in (\ref{anglescase3a}) and (\ref{CPinvcase3a}), that are derived from $U_{PMNS, 3}$ in (\ref{Unucase3}).
However, it is also interesting to consider the case in which this matrix is multiplied by $P_2$ from the left such that we have to replace $m$ and $\theta$
in the formulae in (\ref{anglescase3a}) and (\ref{CPinvcase3a}) by $m-\frac n3$ and $\pi-\theta$. Interestingly enough, in this case a smaller (odd) value of $n$ is sufficient
for achieving a good fit to the reactor and the atmospheric mixing angles. For $n=11$ and $m=3$ we obtain $\sin^{2}\theta_{13}\approx0.0239$ and $\sin^{2}\theta_{23}\approx0.390$ leading to contributions to $\chi^2_{\rm tot}$ of $\chi^{2}_{13}\approx3.91$ and $\chi^{2}_{23}\approx3.86$. As one can see from table
\ref{tab:case3an11shift} that is subject to the same conventions as tables \ref{tab:case3an16} and \ref{tab:case3an17}, also here the solar mixing angle can be fitted for all shown choices of $s$, but one ($s=2$), to $(\sin^2\theta_{12})^{\rm bf}=0.304$. The behavior of the CP invariants and the corresponding CP phases can be described in the same way as for $n=16$, $m=1$ and $n=17$, $m=1$. A value of the atmospheric
mixing angle belonging to the second octant is in this case easily achieved by considering the PMNS mixing matrix with an additional exchange of the second and third rows.
Then $\sin^2\theta_{23}$ is $\sin^{2}\theta_{23}\approx0.610$ meaning that  $\chi^{2}_{23}\approx1.49$. 
The results for the values of $s$ not shown in table \ref{tab:case3an11shift} are easily obtained using the third symmetry found in table 
\ref{tab:case3symmetries}. Similarly, the application of the other two symmetries in this table allows to recover numerical results for a PMNS mixing matrix $U_{PMNS, 3}$
that is the product of $P_1$ and the matrix displayed in (\ref{Unucase3}), i.e. 
the mixing parameters are given by (\ref{anglescase3a}) and (\ref{CPinvcase3a}) replacing $m$ and $\theta$ by $m+\frac n3$ and $\pi-\theta$. If the index $n$ shall be even instead of odd, the smallest possible choice of $n$ that
admits a reasonable fit to the experimental data is $n=22$ and $m=6$. This is clear, since the ratio
$m/n$ is the same as in the case $n=11$ and $m=3$ that we have just discussed. Thus, also the numerical results found in table \ref{tab:case3an11shift} apply in this case.
In addition, there are results originating from odd values of $s$ for $n=22$ and $m=6$ that cannot be

\begin{landscape}  
  
\begin{table}[t!]
\centering
\begin{tabular}{c}
$
\begin{array}{|l||c|c|c|c|c|c|c|c|c|c|c|c|c|}
\hline
 s & 0 & 1 & 2 & 3 & 4 & 5 & 6 & 7 & 8 \\
\hline
 \chi^{2}_{\rm tot} & 13.9 & 13.9  & 13.9  & 15.0 & 13.9 & 13.9  & 13.9 & 13.9 & 23.9   \\
\hline
  \theta _{\rm bf} & \ba 1.93\\(3.10) \ea & \ba 2.00\\(3.09) \ea  & \ba 2.40\\(3.02)\ea & 0.265  & \ba 0.0584\\(1.07) \ea  & \ba 0.0415\\(1.20) \ea  & \ba 0.0441\\(1.18)\ea & \ba 0.0758\\(0.955)\ea & 0 \\
\hline
  \chi^{2}_{12} & 0 & 0 & 0 & 1.16 & 0 & 0 & 0 & 0 & 10.0 \\
\hline
 \sin^{2}\theta_{12} & 0.304 & 0.304  & 0.304  & 0.317 & 0.304  & 0.304  & 0.304 & 0.304 & 0.342 \\
\hline
 J_{CP} & 0 & \ba 0.0159\\(0.0021)\ea  &\ba 0.0348\\(0.0082)\ea  & -0.019 & \ba-0.0031\\(-0.0225)\ea  & \ba-0.00061\\(-0.0050)\ea  & \ba0.0013\\(0.0102)\ea & \ba0.0047\\(0.0296)\ea & 0  \\
\hline
 \sin\delta & 0 & \ba0.458\\(0.0594)\ea & \ba0.9995\\(0.234)\ea  & -0.533 & \ba-0.0896\\(-0.646)\ea & \ba-0.0176\\(-0.143)\ea & \ba0.0367\\(0.293)\ea & \ba0.137\\(0.852)\ea & 0    \\
\hline
 I_1 & 0 & \pm0.189 & \pm0.116 & 0  & \pm0.201 & \pm0.0792 & \mp0.146 & \mp0.177 & 0  \\
\hline
 \sin\alpha & 0 & \pm0.939 & \pm0.579 & 0 & \pm0.998 & \pm0.394 & \mp0.725 & \mp0.882  & 0    \\
\hline
 I_2 & 0 & \ba0.0114\\(0.00066)\ea & \ba0.0135\\(0.0026)\ea & -0.0060  & \ba-0.0010\\(-0.0135)\ea  & \ba-0.00020\\(-0.0044)\ea  & \ba0.00041\\(0.0083)\ea & \ba0.0015\\(0.0144)\ea & 0 \\
\hline
 \sin\beta & 0 & \ba0.662\\(0.0383)\ea & \ba0.784\\(0.152)\ea & -0.357 & \ba-0.0578\\(-0.784)\ea & \ba-0.0113\\(-0.253)\ea & \ba0.0237\\(0.481)\ea & \ba0.0882\\(0.837)\ea & 0   \\
 \hline
\end{array}
$
\end{tabular}
\caption{\label{tab:case3an16}{\small  \textbf{Case 3 a)}. Results of the $\chi^2$ analysis for $n= 16$ and $m=1$ obtained for the mixing angles and CP invariants given in (\ref{anglescase3a}) and (\ref{CPinvcase3a}). The choice $n=16$ is the smallest even $n$ that provides $\chi^{2}_{\rm tot}\lesssim27$. The values of $\sin^2\theta_{13}$ and 
$\sin^2\theta_{23}$ only depend on the ratio $m/n$ and read for $m/n=1/16$: $\sin^{2}\theta_{13}\approx0.0254$ and $\sin^{2}\theta_{23}\approx0.613$. Their contributions
to $\chi^2_{\rm tot}$ are $\chi^{2}_{13}\approx12.1$ and $\chi^{2}_{23}\approx1.81$. The one resulting from the fit of $\sin^2\theta_{12}$ depends on the parameters $s$
and $\theta_{\rm bf}$ and is displayed in the table. As one can see, for most $s$ the experimental best fit value of $\sin^2\theta_{12}$ can be achieved.
The CP invariants $I_{1,2}$ and the sines of the Majorana phases are computed for $k_{1}=k_{2}=0$.
For $s=0$ and $s=8$ (that results in $\theta_{\rm bf}=0$) an accidental CP symmetry is present in the charged lepton and neutrino sectors and thus all CP phases
are trivial, see (\ref{Y1tcase3a})-(\ref{Y1ts0case3a}). The corresponding results for $s>8$ are achieved by exploiting the symmetry transformations reported in table~\ref{tab:case3symmetries}. Furthermore, the choice $n=16$ and $m=15$ leads to the same reactor mixing angle, but the atmospheric mixing angle is smaller than $\pi/4$,
i.e.  $\sin^{2}\theta_{23}\approx0.387$, contributing  $\chi^{2}_{23}\approx4.31$ to the value of $\chi^2_{\rm tot}$. Again, the results for $m=15$ and all possible $s$ can be obtained
from those shown here using the symmetry transformations in table \ref{tab:case3symmetries}.}}
\end{table}

\begin{table}[t!]
\centering
\begin{tabular}{c}
$
\begin{array}{|l||c|c|c|c|c|c|c|c|c|c|c|c|c|}
\hline
 s & 0 & 1 & 2 & 3 & 4 & 5 & 6 & 7 & 8 \\
\hline
 \chi^{2}_{\rm tot} & 1.58 & 1.58  & 1.58 & 8.30 & 1.58 & 1.58 & 1.58 & 1.58 & 1.58  \\
\hline
  \theta _{\rm bf} & \ba1.93\\(3.10)\ea & \ba1.99\\(3.09)\ea & \ba2.31\\(3.05)\ea & 0.134 & \ba0.0675\\(0.995)\ea & \ba0.0426\\(1.18)\ea & \ba0.0403\\(1.20)\ea & \ba0.0543\\(1.09)\ea & \ba0.192\\(0.493)\ea \\
\hline
  \chi^{2}_{12} & 0 & 0 & 0 & 6.72 & 0 & 0 & 0 & 0 & 0 \\
\hline
 \sin^{2}\theta_{12} & 0.304 & 0.304 & 0.304 & 0.335 & 0.304 & 0.304 & 0.304 & 0.304 & 0.304 \\
\hline
 J_{CP} & 0 & \ba0.0141\\(0.0018)\ea & \ba0.0319\\(0.0060)\ea & -0.0095 & \ba-0.0038\\(-0.0261)\ea & \ba-0.0011\\(-0.0091)\ea & \ba0.00053\\(0.0044)\ea & \ba0.0026\\(0.0198)\ea & \ba0.0129\\(0.0287)\ea  \\
\hline
 \sin\delta  & 0 & \ba0.428\\(0.0535)\ea & \ba0.969\\(0.184)\ea & -0.280 & \ba-0.117\\(-0.792)\ea & \ba-0.0334\\(-0.275)\ea & \ba0.0161\\(0.134)\ea & \ba0.0793\\(0.600)\ea & \ba0.391\\(0.871)\ea    \\
\hline
 I_1 & 0 & \pm0.184 & \pm0.144 & 0 & \pm0.189 & \pm0.140 & \mp0.0752 & \mp0.202 & \mp0.0529  \\
\hline
 \sin\alpha & 0 & \pm0.911 & \pm0.712 & 0 & \pm0.936 & \pm0.691 & \mp0.372 & \mp0.9994 & \mp0.262  \\
\hline
 I_2 & 0 & \ba0.0097\\(0.00053)\ea & \ba0.0126\\(0.0018)\ea & -0.0028 & \ba-0.0012\\(-0.0127)\ea & \ba-0.00033\\(-0.0070)\ea & \ba0.00016\\(0.0036)\ea & \ba0.00078\\(0.0116)\ea & \ba0.0039\\(0.0092)\ea \\
\hline
 \sin\beta  & 0 & \ba0.633\\(0.0345)\ea & \ba0.820\\(0.119)\ea & -0.190 & \ba-0.0753\\(-0.828)\ea & \ba-0.0215\\(-0.455)\ea & \ba0.0104\\(0.238)\ea & \ba0.0511\\(0.760)\ea & \ba0.255\\(0.604)\ea  \\
 \hline
\end{array}
$
\end{tabular}
\caption{\label{tab:case3an17}{\small  \textbf{Case 3 a)}. Results for the smallest odd value of the index $n$, $n=17$ together with $m=1$, leading to $\chi^2_{\rm tot}\lesssim27$
for the PMNS mixing matrix $U_{PMNS, 3}$ in (\ref{Unucase3}) that is also considered in table \ref{tab:case3an16}. 
The ratio $m/n=1/17$ fixes the reactor and the atmospheric mixing angles to $\sin^{2}\theta_{13}\approx0.0225$ and $\sin^{2}\theta_{23}\approx0.607$, contributing 
$\chi^{2}_{13}\approx0.371$ and $\chi^{2}_{23}\approx1.21$ to the value of $\chi^2_{\rm tot}$, respectively. The parameters $k_1$ and $k_2$ are set to zero.
As expected, the choice $s=0$ leads to an
accidental CP symmetry that results in trivial CP phases, see (\ref{Y1ts0case3a}). If we consider $n=17$ and $m=16$, the atmospheric mixing angle is
fixed to $\sin^{2}\theta_{23}\approx0.393$ that contributes $\chi^{2}_{23}\approx3.46$ to $\chi^2_{\rm tot}$. Results for this case as well as for $n=17$, $m=1$ and
$s>8$ can be obtained by applying the symmetry transformations given in table~\ref{tab:case3symmetries} to the results presented here.}} 
\end{table}


\end{landscape}

\begin{table}[t!]
\centering
\begin{tabular}{c}
$
\begin{array}{|l||c|c|c|c|c|c|c|c|c|}
\hline
 s & 0 & 1 & 2 & 3 & 4 & 5  \\
\hline
 \chi^{2}_{\rm tot}& 7.77  & 7.77 & 11.7 & 7.77 & 7.77  & 7.77   \\
\hline
  \theta _{\rm bf} & \ba0.0401\\(1.21)\ea & \ba0.0625\\(1.04)\ea  & 2.94 &  \ba2.00\\(3.09)\ea  & \ba1.95\\(3.10)\ea & \ba2.35\\(3.04)\ea \\
\hline
 \chi^{2}_{12}& 0  & 0 & 3.91 & 0 & 0 & 0   \\
\hline
 \sin^{2}\theta_{12} & 0.304 & 0.304 & 0.328 & 0.304 & 0.304 & 0.304  \\
\hline
 J_{CP} & 0 & \ba-0.0035\\(-0.0244)\ea & 0.0143 & \ba0.0150\\(0.0019)\ea & \ba-0.0071\\(-0.00087)\ea & \ba-0.0335\\(-0.0070)\ea   \\
\hline
 \sin\delta  & 0 & \ba-0.102\\(-0.720)\ea & 0.413 & \ba0.442\\(0.0563)\ea & \ba-0.210\\(-0.0256)\ea & \ba-0.989\\(-0.206)\ea     \\
\hline
 I_1 & 0 & \mp0.197 & 0 & \mp0.187& \pm0.112& \pm0.131 \\
\hline
 \sin\alpha & 0 & \mp0.977& 0 & \mp0.926& \pm0.556 & \pm0.651  \\
\hline
 I_2 & 0 & \ba0.0011\\(0.0132)\ea & -0.0044 & \ba-0.0105\\(-0.00059)\ea & \ba0.0059\\(0.00027)\ea & \ba0.0131\\(0.0022)\ea  \\
\hline
 \sin\beta  & 0 & \ba0.0661\\(0.811)\ea & -0.279 & \ba-0.647\\(-0.0363)\ea& \ba0.361\\(0.0165)\ea & \ba0.806\\(0.133)\ea  \\
 \hline
\end{array}
$
\end{tabular}
\caption{\label{tab:case3an11shift}{\small  \textbf{Case 3 a)}. Results of the $\chi^2$ analysis for $n=11$ and $m=3$ that is the smallest value of $n$ leading to 
$\chi^{2}_{\rm tot}\lesssim27$ for mixing angles and CP invariants as given in (\ref{anglescase3a}) and (\ref{CPinvcase3a}) with the replacements $m\to m-\frac n3$ and $\theta\to \pi-\theta$.
The reactor and the atmospheric mixing angles read $\sin^{2}\theta_{13}\approx0.0239$ and $\sin^{2}\theta_{23}\approx0.390$, 
contributing $\chi^{2}_{13}\approx3.91$ and $\chi^{2}_{23}\approx3.86$ to $\chi^2_{\rm tot}$, respectively. 
Again, the CP invariants $I_{1,2}$ and Majorana phases are shown for $k_{1}=k_{2}=0$. Like for the other choices of $n$ and $m$, for $s=0$ all CP 
phases are trivial indicating the presence of an accidental CP symmetry, see  (\ref{Y1ts0case3a}).
Results for $s>5$ are obtained from those given here by using the symmetry transformations in table \ref{tab:case3symmetries}. 
A permutation of the second and third rows of the PMNS mixing matrix leads to an atmospheric mixing angle 
$\sin^{2}\theta_{23}\approx0.610$ giving rise to  $\chi^{2}_{23}\approx1.49$. If we consider the mixing pattern resulting in  (\ref{anglescase3a}) and (\ref{CPinvcase3a})
with the replacements $m\to m+\frac n3$ and $\theta\to\pi-\theta$, we find $m=8$ for $n=11$ and the results of the $\chi^2$ analysis of this case can be deduced
from those presented here by exploiting the symmetry transformations in table~\ref{tab:case3symmetries}. 
If the index $n$ shall be even, the smallest value is $n=22$ (and $m=6$) whose results (for even $s$) coincide with those given here.}}
\end{table}

\noindent obtained for $n=11$ and $m=3$.

Lastly, we comment on the fact that the solar mixing angle is either accommodated to its best fit value $(\sin^2\theta_{12})^{\rm bf}=0.304$, if two different ``best
fitting" points $\theta_{\rm bf}$ exist, or its value is larger and then only one value for $\theta_{\rm bf}$ is given, as displayed in tables~\ref{tab:case3an16}--\ref{tab:case3an11shift}.
Since the reactor and the atmospheric mixing angles, and hence their contributions to the $\chi^2$ function, are fixed by the choice of $m/n$, effectively only the contribution
$\chi^2_{12}$ depends on the variation of $\theta$ for a given value of the parameter $s$. In the case in which two different values of $\theta_{\rm bf}$ are mentioned,
the minimum of the solar mixing angle as function of $\theta$ is smaller than the experimental best fit value, $\sin^2 \theta_{12}\,(\theta_{\rm min}) < 0.304$, and thus it is possible
to obtain $\sin^2 \theta_{12}=(\sin^2\theta_{12})^{\rm bf}$ for some value of $\theta$ (and consequently $\chi^{2}_{12}=0$). 
Since $\sin^2\theta_{12}$ is a symmetric function with respect to $\theta_{\rm min}$  in its
vicinity, we indeed find two such values $\theta_{\rm bf}$, $\theta_{\rm bf, 1} < \theta_{\rm bf, 2}$ that fulfill
 $\theta_{\rm bf,2}-\theta_{\rm min}=\theta_{\rm min}-\theta_{\rm bf,1}$. If the minimum value of $\sin^2\theta_{12}$ attained turns out to be
larger than $(\sin^2\theta_{12})^{\rm bf}$, the choice of $\theta_{\rm bf}=\theta_{\rm min}$ minimizes the $\chi^2$ function (however, $\chi^2_{12}>0$). 
At  $\theta_{\rm min}$ the relation
\begin{equation}
\label{thetamincase3a}
	\tan2\,\theta_{\rm min} \; = \; -2\,\sqrt{2}\,\frac{\cos\, \phi_{m}}{\cos2\, \phi_{m}}\,\cos3 \, \phi_{s}\,,
\end{equation}
is satisfied, assuming $\cos 2\, \theta_{\rm min}\neq 0$. If we plug (\ref{thetamincase3a}) into the expression for $I_1$ in (\ref{CPinvcase3a}), we find that $I_1$ vanishes at $\theta_{\rm min}$, independently
of the choice of $s$, $m$ or $n$ and, hence, the Majorana phase $\alpha$ is trivial. If $\theta_{\rm bf}=\theta_{\rm min}$ we thus find vanishing $I_1$ and $\sin\alpha$,
see $s=3$ in tables \ref{tab:case3an16} and \ref{tab:case3an17} and $s=2$ in table \ref{tab:case3an11shift}. If we instead find two different values of $\theta_{\bf}$,
$\theta_{\rm bf, 1}$ and $\theta_{\rm bf, 2}$, we see that the CP invariant $I_1$ fulfills $I_1 \, (\theta_{\rm bf, 1})= -I_1 \, (\theta_{\rm bf, 2})$, since the expression of $I_{1}$
in (\ref{CPinvcase3a}) can be written as
\begin{equation}
\label{I1withthetamincase3a}
	I_{1} \;=\;\frac{\sqrt{2}}{9}\, (-1)^{k_1+1} \,\frac{\cos\phi_{m}\,\cos2\,\phi_{m}}{\cos2\,\theta_{\rm min}}\,\sin3\,\phi_{s}\,\sin 2\left(\theta-\theta_{\rm min}\right) \,,
\end{equation}
and thus also the Majorana phase $\alpha$ fulfills $\sin\alpha \, (\theta_{\rm bf, 1})= -\sin\alpha \, (\theta_{\rm bf, 2})$. Concerning the other CP phases $\delta$ and $\beta$
no such statement can be made, since they are neither even nor odd functions with respect to $\theta=\theta_{\rm min}$.

\mathversion{bold}
\subsubsection{Case 3 b.1)}
\mathversion{normal}
\label{sec432}

\mathversion{bold}
\subsubsection*{Analytical results}
\mathversion{normal}
\label{sec4321}

This mixing pattern is obtained by using the matrix $U_{PMNS, 3}$ with the columns permuted by the matrix $P_1$, defined in (\ref{P1}),
i.e. we apply this matrix from the right to the PMNS mixing matrix in (\ref{Unucase3}).
The mixing angles read
\begin{eqnarray}
\label{anglescase3b1}
&&\sin^2 \theta_{13} = \frac 13 \, \left( 1+ \cos 2 \, \phi_m \sin^2 \theta + \sqrt{2} \cos \phi_m \cos 3 \, \phi_s \sin 2 \theta \right) \;\; ,
 \\ \nonumber
&&\sin^2 \theta_{23} = \frac 12 \, \left( 1 + \frac{2 \sqrt{3} \sin \phi_m \sin \theta \, [\sqrt{2} \cos 3\, \phi_s \cos \theta -\cos \phi_m \sin \theta]}{2 - \cos 2 \,\phi_m \sin^2 \theta - \sqrt{2} 
\cos \phi_m \cos 3 \,\phi_s \sin 2 \theta} \right) \;\; ,
\\ \nonumber
&&\sin^2 \theta_{12}= 1-\frac{2 \sin^2 \phi_m}{2-\cos 2 \,\phi_m \sin^2 \theta -\sqrt{2} \cos \phi_m \cos 3 \,\phi_s \sin 2 \theta} 
\end{eqnarray}
and the CP invariants are given by
\begin{eqnarray}
\label{CPinvcase3b1}
&&J_{CP}= -\frac{1}{6 \sqrt{6}} \, \sin 3 \,\phi_m \sin 3\, \phi_s \sin 2 \theta \;\; ,
\\ \nonumber
&&I_1=\frac 49 \, (-1)^{k_2+1} \, \sin^2 \phi_m \sin 3 \,\phi_s \sin \theta \, \left(\cos 3 \,\phi_s \sin \theta - \sqrt{2} \cos \phi_m \cos \theta \right)  \;\; ,
\\ \nonumber
&&I_2= \frac 49 \, (-1)^{k_1+k_2+1} \, \sin^2 \phi_m \sin 3 \,\phi_s \,  \cos \theta \, \left( \cos 3 \,\phi_s \cos \theta + \sqrt{2} \cos \phi_m \sin \theta \right) \; .
\end{eqnarray}

Again, twelve permutations lead to this mixing, if possible shifts in $\theta$, but also in $m$, like in case 3 a), are taken into account.
Indeed, the permutations that allow us to generate mixing pattens with $m$ being replaced by $n-m$, $m-\frac{n}{3}$ or $m+\frac{n}{3}$ are the same
as in case 3 a). Furthermore, the symmetries found in table \ref{tab:case3symmetries} are also symmetries of the formulae in (\ref{anglescase3b1})
and (\ref{CPinvcase3b1}) and, if $m$ is replaced by $m-\frac{n}{3}$ or $m+\frac{n}{3}$, we find the same modifications to these symmetries as in case 3 a).
Eventually, also the formulae in (\ref{anglescase3b1}) and (\ref{CPinvcase3b1}) exhibit for $s=\frac{n}{2}$ (for $n$ even) [and independent of the value of $m$]
 a well-defined transformation behavior, if $\theta$ is replaced 
 by $\pi-\theta$, i.e. the expressions of the mixing angles are even functions in $\theta$, whereas the CP invariants are odd functions
and thus change sign, if $\theta$ is changed into $\pi-\theta$. 

In order to study this case analytically we define the following quantities
\begin{eqnarray}
\label{pcase3b}
&&p=\cos 2 \,\phi_m \sin^2 \theta + \sqrt{2} \, \cos \phi_m \cos 3 \,\phi_s \sin 2 \theta \; ,
\\
\label{qcase3b}
&&q=2 \sin \phi_m \sin \theta \, (\cos \phi_m \sin \theta - \sqrt{2} \, \cos 3 \,\phi_s \cos \theta)
\end{eqnarray}
and see that we can write the formulae for the mixing angles as
\begin{equation}
\label{anglespqcase3b}
\sin^2 \theta_{13}= \frac 13 \, (1+p) \;\; , \;\; \sin^2 \theta_{12} = 1- \frac{2 \, \sin^2 \phi_m}{2-p} \;\; , \;\; \sin^2 \theta_{23} = \frac 12 \, \left( 1 - \frac{\sqrt{3} \, q}{2-p} \right) \; .
\end{equation}
We can express $\sin^{2}\theta_{12}$ and  $\sin^2 \theta_{23}$ in terms of $\sin^{2}\theta_{13}$
\begin{equation}
\label{sinth12sqsinth23sqinsinth13sqcase3b1}
\sin^2 \theta_{12} = 1- \frac{2 \, \sin^2 \phi_m}{3 \, (1-\sin^2 \theta_{13})} \;\;\; \mbox{and} \;\;\;
\sin^2 \theta_{23} = \frac 12 \left( 1- \frac{q}{\sqrt{3} \, (1-\sin^2 \theta_{13})} \right)\,.
\end{equation}
Since the solar mixing angle is to good approximation $\sin^2 \theta_{12} \approx 1/3$, the ratio $m/n$ is constrained to fulfill
\begin{equation}
\label{con3b11}
\sin^2 \phi_m \approx 1 \, ,
\end{equation}
i.e.
\begin{equation}
\label{mnallowedcase3b1}
m \approx \frac n2 \;\;\; \mbox{for} \;\;\; n \;\; \mbox{even} \;\;\; \mbox{and} \;\;\; m \approx \frac{n\pm1}{2} \;\;\; \mbox{for} \;\;\; n \;\; \mbox{odd} \; .
\end{equation}
Then 
\begin{equation}
\label{con3b12}
\cos \phi_m \approx 0 \;\; , \;\;\; \cos 2 \,\phi_m \approx -1 \;\; , \;\;\; \sin 2 \,\phi_m \approx 0 \;\; , \;\;\;  
\end{equation}
and consequently
\begin{equation}
\label{con3b12bis}
-\sin^2 \theta \approx p = 3 \sin^2 \theta_{13}-1 \; .
\end{equation} 
This relation determines  the value of $\theta$ to be
\begin{equation}
\label{con3b13}
\theta_0 \approx 1.31 \;\;\; \mbox{or} \;\;\; \theta_0 \approx 1.83 \; ,
\end{equation}
if $\sin^2\theta_{13}$ is set to its experimental best fit value, $(\sin^2\theta_{13})^{\mathrm{bf}}=0.0219$. These two values of $\theta_0$ are related by the transformation 
$\theta\;\rightarrow\;\pi-\theta$, see also the first symmetry transformation in table \ref{tab:case3symmetries}. Indeed, the reactor and the solar mixing angles only
depend on the continuous parameter $\theta$ and not on $s$ (the choice of the CP transformation $X$) for $m=n/2$, see left panel in figure \ref{Fig:7}, and fulfill the sum rule
\begin{equation}
\label{sumrulemn2case3b1}
\sin^2\theta_{12}=\frac{1-3\, \sin^2 \theta_{13}}{3 \, (1-\sin^2\theta_{13})} 
\end{equation}
that has also been found in \cite{sumruleBranco}. For $\theta_0$ in (\ref{con3b13}) the solar mixing angle takes the value $\sin^2\theta_{12}\approx0.318$, see also
table \ref{tab:case3b1neven} in the numerical analysis. This value is well within the experimentally preferred $3\,\sigma$ range \cite{nufit}.
Note that if we had neglected non-zero $\theta_{13}$ in (\ref{con3b12bis}), the solution 
would have been $\theta_0=\frac{\pi}{2}$. Indeed, for $m=n/2$ ($\phi_m=\pi/2$) and $\theta=\pi/2$ mixing is TB. 
Using (\ref{con3b11}, \ref{con3b12}, \ref{con3b13}) we find
\begin{equation}
\sin^2 \theta_{23} \approx \frac 12 \, \left( 1+ \sqrt{\frac 23} \, \frac{\cos 3 \, \phi_s \sin 2 \theta_0}{1-\sin^2 \theta_{13}} \right)
\end{equation}
that tells us that the allowed values of $\phi_s$ ($s/n$) are constrained by the request to accommodate the atmospheric mixing angle well, e.g.
for $\theta_0 \approx 1.31$ we find as allowed intervals
\begin{equation}
\label{con3b14}
0.09 \lesssim s/n \lesssim 0.23 \;\; , \;\; 0.44 \lesssim s/n \lesssim 0.58 \;\;\; \mbox{and}\;\;\; 0.75 \lesssim s/n \lesssim 0.90 \; .
\end{equation}
The constraints derived with $\theta_0 \approx 1.83$ are very similar. Thus, the mixing angle $\theta_{23}$ can be accommodated well for a large range of $s/n$. 
If we refine our analysis and consider $m \neq n/2$, i.e. $m=n \, (1/2 + \kappa/\pi)$, $\kappa\ll 1$, we can derive a relation between the deviation of $\theta$ from $\theta_0$,
$\theta=\theta_0+\epsilon$ and $\kappa$
\begin{equation}
\label{approxdevcase3b1}
\epsilon \approx - \sqrt{2} \, \cos 3 \, \phi_s \, \kappa \; .
\end{equation}
For $n=20$ and $m=11$, $\kappa \approx 0.16$ we find $\epsilon \approx -0.22 \, \cos 3 \, \phi_s$. This approximation is displayed as dotted lines in 
the right panel of figure \ref{Fig:7} and fits the exact result reasonably well.

If $m=n/2$, the CP invariants in (\ref{CPinvcase3b1}) read

\vspace{-0.15in}
\small
\begin{equation}
\label{CPinvmn2case3b1}
J_{CP}=\frac{1}{6 \, \sqrt{6}} \, \sin 3 \, \phi_s \, \sin 2\theta \; , \;\; I_1= \frac 29 \, (-1)^{k_2+1} \, \sin 6 \, \phi_s \, \sin^2 \theta \; , \;\;  I_2= \frac 29 \, (-1)^{k_1+k_2+1} \, \sin 6 \, \phi_s \, \cos^2 \theta \; .
\end{equation}
\normalsize
For $\theta_0$ and $s/n$ as chosen in (\ref{con3b13}) and (\ref{con3b14}) the Dirac phase attains 
a lower value of 
\begin{equation}
\label{sindeltaboundcase3b1}
|\sin\delta| \gtrsim 0.71 
\end{equation}
and a maximal value can be obtained, if $s/n=1/6$, $s/n=1/2$ or $s/n=5/6$. The first and the third possibilities are excluded, since we do not consider the case $3 \mid n$. However,
the case $s=n/2$ is allowed for all even $n$.
Interestingly enough, the two Majorana phases $\alpha$ and $\beta$ depend for $m=n/2$ only on the parameter $s/n$, i.e. the absolute value of both reads
\begin{equation}
\label{sinasinbcase3b1}
|\sin\alpha| = |\sin\beta| = |\sin 6 \, \phi_s| \; .
\end{equation}

We can use the results obtained for $\tilde{Y}_1$ in the case 3 a), see (\ref{Y1tcase3a})-(\ref{Y1ts0case3a}), applying the permutation $P_1$
to the rows and columns of $\tilde{Y}_1$, $P_1^T \tilde{Y}_1 P_1$. Thus, the conclusions regarding the CP phases are very similar to those above. The only difference 
is that now the Majorana phases for $\tilde{Y}_1 \left(\theta=0\right)$ read 
\begin{equation}
\sin\alpha=0 \;\;\; \mbox{and} \;\;\; |\sin\beta|=|\sin 6\,\phi_s| \; ,
\end{equation}
while for $\tilde{Y}_1 \left(\theta=\frac{\pi}{2}\right)$ they read
\begin{equation}
|\sin\alpha|=|\sin 6\,\phi_s|  \;\;\; \mbox{and} \;\;\; \sin\beta=0 \; .
\end{equation}
As in case 3 a), all CP phases are trivial for $s=0$, see (\ref{Y1ts0case3a}).

\begin{figure}[t!]
\begin{center}
\begin{tabular}{cc}
\includegraphics[width=0.48\textwidth]{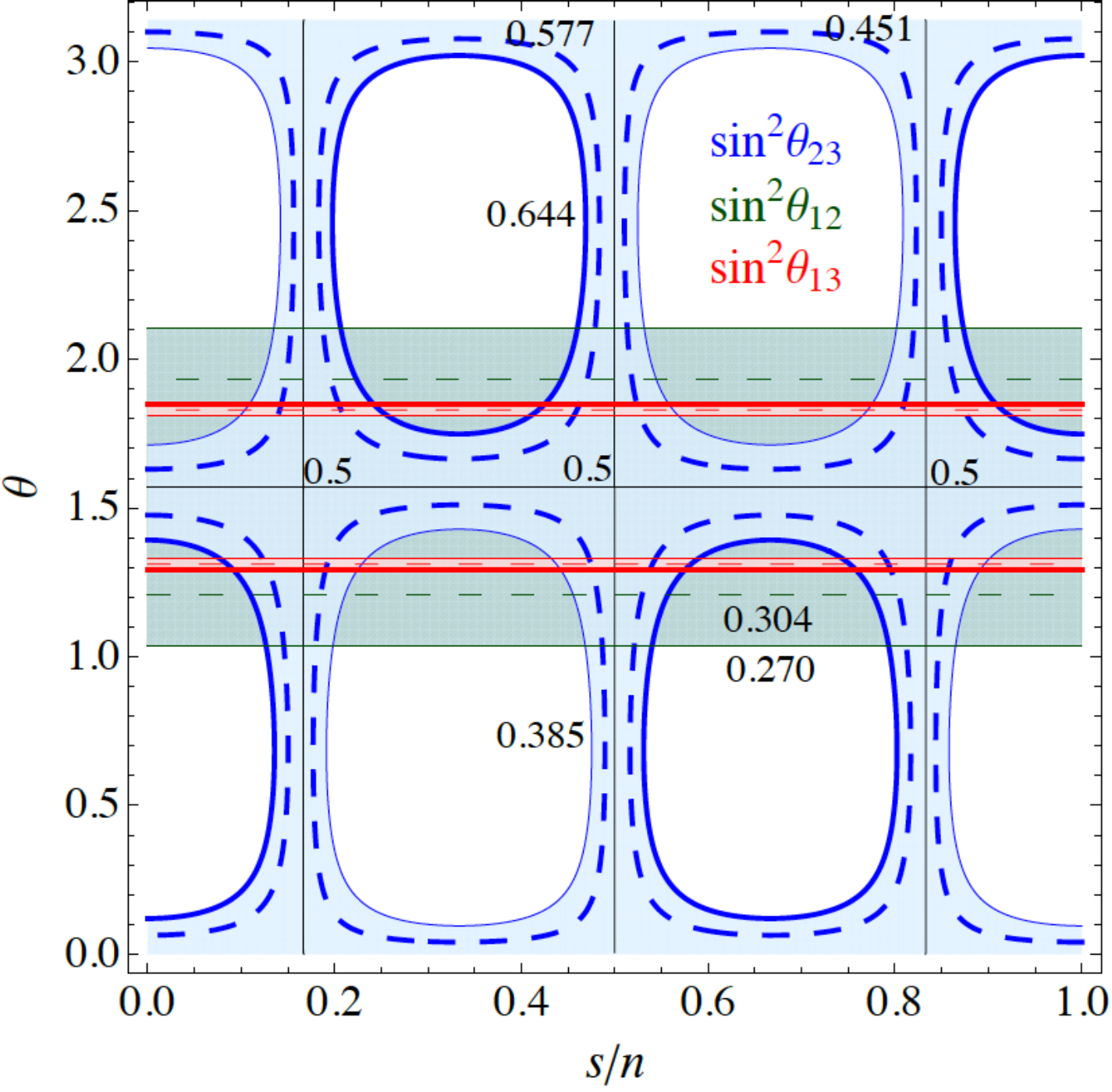}&
\includegraphics[width=0.48\textwidth]{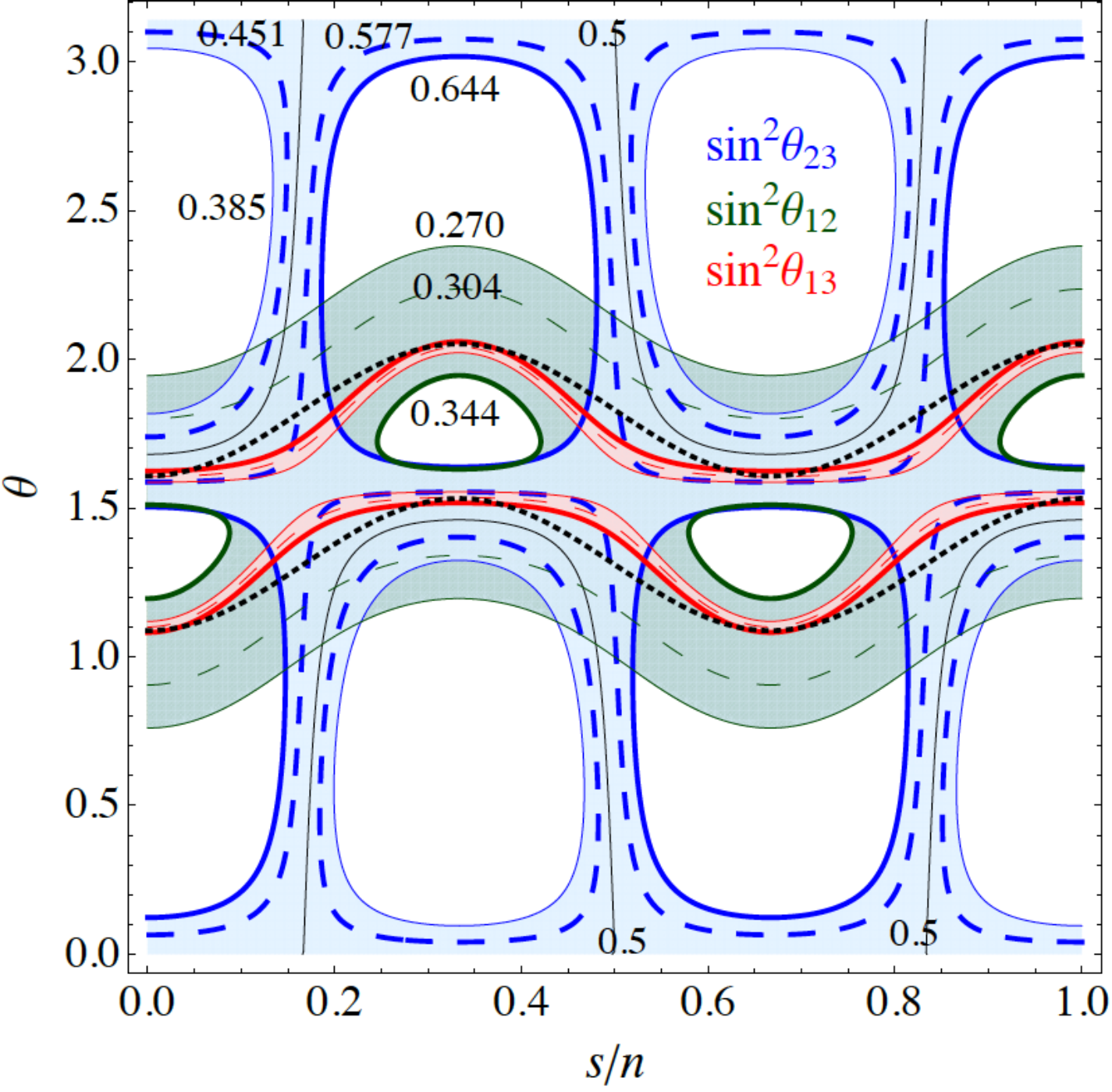}
\end{tabular}
\caption{\small{\textbf{Case 3 b.1)}. Contour plots of $\sin^{2}\theta_{ij}$, obtained from (\ref{anglescase3b1}), in the plane  $\theta$ versus $s/n$ for $n=20$ and $m=n/2=10$ (left panel) and $m=11$ (right panel). We use the same conventions and color coding as in figure \ref{Fig:1}.
The dotted lines in the right panel indicate the approximation given in (\ref{approxdevcase3b1}). The figure corresponding to $n=20$ and $m=9$
can be obtained from the one in the right panel by applying the first transformation shown in table \ref{tab:case3symmetries}, i.e.
we reflect the $3\,\sigma$ contour regions in the line defined by $\theta=\pi/2$ and take into account that $\sin^2\theta_{23}$ becomes
$\cos^2\theta_{23}$ so that the blue colored area represents the region in which $0.385\leq\cos^{2}\theta_{23}\leq0.644$ holds.}}
\label{Fig:7}
\end{center}
\end{figure}
%

\mathversion{bold}
\subsubsection*{Numerical results}
\mathversion{normal}
\label{sec4322}

 We proceed with the presentation of our numerical results for this case. As has been noted, in this case the index $n$ of the group $\Delta (6 \, n^2)$
 can be even as well as odd. Furthermore, the parameter $m$ is constrained by the condition in  (\ref{mnallowedcase3b1}). 
In figures \ref{Fig:7}-\ref{Fig:9} we display the results obtained for mixing angles and CP invariants using the formulae in  (\ref{anglescase3b1}) and (\ref{CPinvcase3b1})
for the choice $n=20$. We do so, since for this value of $n$ not only the choice $m=n/2=10$, but also $m=11$ ($m=9$ as well) allow a reasonably
good fit to the experimental data with $\chi^2_{\rm tot}\lesssim27$ for certain values of  $s$ and the continuous parameter $\theta$. This can be clearly
seen from figure \ref{Fig:7} where we show the $3 \, \sigma$ contour regions of $\sin^2\theta_{ij}$ in the $s/n$-$\theta$ plane
for the case $m=n/2=10$ in the left panel and for $m=11$ in the right one (the color coding is the same as in figure \ref{Fig:1}). 
Since the solar mixing angle  fulfills $\sin^2 \theta_{12} \lesssim 1/3$ for
$m/n=1/2$, see above, no contour line associated with the $3\,\sigma$ upper limit of $\sin^{2}\theta_{12}$ is present in the left panel of figure \ref{Fig:7}.
As has been estimated in (\ref{con3b13}) for $m/n=1/2$ and as is obvious from figure \ref{Fig:7} (as well as confirmed by the results found in table \ref{tab:case3b1neven}), 
the parameter $\theta$ is practically fixed by the requirement to accommodate the reactor mixing angle well. For $m=n/2$
it takes the values in (\ref{con3b13}) independent of $s$, while for the choice $n=20$ and $m=11$ the ``best fitting" $\theta_{\rm bf}$ reveals a certain dependence  
on the parameter $s$ which can be approximated by the expression in (\ref{approxdevcase3b1}). This approximation is presented as dotted curves in the right panel of 
figure \ref{Fig:7} and agrees with the exact result to a certain extent. In the case $m=n/2$ both best fit values of the atmospheric mixing angle,
 $(\sin^{2}\theta_{23})^{\rm bf}=0.451$ and $(\sin^{2}\theta_{23})^{\rm bf}=0.577$, overlap with the red areas, whereas for $n=20$ and $m=11$ only the value 
 $(\sin^{2}\theta_{23})^{\rm bf}=0.577$ has a non-vanishing overlap.
 The figure corresponding to the choice $m=9$ and $n=20$ can be obtained from the one shown in the right panel of figure \ref{Fig:7} by applying the first symmetry
 transformation in table \ref{tab:case3symmetries}, i.e. by reflecting the $3\,\sigma$ contour regions for $m=11$ and $n=20$ in the line defined by $\theta=\pi/2$
where now the blue region indicates $0.385\leq\cos^2 \theta_{23}\leq 0.644$, since $\sin^2\theta_{23}$ is replaced by $\cos^2\theta_{23}$. Consequently, 
values for $\theta_{23}$ smaller than $\pi/4$ (and thus close to $(\sin^{2}\theta_{23})^{\rm bf}=0.451$) are accommodated well.
    
Turning to the CP phases we first discuss them for $n=20$ and $m=n/2=10$. As has been shown in (\ref{sinasinbcase3b1}), the Majorana phases $\alpha$ and $\beta$
do not depend on the parameter $\theta$ in this case. Thus, we only plot  $\sin\delta$ in the $s/n$-$\theta$ plane in the left panel of figure \ref{Fig:8}.
The black areas indicate the regions in which all three lepton mixing angles are within their experimentally preferred $3\,\sigma$ ranges. 
As estimated in (\ref{sindeltaboundcase3b1}), the absolute value of the Dirac phase has a non-trivial lower limit in this case and can also attain a maximal value. We note some peculiarities of the plot
in the left panel of figure \ref{Fig:8}: for $\theta=0, \pi/2$ and $\pi$ the Dirac phase is not physical, since for these values either the reactor or the solar mixing angle
vanishes, as can be seen from
 \begin{equation}
\label{sinsqtheta1312mn2case3b1}
	\sin^{2}\theta_{13} = \frac{1}{3}\,\cos^{2}\theta \;\;\; \mbox{and} \;\;\;
	\sin^{2}\theta_{12} = \frac{\sin^{2}\theta}{2\,+\,\sin^{2}\theta}  \; .
\end{equation}
We can expand $\sin\delta$ around these particular values of $\theta$,  $\theta=\overline{\theta}+\varepsilon$ with $\overline{\theta}=0,\,\pi/2,\,\pi$ and $|\varepsilon|\ll1$,
and find at leading order in $\varepsilon$
\begin{equation}
	\sin\delta \; =\; (-1)^k\; \text{sgn}(\varepsilon)\,\sin3\,\phi_{s} 
\end{equation}
with $k=0$ for $\bar{\theta}=0, \, \pi$ and $k=1$ for $\bar{\theta}=\pi/2$, respectively.
Also the points in the left panel of figure \ref{Fig:8} in which all contour lines converge correspond to unphysical values of the Dirac phase, since in these points
the atmospheric mixing angle either vanishes or becomes $\pi/2$.

\begin{figure}[t!]
\begin{center}
\begin{tabular}{cc}
\includegraphics[width=0.48\textwidth]{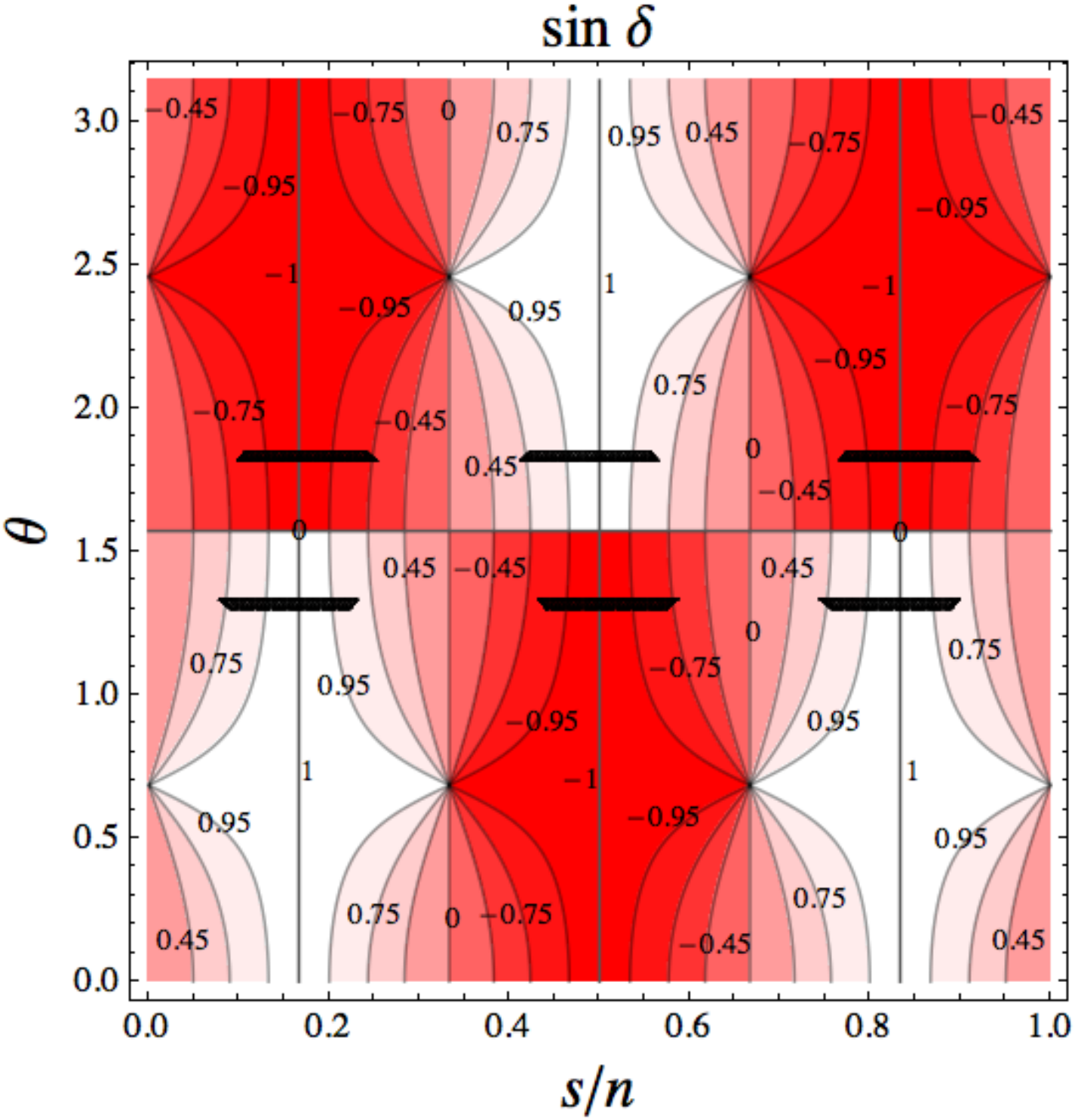}&
\includegraphics[width=0.48\textwidth]{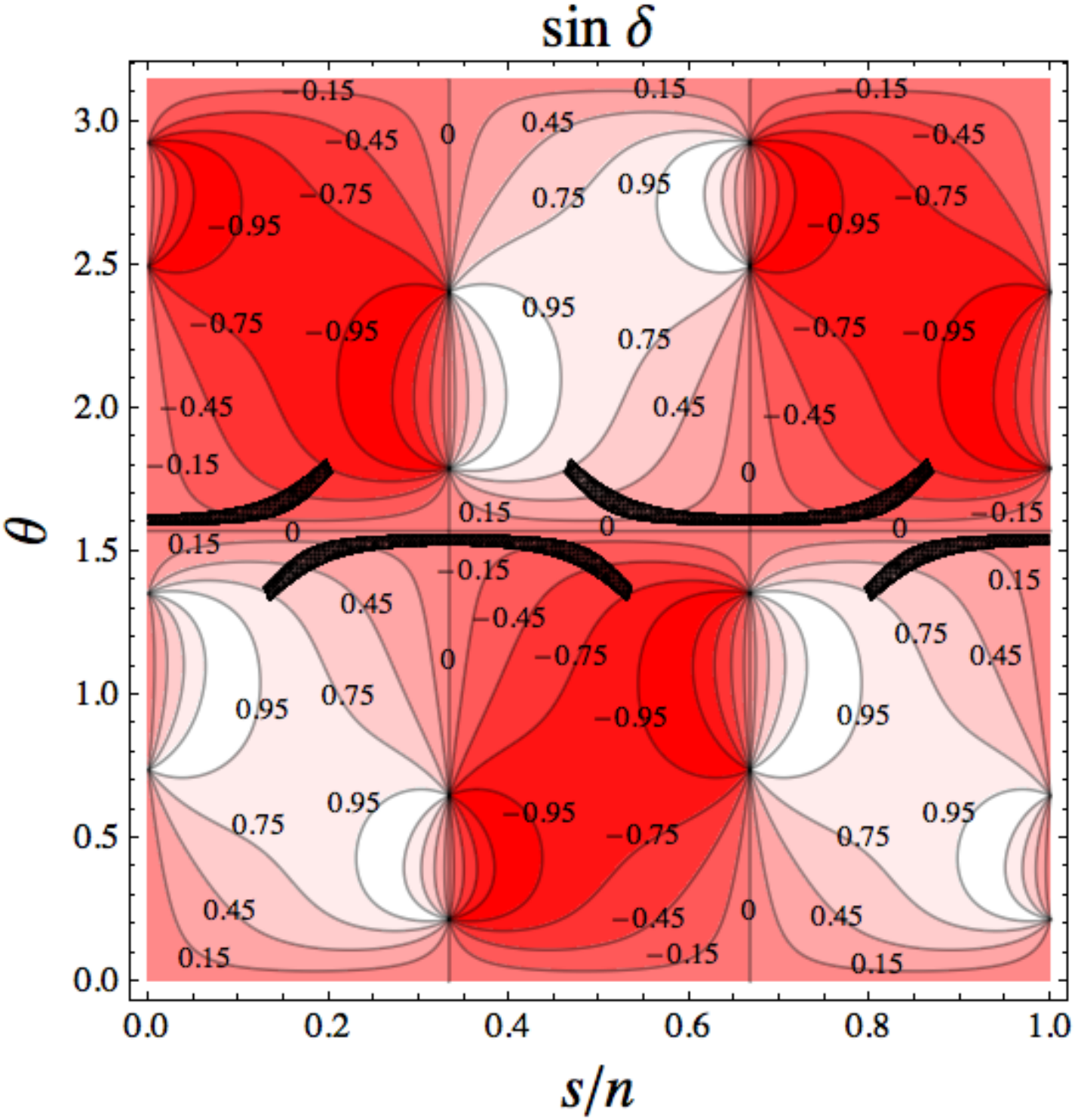}
\end{tabular}
\caption{\small{\textbf{Case 3 b.1)}. Predictions for the Dirac phase $\sin\delta$ for $n=20$, $m=n/2=10$ (left panel) and
and $m=11$ (right panel). The black areas represent the regions in $\theta$ and $s$ for which the lepton mixing angles are compatible with experimental data 
at the $3\,\sigma$ level or better, compare figure~\ref{Fig:7}.}}
\label{Fig:8}
\end{center}
\end{figure}

If we consider $n=20$ and $m=11$ the results for the Dirac phase are different, as can be seen in the right panel of figure \ref{Fig:8}. 
In particular, this phase cannot attain maximal values anymore in the regions in which all three mixing angles are within their experimentally preferred $3\,\sigma$
ranges (black areas in the figure). Instead, its maximal value is $|\sin\delta|\approx 0.75$. Also here the points in which all the contour lines converge indicate
unphysical values of the Dirac phase, since either the reactor, solar or atmospheric mixing angle vanishes or $\theta_{23}=\pi/2$ holds.
As regards the predictions for the Majorana phases $\alpha$ and $\beta$ for $n=20$ and $m=11$, these are displayed in the $s/n$-$\theta$ plane 
in figure \ref{Fig:9}. Again, the black areas indicate the regions in which all three lepton mixing angles are within their $3\,\sigma$ intervals.
Note that we have set $k_1=k_2=0$, when computing $\sin\alpha$ and $\sin\beta$ from (\ref{CPinvcase3b1}). Also in these figures points in which all contour
lines converge correspond to unphysical values of the CP phases, because either the solar (relevant for $\sin\alpha$) or the reactor mixing angle (relevant for $\sin\beta$)
vanishes. The figures of the CP phases for the choice $n=20$ and $m=9$ can be easily deduced from those for $n=20$ and $m=11$, if we apply the
first symmetry transformation in table \ref{tab:case3symmetries}, i.e. for $\sin\alpha$ and $\sin\beta$ the plots are the same as in figure \ref{Fig:9}, replacing only
$\theta$ with $\pi-\theta$, 
whereas for $\sin\delta$ we must not only replace $\theta$ with $\pi-\theta$ in the right panel of figure \ref{Fig:8}, but also change the sign of $\sin\delta$.
\begin{figure}[t!]
\begin{center}
\begin{tabular}{cc}
\includegraphics[width=0.48\textwidth]{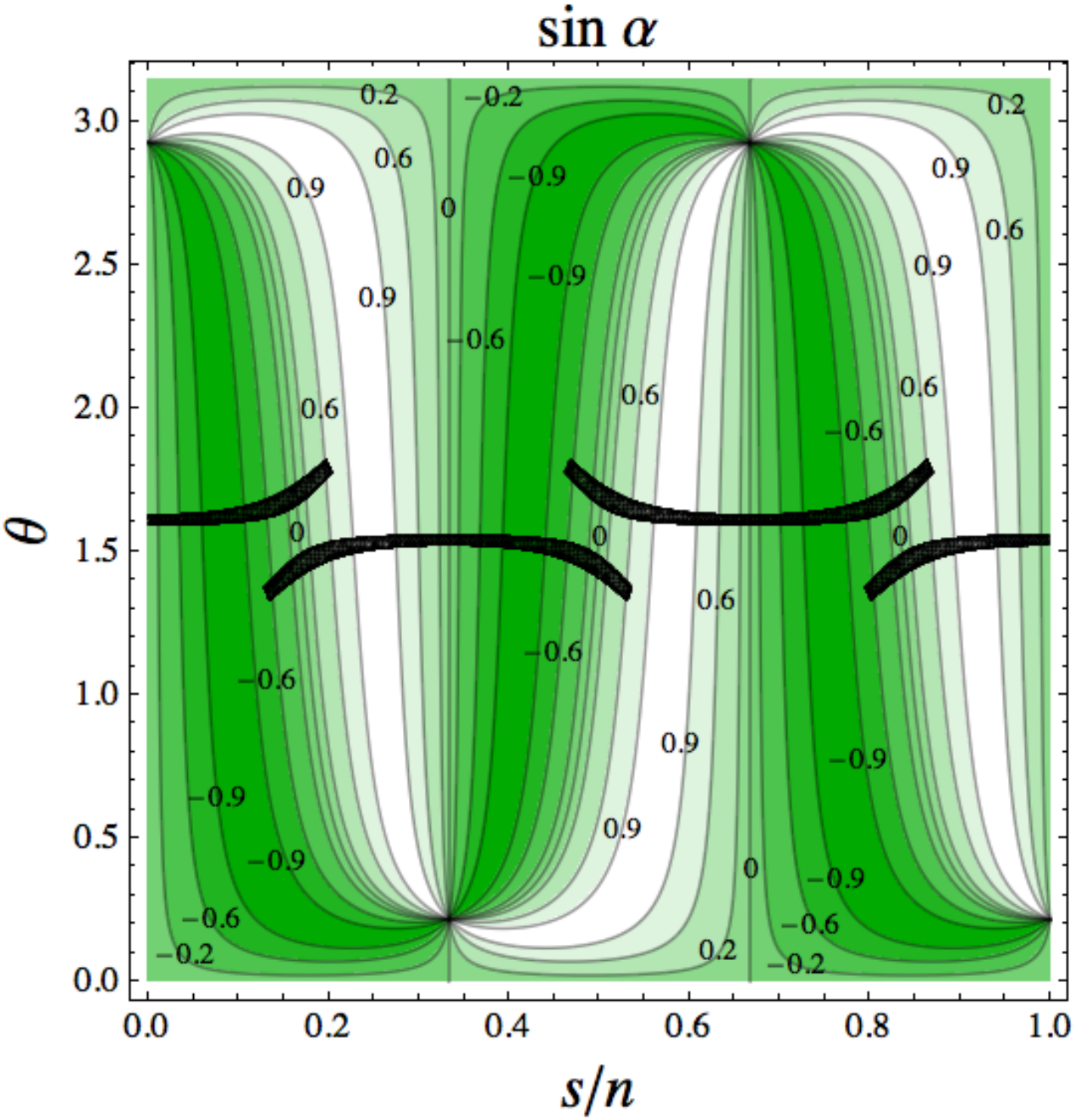}&
\includegraphics[width=0.48\textwidth]{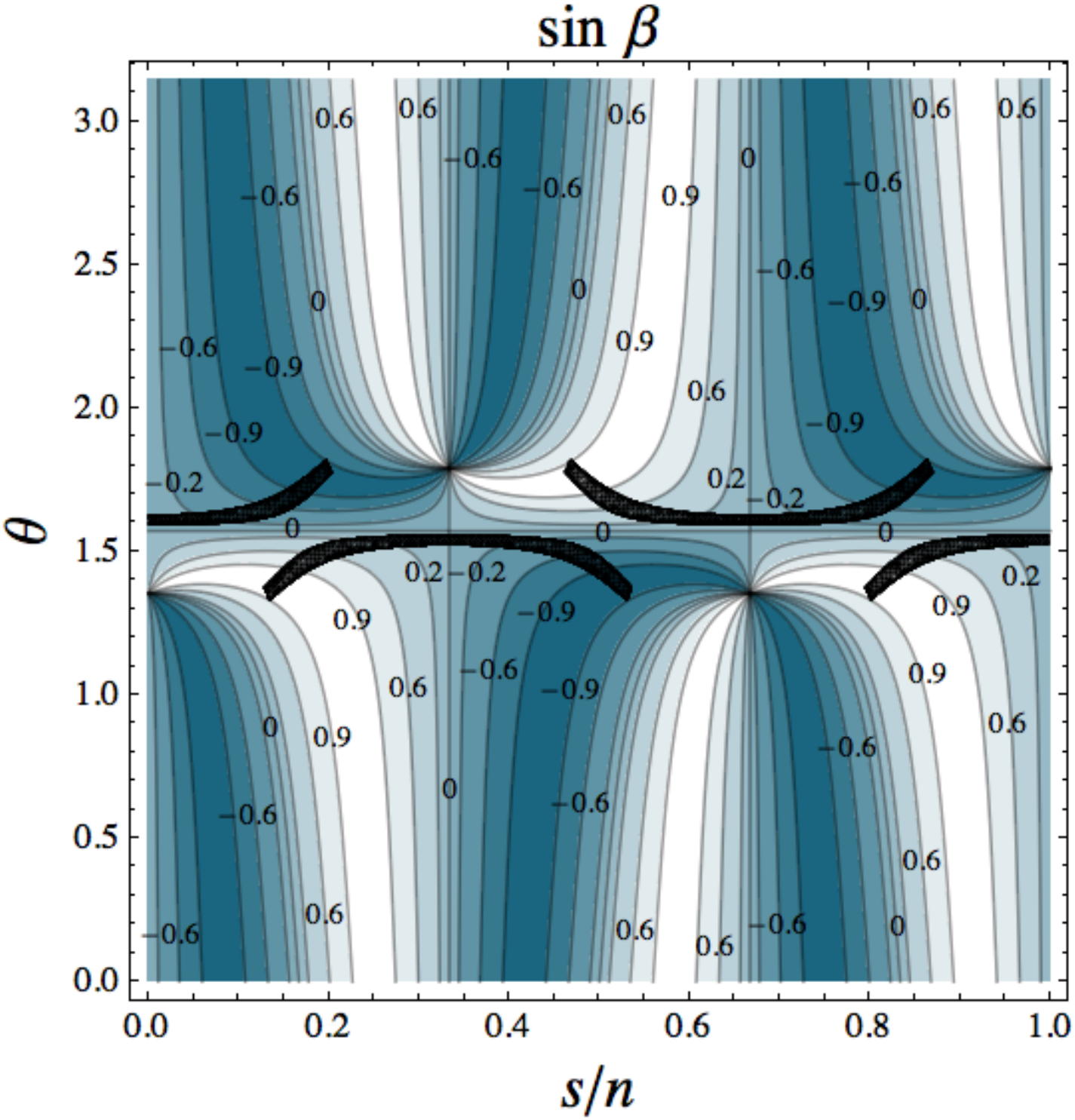}
\end{tabular}
\caption{\small{\textbf{Case 3 b.1)}. Predictions for the Majorana phases $\sin\alpha$ and $\sin\beta$ for $n=20$ and $m=11$ obtained 
from the CP invariants $I_{1,2}$ in (\ref{CPinvcase3b1}) for $k_{1}=k_{2}=0$. Again,
in the black regions all three lepton mixing angles are within their experimentally preferred $3\,\sigma$ intervals, compare figure~\ref{Fig:7}. }}
\label{Fig:9}
\end{center}
\end{figure}
\begin{table}[t!]
\centering
\begin{tabular}{c}
$
\begin{array}{|l||c|c||c|c||c|}
\hline
n & \multicolumn{2}{c||}{8} &  \multicolumn{2}{c||}{10} & \text{even}  \\
 m & \multicolumn{2}{c||}{4} & \multicolumn{2}{c||}{5} & n/2\\\hline
 s & 1 & 2 & 1 & 2 & n/2 \\
\hline
 \chi^{2}_{\rm tot} & 1.44~(2.39) & 7.95 & 4.16~(6.90) & 1.67~(1.57) & 2.13  \\
\hline
  \theta _{\rm bf} & 1.31~(1.83) & 1.83 & 1.31~(1.83) & 1.31~(1.83) & 1.31~(1.83) \\
\hline
 \chi^{2}_{23} & 0.0070~(0.953) & 6.39  & 2.69~(5.43) & 0.234~(0.138) & 0.690 \\
\hline
 \sin^{2}\theta_{23} & 0.579~(0.421) & 0.645 & 0.621~(0.379) & 0.436~(0.564) & 1/2 \\
\hline
 \sin^{2}\theta_{12} & 0.318 & 0.319 & 0.319 & 0.318 & 0.318 \\
\hline
 \sin^{2}\theta_{13} & 0.0220 & 0.0216 & 0.0218 & 0.0220 & 0.0220 \\
\hline
 J_{CP} & \pm 0.0312 & -0.0237 & \pm0.0272 & \pm0.0321 & \mp0.0338 \\
\hline
 \sin\delta  &  \pm0.936 & -0.739 & \pm0.834  & \pm0.959  & \mp1 \\
\hline
 I_1 &  -0.147 & 0.208 & -0.198 & 0.122 & 0\\
\hline
 I_2 &   -0.0104 & 0.0144 & -0.0138 & 0.0086 & 0\\
\hline
 \sin\alpha=\sin\beta & -1/\sqrt{2}\approx-0.707 & 1 & -0.951 & 0.588 & 0\\
 \hline
\end{array}
$
\end{tabular}
\caption{\label{tab:case3b1neven}{\small  \textbf{Case 3 b.1)}. Results of the $\chi^2$ analysis for the smallest even values of $n$ that allow $\chi^2_{\rm tot} \lesssim 27$.
These results are obtained using the formulae in (\ref{anglescase3b1}) and (\ref{CPinvcase3b1}). The integers $k_{1,2}$ are set to zero. 
The fit of the reactor and solar mixing angles contribute $\chi^{2}_{13}\lesssim0.08$ and $\chi^{2}_{12}\lesssim1.5$ to $\chi^2_{\rm tot}$, respectively.
The (absolute) values of $\sin\alpha$ and $\sin\beta$ are always equal for $m=n/2$, see (\ref{sinasinbcase3b1}), and for $n=8$, $m=4$ and $s=2$ we find maximal 
Majorana phases. Note that this case can be ``reduced" to $n=4$, $m=2$ and $s=1$. For choices of the parameter $s>n/2$ results can be obtained by applying the symmetry transformations
in table \ref{tab:case3symmetries} to those presented here. The fact that in most cases two different values of $\theta_{\rm bf}$ lead to a reasonable fit
with $\theta_{23} \gtrless \pi/4$ (and opposite sign for $J_{CP}$) is also observed in the left panel in figure \ref{Fig:7}. The choice of parameters mentioned in the last column
always allows for a good fit of the experimental data, if $n$ is even. Thus, the smallest value of the index $n$ for that this choice can be realized is $n=2$.} }
\end{table}

In  tables~\ref{tab:case3b1neven} and  \ref{tab:case3b1nodd} we present the results of our $\chi^2$ analysis for the smallest 
even and odd values of the index $n$ that allow for $\chi^2_{\rm tot}\lesssim27$
and all lepton mixing angles within their experimentally preferred $3\,\sigma$ ranges, 
using the formulae in (\ref{anglescase3b1}) and (\ref{CPinvcase3b1}) with $k_1=k_2=0$. 
As can be read off from table \ref{tab:case3b1neven},
the smallest even value of $n$ is $n=2$ with $m=1$ and $s=1$ (last column of the table). This case has already been studied in the literature
\cite{S4CPgeneral} and it leads to a maximal Dirac phase and trivial Majorana phases. The smallest even $n$ that also permits non-trivial
Majorana phases is $n=4$ with $m=2$ and $s=1$. This case is implicitly contained in table \ref{tab:case3b1neven}, since the result for $n=8$,
$m=4$ and $s=2$ can be ``reduced" to the former set of $n$, $m$ and $s$ by dividing out the common factor two of all parameters $n$, $m$ and $s$.
This is very similar to what has been described in case 2) (there for the parameters $n$, $u$ and $v$), see discussion in the paragraph below (\ref{I1u0case2}). 
Indeed, in this case both Majorana phases are maximal, while the Dirac phase is large. This is consistent with the findings in \cite{Delta96CP}.
However, in this case the value of the atmospheric mixing angle is very close to the upper $3\,\sigma$ limit \cite{nufit}.
As shown for $n=20$ and $m=n/2=10$ in the left panel in figure \ref{Fig:7}, there are (mostly) two ``best fitting" values $\theta_{\rm bf}$ one leading to 
$\theta_{23}$  smaller than $\pi/4$ and one larger than $\pi/4$. If we consider the particular choice $m=n/2$ and $s=n/2$, the atmospheric mixing
angle is maximal, see (\ref{anglescase3b1}). The value of the reactor mixing angle is accommodated very close to $(\sin^2\theta_{13})^{\rm bf}=0.0219$ in all cases. This
value entails, as explained in the analytical study, $\sin^2\theta_{12}\approx0.318$. Since the two best fitting values $\theta_{\rm bf}$ are related by 
$\theta_{\rm bf, 2}=\pi-\theta_{\rm bf, 1}$, compare to the first symmetry in table \ref{tab:case3symmetries}, the CP invariant $J_{CP}$ (and $\sin\delta$) has opposite signs for
the two values, while the Majorana invariants $I_1$ and $I_2$ are the same. The estimated lower bound for $|\sin\delta|$ mentioned in (\ref{sindeltaboundcase3b1}) is clearly
fulfilled in the cases in table \ref{tab:case3b1neven}. As already observed in (\ref{sinasinbcase3b1}), the sines of the Majorana phases $\alpha$ 
and $\beta$ have the same absolute value in these cases (the sign depends on $k_1$ and $k_2$). In order to obtain numerical results for values of $s$ that are 
not shown in table \ref{tab:case3b1neven} we can make use of the third symmetry transformation in table \ref{tab:case3symmetries}.

\begin{table}[t!]
\centering
\begin{tabular}{c}
$
\begin{array}{|l||c|c|c|c|c|c|c|}
\hline
 s & 0 & 1 & 2 & 3 & 4 & 5 \\
\hline
 \chi^{2}_{\rm tot} & 5.55 & 5.53 & 5.80~(8.49) & 5.54 & 5.55 & 5.51    \\
\hline
  \theta _{\rm bf} & 1.52 & 1.50 & 1.71~(1.38) & 1.64 & 1.63 & 1.67  \\
\hline
 \chi^{2}_{23} & 0.0438 & 0.0157 & 0.290~(2.98) & 0.0321 & 0.0411 & 0.000010  \\
\hline
 \sin^{2}\theta_{23} & 0.463 & 0.458 & 0.434~(0.398) & 0.462 & 0.463 & 0.451 \\
\hline
 \sin^{2}\theta_{12} & 0.332 & 0.332 & 0.332 & 0.332 & 0.332 & 0.332 \\
\hline
 \sin^{2}\theta_{13} & 0.0220 & 0.0220 & 0.0220 & 0.0220 & 0.0220 & 0.0220  \\
\hline
 J_{CP} & 0 & 0.0071 & -0.0166~(0.0234) & -0.0043 & 0.0020 & 0.0110 \\
\hline
 \sin\delta & 0 & 0.209 & -0.489~(0.700) & -0.125 & 0.0592 & 0.323 \\
\hline
 I_1 & 0 & -0.209 & 0.0485~(0.0756) & 0.194 & -0.116 & -0.155 \\
\hline
 \sin\alpha & 0 & -0.986 & 0.228~(0.356) & 0.915 & -0.546 & -0.731\\
\hline
 I_2 & 0 & -0.0063 & 0.0129~(-0.0142) & 0.0038 & -0.0018 & -0.0094 \\
\hline
 \sin\beta  & 0 & -0.436 & 0.894~(-0.992) & 0.267 & -0.127 & -0.652 \\
 \hline
\end{array}
$
\end{tabular}
\caption{\label{tab:case3b1nodd}{\small  \textbf{Case 3 b.1)}. Results for the smallest odd value of $n$ that allows for $\chi^2_{\rm tot}\lesssim27$, namely $n=11$.
Also here the mixing angles and CP invariants are computed using the expressions in (\ref{anglescase3b1}) and (\ref{CPinvcase3b1}) with $k_{1}=k_{2}=0$.
Two values of the parameter $m$, $m=5$ and $m=6$, are admitted by the fit. Here we only display $m=5$, since results for $m=6$ can be obtained via the 
symmetry transformations in table \ref{tab:case3symmetries}. The most notable difference lies in the fact that $m=6$ usually leads to  $\sin^{2}\theta_{23}>1/2$
at $\theta_{\rm bf}$ and that $\chi^2_{\rm tot}$ is slightly larger than for $m=5$, i.e. $\chi^{2}_{\rm tot}\gtrsim 6$. 
With the help of table \ref{tab:case3symmetries} also the results for $s>n/2=11/2$ can be obtained, showing that for all
values of the parameter $s$, $0\leq s\leq 10$, reasonable fits are possible. 
The fit of reactor and solar mixing angles contributes $\chi^{2}_{13}\lesssim0.02$ and $\chi^{2}_{12}\lesssim5.5$ to $\chi^2_{\rm tot}$, respectively.
As explained, the choice $s=0$ implies the presence of an accidental CP symmetry entailing trivial CP phases.}}
\end{table}
\begin{table}[t!]
\centering
\begin{tabular}{c}
$
\begin{array}{|l||c|c|c||c|c|c|c|c|c|}
\hline
n & \multicolumn{3}{c||}{5}  & \multicolumn{5}{c|}{8}  \\
 m &  \multicolumn{3}{c||}{4} & \multicolumn{5}{c|}{7} \\\hline
 s & 0 & 1 & 2 & 0 & 1 & 2 & 3 & 4 \\
\hline
 \chi^{2}_{\rm tot} & 5.06 & \ba3.61\\(8.85)\ea & 5.76  & 7.68  & \ba6.02\\(12.4)\ea  & 7.10  & 7.56 & 4.69  \\
\hline
  \theta _{\rm bf} & 1.68 & \ba1.40\\(1.84)\ea & 1.45  & 1.50 & \ba1.45\\(1.84)\ea & 1.66 & 1.65 & \ba1.39\\(1.75)\ea\\
\hline
 \chi^{2}_{23} & 1.72 & \ba0.308\\(5.53)\ea & 2.42 & 2.93 &  \ba1.33\\(7.65)\ea & 2.38 & 2.82 & 0.0080  \\
\hline
 \sin^{2}\theta_{23} & 0.531 & \ba0.484\\(0.378)\ea & 0.523  & 0.517 & \ba0.537\\(0.652)\ea  & 0.523 & 0.518 & 0.574\\
\hline
 \sin^{2}\theta_{12} & 0.326 & 0.326 & 0.326  & 0.330 & 0.330 & 0.330 & 0.330 & 0.330\\
\hline
 \sin^{2}\theta_{13} & 0.0222 & \ba0.0220\\(0.0219)\ea & 0.0222  & 0.0218 & \ba0.0219\\(0.0218)\ea & 0.0218 & 0.0218 & 0.0220 \\
\hline
 J_{CP} & 0 & \ba-0.0208\\(0.0311)\ea & 0.0094  & 0 & \ba-0.0141\\(0.0296)\ea & 0.0080 & -0.0036 & \pm0.0225\\
\hline
 \sin\delta  & 0 & \ba-0.612\\(0.945)\ea & 0.276  & 0 & \ba-0.416\\(0.914)\ea & 0.236 & -0.107 & \pm0.667\\
\hline
 I_1 & 0 & \ba0.115\\(0.136)\ea & -0.201  & 0 & \ba-0.143\\(-0.163)\ea & 0.212 & -0.151 & \mp0.0144\\
\hline
 \sin\alpha & 0 & \ba0.547\\(0.647)\ea & -0.958  & 0 & \ba-0.676\\(-0.769)\ea & 0.9997 & -0.715 & \mp0.0683\\
\hline
 I_2 & 0 & \ba0.0142\\(-0.0068)\ea & -0.0080 & 0 & \ba-0.0114\\(0.0082)\ea & 0.0069 & -0.0032 &\pm 0.0144\\
\hline
 \sin\beta & 0 & \ba0.981\\(-0.469)\ea & -0.544 & 0 & \ba-0.793\\(0.575)\ea & 0.484 & -0.225 & \pm1\\
 \hline
\end{array}
$
\end{tabular}
\caption{\label{tab:case3b1shift}{\small  \textbf{Case 3 b.1)}. Results of the $\chi^2$ analysis for $n=5$ ($m=4$) and $n=8$ ($m=7$) that are the smallest
odd and even values of the index $n$ for which $\chi^2_{\rm tot}\lesssim 27$, if we consider the formulae in (\ref{anglescase3b1}) and (\ref{CPinvcase3b1})
for $m$ and $\theta$ being replaced by $m-\frac n3$ and $\pi-\theta$, respectively. The parameters $k_{1,2}$ in (\ref{CPinvcase3b1}) are taken to be zero.
The contributions to $\chi^2_{\rm tot}$ arising from the fit of the reactor and the solar mixing angles 
are $\chi^{2}_{13}\lesssim 0.08~(0.02)$ and $\chi^{2}_{12}\lesssim 3.3~(4.7)$ for $n=5~(8)$, respectively.
For $s=0$ all CP phases are trivial due to the presence of an accidental CP symmetry, see (\ref{Y1ts0case3a}).
Taking into account the third symmetry transformation in table \ref{tab:case3symmetries} we see that all admitted values of $s$, $0\leq s\leq n-1$, allow
for a good fit. The other symmetries in that table show that for $n=5$, $m=1$ and $n=8$, $m=1$ also reasonable fits are obtained, belonging 
to a mixing pattern given by the formulae in (\ref{anglescase3b1}) and (\ref{CPinvcase3b1}) with $m$ and $\theta$ replaced by $m+\frac n3$
and $\pi-\theta$, respectively. In the case $n=8$, $m=7$ and $s=4$ the first symmetry in table \ref{tab:case3symmetries} explains the presence of two different values for 
$\theta_{\rm bf}$ (given by $\theta$ and $\pi-\theta$) leading to the same best fitted values of the mixing angles and opposite signs for the three CP invariants. 
In addition, $s=n/2$ explains why the (absolute) values of $I_1$ and $I_2$ are identical.}} 
\end{table}

The smallest value of an odd index $n$ that allows for $\chi^2_{\rm tot}\lesssim27$ is $n=11$ and we display results for $m=5$ (remember $m/n\approx1/2$ is required)
and $n=11$ in table \ref{tab:case3b1nodd}. In this case for all admitted values of the parameter $s$, $0 \leq s \leq 10$, (at least) one value of $\theta$ can be found 
for which all lepton mixing angles are fitted reasonably well. In table \ref{tab:case3b1nodd} only values $s<n/2=11/2$ are presented, since the results for those larger than $s=5$
can be obtained from table \ref{tab:case3b1nodd} by exploiting the symmetry transformations in table \ref{tab:case3symmetries}. Only in the case $s=0$ all CP phases vanish,
since in this case an accidental CP symmetry is present in the charged lepton and neutrino sectors, see (\ref{Y1ts0case3a}). As observed for $n=20$ and $m=11$, see right
panel in figure \ref{Fig:8}, also here the value of $\delta$ cannot be maximal, $|\sin\delta| \lesssim 0.7$. For $n=11$ not only $m=5$, but also the choice $m=6$
leads to a good agreement with the experimental data on lepton mixing angles. Results for this case can be obtained, as before, by applying the symmetry transformations
in table \ref{tab:case3symmetries}. For the choice $m=6$ in general the value of the atmospheric mixing angle is larger than $\pi/4$ at the best fitting point(s) $\theta_{\rm bf}$.
The values of $\chi^2_{23}$ obtained in these cases fulfill $\chi^2_{23} \gtrsim 0.6$.

If we consider instead the formulae in (\ref{anglescase3b1}) and (\ref{CPinvcase3b1}) with the replacements $m\;\rightarrow\;m-\frac n3$ and $\theta\;\rightarrow\;\pi-\theta$,
i.e. we want to study the results for a different permutation of the PMNS mixing matrix in (\ref{Unucase3}) than before, the smallest odd and even values of $n$ leading to a good fit are
$n=5$ and $n=8$, respectively. In particular, the case $n=5$ that requires the choice $m=4$ is interesting, since the associated flavor group $\Delta (150)$ 
is quite small.\footnote{This group has also been discussed in \cite{D150D600,Lamscan}.} The results are shown in table \ref{tab:case3b1shift}. For all values of the parameter $s$, $0\leq s\leq4$, a reasonably
good fit to the experimental data can be achieved and we choose as representatives $s\leq2$, since the results for the other two values $s=3$ and $s=4$ can be 
straightforwardly deduced from table \ref{tab:case3b1shift} using table \ref{tab:case3symmetries}. Similarly, the choice $n=8$ and $m=7$ allows to accommodate
the mixing angles well for all possible choices of the parameter $s$ for a certain value $\theta_{\rm bf}$. The results for the values $s>4$ are not displayed in table \ref{tab:case3b1shift}, but can be obtained from the latter with the help of the symmetry transformations in table \ref{tab:case3symmetries}. Note that for $n=8$, $m=7$
and $s=4$ due to the choice $s=n/2$ two values of $\theta_{\rm bf}$ lead to the same reasonable fit to the experimental data. These two values $\theta_{\rm bf, 1}$ 
and $\theta_{\rm bf, 2}$ are related by $\theta_{\rm bf, 2}=\pi-\theta_{\rm bf, 1}$, see third symmetry transformation in table \ref{tab:case3symmetries}. Furthermore,
we see that in this case all CP invariants for $\theta_{\rm bf, 1}$ have opposite sign as those for $\theta_{\rm bf, 2}$. In addition, the choice $s=n/2=4$ tells
us that the CP invariants $I_1$ and $I_2$ have to have the same absolute values (their signs, obviously, depend on the choice of $k_1$ and $k_2$, see (\ref{CPinvcase3b1})).
This observation is independent from the other parameters $n$, $m$ and $\theta$.  
As expected in both cases, $n=5$ and $n=8$, the choice $s=0$ leads to an accidental CP symmetry that enforces trivial CP phases, see (\ref{Y1ts0case3a}). 
Finally, we notice that exploiting the relation between results for $m$ and $n-m$, see table \ref{tab:case3symmetries}, 
we find that also $n=5$ and $m=1$ as well as $n=8$ and $m=1$ allow us to accommodate the experimental data well.
(Note that these solutions correspond to a different permutation of the PMNS mixing matrix in (\ref{Unucase3}), i.e. the one that leads to formulae for mixing angles and 
CP invariants in (\ref{anglescase3b1}) and (\ref{CPinvcase3b1}) with $m$ and $\theta$ replaced by $m+\frac n3$ and $\pi-\theta$.)
While the results for the reactor and the solar mixing angles are the same as for the displayed cases, the atmospheric mixing angle takes values in the opposite octant, according
to the fact that $\sin^2\theta_{23}$ becomes replaced by $\cos^2\theta_{23}$. 
In this case, the value of $\chi^2_{\rm tot}$ is for almost all values of $s$ smaller than the one reported in table~\ref{tab:case3b1shift} -- except for $s=1$.

\newpage
\mathversion{bold}
\subsubsection{Case 3 b.2)}
\mathversion{normal}
\label{sec433}

The last type of mixing pattern can be obtained from the PMNS mixing matrix $U_{PMNS, 3}$ in (\ref{Unucase3}) by exchanging its second and third columns.
Since this corresponds to taking the PMNS mixing matrix of case 3 b.1) and exchanging its first and second columns, we find the same results
for  $\sin^2 \theta_{13}$ and $\sin^2 \theta_{23}$ as in (\ref{anglescase3b1}), while $\sin^2 \theta_{12}$ becomes $\cos^2 \theta_{12}$ in this case. Furthermore,
the signs of $J_{CP}$ and of $I_1$ are changed with respect to those in (\ref{CPinvcase3b1}), whereas $I_2$ has now a different dependence 
on the parameters 
\begin{equation}
\label{I2case3b2}
I_2=\frac{1}{9} \, (-1)^{k_1+1} \cos \phi_m \, \sin 3 \,\phi_s \, \left( 4 \cos \phi_m \cos 3 \,\phi_s \cos 2 \theta + \sqrt{2} \, \cos 2 \,\phi_m \sin 2\theta \right) \; . 
\end{equation}
Again, the results for mixing angles and CP invariants obtained for the remaining eleven permutations are related through shifts in $\theta$ and/or in the parameter $m$
to those presented here. 
The crucial difference between this case and case 3 b.1) is the change of $\sin^2 \theta_{12}$ into $\cos^2 \theta_{12}$, i.e.
$\sin^2 \theta_{12}$ can now be written as
\begin{equation}
\label{sinth12sqinsinth13sqcase3b2}
\sin^2 \theta_{12} = \frac{2 \, \sin^2 \phi_m}{3 \, (1-\sin^2 \theta_{13})} \; .
\end{equation}
Again, its value should be close to $1/3$ meaning that
\begin{equation}
\label{cond13b2}
\sin^2 \phi_m \approx \frac 12
\end{equation}
has to be fulfilled that requires in turn\footnote{The index $n$ should be divisible by four or the formulae have to be modified in such a way that $n$ on the right-hand side
of the equations is replaced by $n\pm k$ with $k$ chosen so that $n\pm k$ is divisible by four.} 
\begin{equation}
\label{mnallowedcase3b2}
m \approx \frac n4 \;\;\; \mbox{or} \;\;\; m \approx \frac{3\, n}{4} \; .
\end{equation}
Furthermore, we see 
\begin{equation}
\sin \phi_m \approx \frac{1}{\sqrt{2}} \;\; , \;\; \cos \phi_m \approx \pm \frac{1}{\sqrt{2}} \;\; , \;\;  \cos 2 \,\phi_m \approx 0 \;\; , \;\; \sin 2 \,\phi_m \approx \pm 1 \;\; , \;\; 
\end{equation}
with ``+", if $m/n\approx1/4$, and ``-" for $m/n\approx3/4$. The parameter $p$ in (\ref{pcase3b}) is required to fulfill $p \approx -1$, as can be derived from (\ref{anglespqcase3b}) when neglecting $\theta_{13}$, and it implies here
\begin{equation}
\label{cond23b2}
\cos 3 \,\phi_s \sin 2 \theta \approx \mp 1
\end{equation}
with ``-" for $m/n\approx1/4$ and ``+" for $m/n\approx3/4$. At the same time, this condition tells us that $\sin 2\theta \approx \pm 1$ and hence also 
$\sin \theta \approx \frac{1}{\sqrt{2}}$ for $0 \leq \theta < \pi$ and consequently we find that $q$ in (\ref{qcase3b}) is determined
\begin{equation}
\label{cond33b2}
q \approx \pm \, \frac 32 \;\;\; \mbox{with} \;\;\; ``+" \;\;\; \mbox{for} \;\;\; m/n\approx1/4 \;\;\; \mbox{and} \;\;\; ``-"   \;\;\; \mbox{for} \;\;\; m/n\approx3/4 \; ,
\end{equation}
such that the atmospheric mixing angle in (\ref{anglespqcase3b}) results to be
\begin{equation}
\label{sinth23sqcase3b2}
\sin^2 \theta_{23} \approx \frac 12 \left( 1 \mp \frac{\sqrt{3}}{2} \right) \approx \left\{ \begin{array}{c} 0.067 \;\;\; \mbox{for} \;\;\; m/n\approx1/4 \\ 0.933 \;\;\; \mbox{for} \;\;\; m/n\approx3/4 
\end{array} \right. \; ,
\end{equation}
i.e. this mixing angle cannot be in accordance with experimentally measured values, if $\theta_{13}$ and $\theta_{12}$ are accommodated well.
A refined numerical analysis, e.g. taking into account the non-zero value of $\theta_{13}$, confirms this result as can be clearly seen from figure \ref{Fig:10}
in which we display the $3\,\sigma$ contour regions for $\sin^2 \theta_{ij}$ in the $s/n$-$\theta$ plane,
using the expressions of the lepton mixing angles in case 3 b.2) for $m/n=1/4$. As can be checked, also for the choice $m/n=3/4$
the experimentally preferred $3 \, \sigma$ ranges of the three different mixing angles do not overlap.

\begin{figure}[t!]
\begin{center}
\begin{tabular}{cc}
\includegraphics[width=0.48\textwidth]{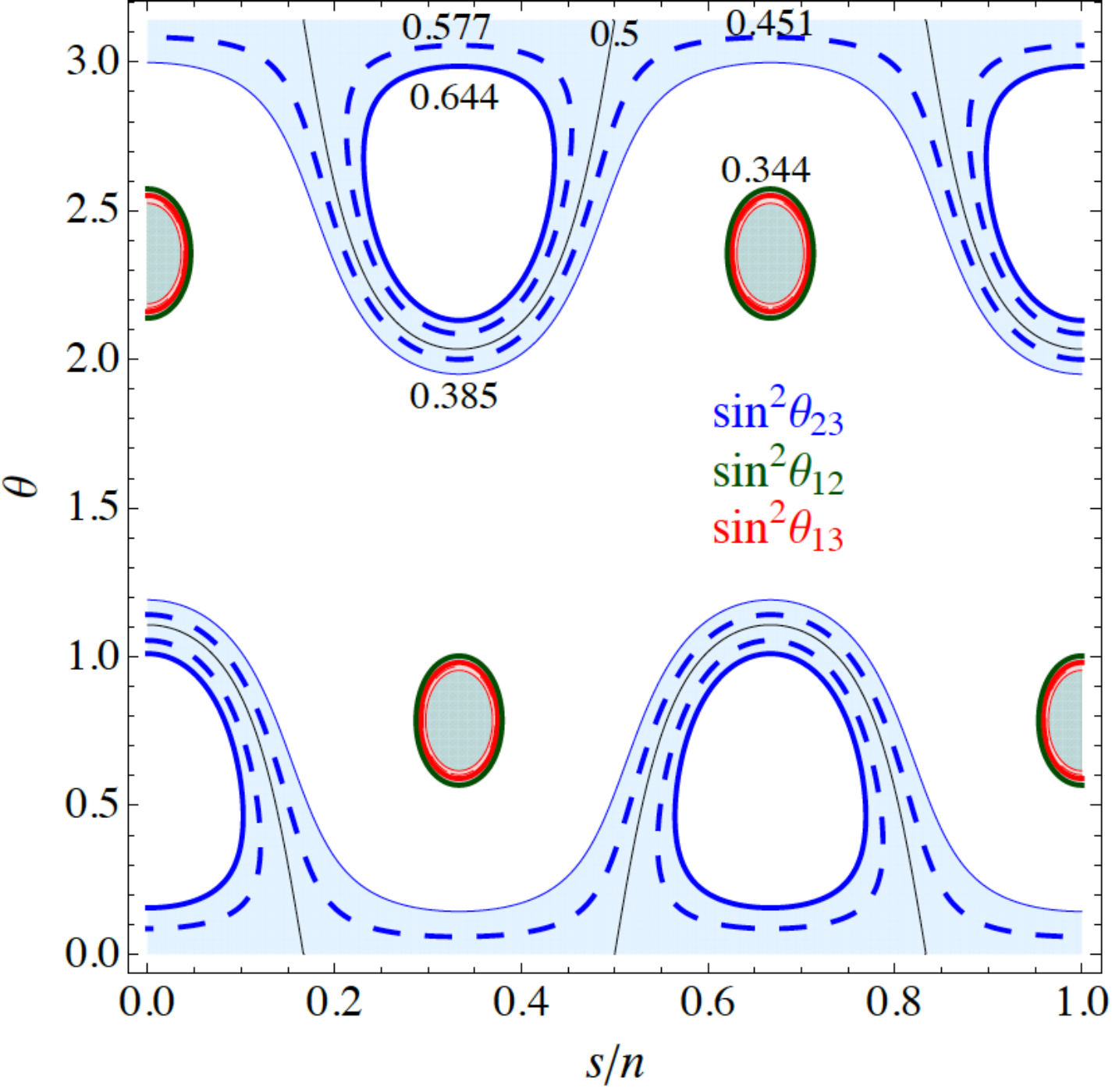}
\end{tabular}
\caption{\small{\textbf{Case 3 b.2)}. Similar to figure~\ref{Fig:7} we plot the $3\,\sigma$ contour regions of $\sin^{2}\theta_{ij}$ in the $s/n$-$\theta$ plane for $m/n=1/4$. 
As one can clearly see, it is impossible to accommodate simultaneously all three lepton mixing angles well in case 3 b.2).}}
\label{Fig:10}
\end{center}
\end{figure}
%
\section{Summary and conclusions}
\label{concl}

We have analyzed in detail lepton mixing patterns that arise from a theory in which a flavor symmetry $G_f=\Delta (3 \, n^2)$ or $G_f=\Delta (6 \, n^2)$ (with an index $n$
not divisible by three)
and a CP symmetry are broken to residual groups $G_e=Z_3$ and $G_\nu=Z_2 \times CP$ in the charged lepton and neutrino sectors, respectively. 
All mixing angles and CP phases are determined by
group theoretical indices (characterizing the flavor group and the generators of the residual symmetries as well as the CP transformation $X$,
representing the CP symmetry) and by one continuous parameter $\theta$, that can take values between $0$ and $\pi$.
We have studied all possible $Z_3$ and $Z_2$ subgroups of $\Delta (3\, n^2)$ and $\Delta (6 \, n^2)$ that can function as residual
symmetries. As regards the CP symmetry, we have focussed on a set of CP transformations that can be consistently combined with $G_f$ as well as with the residual $Z_2$
group in the neutrino sector. Furthermore, we have dealt with the question whether
these CP transformations can correspond to `class-inverting' automorphisms for $\Delta (3 \, n^2)$ and $\Delta (6 \, n^2)$.

We have shown that it is sufficient to 
 consider only three types of combinations of residual symmetries in the charged lepton and neutrino 
 sectors, represented by case 1), case 2), case 3 a) and case 3 b.1), in order to comprehensively discuss lepton mixing.  Especially, the generator of the residual 
symmetry $G_e$ can always be fixed to the generator $a$ of $\Delta (3 \, n^2)$ and $\Delta (6 \, n^2)$. 
Due to the choice of the $Z_2$ symmetry the first two types of combinations, case 1) and case 2), can be realized for $G_f=\Delta (3 \, n^2)$ as well as $G_f=\Delta (6 \, n^2)$, whereas the third type of combination, case 3 a) and case 3 b.1),
is only admitted for $\Delta (6 \, n^2)$. Furthermore, the choice of the $Z_2$ group constrains the index $n$ of the flavor group to be even for the first two types.
Interestingly enough, this choice is also responsible for the fact that the second column of the PMNS mixing matrix has to be trimaximal in case 1) and case 2). 

The mixing angles derived from the first type of combination only depend on the continuous parameter $\theta$ and their experimentally preferred values can be accommodated well
for certain choices of $\theta$. The Dirac phase and one of the Majorana phases vanish, while the value of the other Majorana phase
depends on the chosen CP transformation $X$. The second type of combination instead 
leads to mixing angles and two CP phases, $\delta$ and $\beta$, that depend on two parameters,  $\theta$ as well as on an integer related to the choice of the CP transformation. 
 The Majorana phase $\alpha$ is fixed not only by these two, but in addition by a third parameter that also characterizes 
the CP transformation $X$. As a consequence, for each set of parameters that leads to mixing angles in good agreement with the experimental data, see tables   
 \ref{tab:caseu}-\ref{tab:caseushiftpminus2}, we can obtain 
a variety of different values of the Majorana phase $\alpha$,
see figure \ref{Fig:4}.  We find
that for small and moderate values of the index $n$, $2 \leq n \leq 20$, the data on lepton mixing angles can be accommodated very well
for certain choices of the CP transformation $X$ and the parameter~$\theta$. 

The third type of combination allows for a richer structure of mixing patterns and, indeed, we can divide the resulting
mixing patterns in two categories, case 3 a) and case 3 b.1): for case 3 a) 
the reactor and the atmospheric mixing angles are determined by the choice of the residual $Z_2$
symmetry in the neutrino sector (characterized by the parameter $m$) 
and by the index $n$ of the flavor group, while the solar mixing angle as well as the CP invariants depend, in general,
also on the continuous parameter $\theta$ and the choice of the CP transformation $X$. We find that for a good agreement with the experimental
data the index $n$ has to be at least $n=11$. The solar mixing angle can be fitted to its best fit value in most cases, see tables \ref{tab:case3an16}-\ref{tab:case3an11shift}. 
The CP phases are in general all non-trivial (unless a certain CP transformation $X$ is employed). The Dirac as well as the Majorana phase 
$\alpha$ can obtain (close to) maximal values, while the absolute value of the other Majorana phase has a non-trivial upper bound, $|\sin\beta| \lesssim 0.87$, 
see figure \ref{Fig:6}. The mixing pattern 
belonging to the second category, case 3 b.1), reveals the most complex structure, since all mixing angles and CP invariants depend on the parameter $\theta$, the choice
of the residual $Z_2$ symmetry as well as on the choice of the CP transformation $X$. However, the condition to accommodate the mixing angles well strongly 
constrains the choice of the $Z_2$ group, i.e. the parameter $m$, as well as the value of $\theta$, see (\ref{mnallowedcase3b1}), (\ref{con3b13}) and 
figure \ref{Fig:7}. For the various choices of $X$
different predictions for the CP phases, as shown in figures \ref{Fig:8} and \ref{Fig:9}, are obtained. In particular, for the choice $m=n/2$ the sines of the 
Majorana phases turn out to be equal up to a sign and to depend only on the choice of CP transformation, while the Dirac phase 
has in general a non-trivial  lower bound, $|\sin\delta| \gtrsim 0.71$.
If $m$ is not chosen as $n/2$, also smaller values are obtained for $|\sin\delta|$ and the Majorana phases mildly depend on the continuous
parameter $\theta$. As shown in tables \ref{tab:case3b1neven}-\ref{tab:case3b1shift}, reasonably good agreement with the experimental 
data is achieved for small values of the index $n$, corresponding to a moderately sized flavor group $\Delta (6 \, n^2)$. 
 In particular, we find that $n=5$, i.e. $\Delta (150)$, admits a very good fit to the mixing angles together with non-vanishing and also non-maximal
values of all three CP phases, see table \ref{tab:case3b1shift}.

Given the promising results obtained here it is worth to extend our study. For example, we could consider 
other choices for the residual symmetry $G_e$ in the charged lepton sector for $G_f=\Delta (6 \, n^2)$ or we could 
employ a different set of CP transformations. It would also be interesting to exploit the presented results  in studies of phenomena
that  involve CP phases,
such as  neutrinoless double beta decay and leptogenesis. Furthermore, the construction of concrete models in which the breaking pattern of the flavor and CP symmetry
is achieved dynamically, see e.g. \cite{A4S4CPmodels,Delta48CP,Delta96CP}, is another interesting direction, since in such models also constraints on the 
lepton mass spectrum can be achieved.

\section*{Acknowledgements}

CH thanks Michael A. Schmidt and Thomas Schwetz for discussions as well as the Excellence Cluster `Universe' at the Technische
Universit\"{a}t M\"{u}nchen for the generous support of a visit of AM in Munich. CH would also like to thank the organizers of the 
``Invisibles 14 Workshop'' where parts of these results have been presented. 
EM thanks Maximilian Fallbacher and Andreas Trautner for discussions.
AM acknowledges MIUR (Italy)  for financial support under
the program ``Futuro in Ricerca 2010'', (RBFR10O36O) and the Excellence Cluster `Universe' at the Technische
Universit\"{a}t M\"{u}nchen. The work of EM is supported by the ERC Advanced Grant project ``FLAVOUR'' (267104).
EM acknowledges the kind hospitality of the theory group of Roma Tre University at the final stages of the work.

\appendix

\mathversion{bold}
\section{Conventions for mixing angles and CP invariants, global fit results and $\chi^2$ analysis}
\mathversion{normal}
\label{app1}

In this appendix we fix our conventions for mixing angles and the CP invariants $J_{CP}$, $I_1$ and $I_2$, list the latest global
fit results \cite{nufit} and describe our $\chi^{2}$ analysis.  

\subsection{Conventions for mixing angles and CP invariants}
\label{app11}

As parametrization of the PMNS mixing matrix we use  
\be
U_{PMNS} = \tilde{U} ~{\rm diag}(1, e^{i \alpha/2}, e^{i (\beta/2 + \delta)})~~~, 
\ee
with $\tilde{U}$ being of the form of the Cabibbo-Kobayashi-Maskawa (CKM) matrix $V_{CKM}$ \cite{pdg}
\be
\tilde{U} =
\begin{pmatrix}
c_{12} c_{13} & s_{12} c_{13} & s_{13} e^{- i \delta} \\
-s_{12} c_{23} - c_{12} s_{23} s_{13} e^{i \delta} & c_{12} c_{23} - s_{12} s_{23} s_{13} e^{i \delta} & s_{23} c_{13} \\
s_{12} s_{23} - c_{12} c_{23} s_{13} e^{i \delta} & -c_{12} s_{23} - s_{12} c_{23} s_{13} e^{i \delta} & c_{23} c_{13}
\end{pmatrix}
\ee 
and $s_{ij}=\sin\theta_{ij}$ and $c_{ij}=\cos\theta_{ij}$.  The mixing angles $\theta_{ij}$ range from $0$ to $\pi/2$, while the Majorana phases 
$\alpha, \beta$ as well as the Dirac phase $\delta$ take values between $0$ and $2 \, \pi$. The Jarlskog invariant 
$J_{CP}$ reads \cite{jcp}
\begin{eqnarray} \nonumber
J_{CP} &=&  {\rm Im} \left[ U_{PMNS,11} U_{PMNS,13}^* U_{PMNS,31}^* U_{PMNS,33}  \right] 
 \\ \label{JCP}
 &=& \frac 18 \sin 2 \theta_{12} \sin 2 \theta_{23} \sin 2 \theta_{13} \cos \theta_{13} \sin \delta \, .
\end{eqnarray}
Similar invariants, called $I_1$ and $I_2$, can be defined which depend on the Majorana phases $\alpha$ and $\beta$ 
\cite{Jenkins_Manohar_invariants} (see also \cite{rephasing_invariants_original,Majorana_invariants_other,Jenkins_Manohar_Hilbert})
\begin{eqnarray}
&&I_1 = {\rm Im} [U_{PMNS,12}^2 (U_{PMNS,11}^*)^2] = s^2_{12} c^2 _{12} c^4_{13} \sin \alpha~~~,
\\ \label{I1I2}
&&I_2 =  {\rm Im} [U_{PMNS,13}^2 (U_{PMNS,11}^*)^2] = s^2 _{13} c^2 _{12} c^2_{13} \sin \beta~~~.
\end{eqnarray}
Notice that the Dirac phase has a physical meaning only if all mixing angles are different from $0$ 
and $\pi/2$, as indicated by the data. Analogously, the vanishing of the invariants $I_{1,2}$ only implies $\sin \alpha=0$, $\sin \beta=0$, if  solutions with $\sin2 \theta_{12} =0$, $\cos \theta_{13}= 0$ 
or $\sin 2 \theta_{13}=0$, $\cos \theta_{12}=0$ are discarded. Furthermore, notice that one of the Majorana phases becomes unphysical, if the lightest neutrino mass vanishes.

\subsection{Global fit results}
\label{app12}

We use in our numerical analysis the results of mixing angles taken from \cite{nufit} as given in the left table
of table 1, i.e. the results obtained by including the short baseline reactor data (called RSBL in \cite{nufit}) and leaving reactor fluxes free in
the fit (see free fluxes in \cite{nufit}). The best fit values of $\sin^2 \theta_{ij}$, the $1\, \sigma$ errors as well as $3\, \sigma$ ranges are 
\begin{eqnarray}\nonumber
&& \sin^2 \theta_{13} = 0.0219^{+ 0.0010} _{-0.0011} \;\;\;\;\;\;\;\;\,\, \mbox{and} \;\;\; 0.0188 \leq \sin^2 \theta_{13} \leq 0.0251
\\ \label{anglesbfapp}
&& \sin^2 \theta_{12} = 0.304^{+ 0.012} _{-0.012} \;\;\;\;\;\;\;\;\;\;\;\;\,\mbox{and} \;\;\; 0.270 \leq \sin^2 \theta_{12} \leq 0.344
\\ \nonumber
&& \sin^2 \theta_{23} = \left\{ \begin{array}{c} [0.451^{+ 0.06} _{-0.03}] \\[0.05in]
0.577^{+ 0.027} _{-0.035}
\end{array} \right. \;\;\;\,\; \mbox{and} \;\;\; 0.385 \leq \sin^2 \theta_{23} \leq 0.644
\end{eqnarray}
where the value $\sin^2 \theta_{23} < 0.5$ is a local minimum. The $1 \,\sigma$ errors of this best fit value refer to itself and not to the global minimum, as done in \cite{nufit}. 
We have read these errors off the figure given in \cite{nufit}. 

\noindent In addition, the CP phase $\delta$, here given in radian, is constrained at the $1\, \sigma$ level \cite{nufit}
\begin{equation}
\label{deltabfapp}
\delta =4.38 ^{+1.17}_{-1.03} \;\;\; \mbox{and} \;\;\; 0 \leq \delta \leq 2 \pi \;\;\; \mbox{at} \;\;\; 3\,\sigma\,.
\end{equation}
\mathversion{bold}
\subsection{$\chi^2$ analysis}
\mathversion{normal}
\label{app13}

In our numerical analysis we use a $\chi^2$ function in order to evaluate which mixing patterns agree well with the experimental data on the mixing angles.
This function is defined in the usual way
\begin{eqnarray}
\label{chisqtot}
&&\chi^2_{\mathrm{tot}} = \chi^2_{12} + \chi^2_{13} + \chi^2_{23}
\\ \label{chisqsingle}
\mbox{with}\;\;\;&&\chi^2_{ij} = \left(\frac{\sin^2 \theta_{ij} - (\sin^2 \theta_{ij})^{\mathrm{bf}}}{\sigma_{ij}} \right)^2 \;\;\; \mbox{for} \;\;\; ij=12, 13, 23 \; .
\end{eqnarray}
$\sin^2 \theta_{ij}$ are the mixing angles derived in the different cases, e.g.
(\ref{anglescase1}, \ref{anglescase2}, \ref{anglescase3a}, \ref{anglescase3b1}), that depend in general on several discrete parameters $n$, $u$, $v$, $s$, $m$ as well as on
the continuous parameter $\theta$, $0 \leq \theta < \pi$, $\left(\sin^2 \theta_{ij}\right)^{\mathrm{bf}}$ are the best fit values and 
$\sigma_{ij}$ the $1 \, \sigma$ errors given in (\ref{anglesbfapp}). Note that these errors also depend on whether $\sin^2 \theta_{ij}$ is larger or smaller than the best fit
 value. Since the atmospheric mixing angle has a global minimum at $(\sin^2 \theta_{23})^{\mathrm{bf}}=0.577$ as well as a local one at 
 $(\sin^2 \theta_{23})^{\mathrm{bf}}=0.451$, we compute $\chi^2_{23}$ using $(\sin^2 \theta_{23})^{\mathrm{bf}}=0.451$, if $\sin^2 \theta_{23}$ 
 for $n$, $u$, $v$, $s$, $m$ and
 $\theta$ is smaller or equal 0.5, and use $(\sin^2 \theta_{23})^{\mathrm{bf}}=0.577$ otherwise. A mixing pattern is considered to agree reasonably well with the experimental
 data, if $\chi^2_{\mathrm{tot}} \lesssim 27$ and all mixing angles $\sin^2 \theta_{ij}$ are within the $3\, \sigma$ intervals in (\ref{anglesbfapp}).
For the different mixing patterns that allow for such a situation we present in tables \ref{tab:caseu}--\ref{tab:caseushiftpminus2} and 
\ref{tab:case3an16}--\ref{tab:case3b1shift} values of $n$, $u$, $v$, $s$, $m$ and $\theta=\theta_{\mathrm{bf}}$ for which 
 the $\chi^2$ function is minimized. Since the indication of a preferred value of the Dirac phase $\delta$ coming 
 from global fit analyses, see (\ref{deltabfapp}), is rather weak, i.e. below the $3\, \sigma$ significance, we do not include any information on $\delta$ in the $\chi^2$ function in (\ref{chisqtot}).

\mathversion{bold}
\section{Relations among the different choices $(Q, Z, X)$}
\mathversion{normal}
\label{app2}

First, we show that we can reduce all possible combinations of $(Q, Z, X)$ to $(\tilde{Q}=a, \tilde{Z}, \tilde{X})$ with $\tilde{Z}$ and $\tilde{X}$
being of the same type as $Z$ and $X$, respectively, if the residual symmetry 
$G_e$ of the charged lepton sector is a $Z_3$ group, i.e. it is generated by $Q= a \, c^\gamma d^\delta$ or by
$Q= a^2 \, c^\gamma d^\delta$ with $0 \leq \gamma , \delta \leq n-1$. Then, we prove that the twelve types of combinations $(Q=a, Z, X)$, given
by the twelve possible combinations of $Z$ and $X$, collected in table \ref{tab:ZXchoices},  can be reduced to three distinct
types $(Q=a, Z, X)$, either by applying the similarity transformations $\tilde{\Omega}=a$ and $\tilde{\Omega}=a^2$ or by exploiting the fact
that also $Y=Z X$ is an admissible CP transformation (in the neutrino sector), if $X$ is such a transformation and $Z$ is the generator
of a $Z_2$ symmetry fulfilling the condition in (\ref{XZ}), see end of subsection \ref{sec34}.

\mathversion{bold}
\subsection{Relations among different $(Q=a \, c^\gamma d^\delta, Z, X)$ and $(Q=a^2 c^\gamma d^\delta, Z, X)$}
\mathversion{normal}
\label{app21}

Here we argue that choosing $G_e=Z_3$ always allows us to reduce all combinations of $(Q, Z, X)$ to the triple $(\tilde{Q}=a, \tilde{Z}, \tilde{X})$
(or $(\tilde{Q}=a^2,\tilde{Z}, \tilde{X})$ that leads to the same $Z_3$ symmetry in the charged lepton sector).
As noted in subsection \ref{sec31}, all $Z_3$ subgroups of $\Delta (3 \, n^2)$ and $\Delta (6 \, n^2)$ with $3 \nmid n$, are generated by
elements of the form
\begin{equation}\nonumber
Q= a \, c^\gamma d^\delta \;\;\; \mbox{or} \;\;\; Q=a^2 c^\gamma d^\delta \;\;\; \mbox{with} \;\;\; 0 \leq \gamma, \delta \leq n-1 \; .
\end{equation}
Since all elements of the form $a \, c^\gamma d^\delta$ belong to the same class in $\Delta (3 \, n^2)$ and $\Delta (6 \, n^2)$,
we know that a similarity transformation $\tilde{\Omega}$ must exist which relates $Q=a c^\gamma d^\delta$ to $\tilde{Q}=a$. As can be checked such a transformation is of the form
$\tilde{\Omega} = c^f  d^h$ with $0 \leq f, h \leq n-1$. One can compute $f$ and $h$ which should lead to the correct transformation from the two conditions
\begin{equation}\nonumber
\gamma + f + h = 0 \; (\mbox{mod} \; n) \;\;\; \mbox{and} \;\;\; \delta -f + 2 \, h=0 \; (\mbox{mod} \; n)  \; .
\end{equation}
These equations can be solved for any combination of $\gamma$ and $\delta$. The same type of transformation $\tilde{\Omega}$ also
relates $Q=a^2 c^\gamma d^\delta$ to $\tilde{Q}=a^2$ (these elements also always belong to the same class of the groups $\Delta (3 \, n^2)$ and $\Delta (6 \, n^2)$).
The conditions that determine $f$ and $h$ in this case are
\begin{equation}\nonumber
\gamma + 2 \, f -h=0  \; (\mbox{mod} \; n) \;\;\; \mbox{and} \;\;\; \delta + f + h = 0  \; (\mbox{mod} \; n) \; .
\end{equation}
In the next step we apply $\tilde{\Omega} = c^f  d^h$ with arbitrary $f$ and $h$ to all the twelve pairs $(Z, X)$ that we have collected in table \ref{tab:ZXchoices} in subsection \ref{sec35}.
Clearly, the form of $Z$ does not change when $\tilde{\Omega}$ is applied, if it is an element containing only $c$ and $d$, i.e. $\tilde{Z}=Z$ for $Z=c^{n/2}$, $Z=d^{n/2}$ and 
$Z=(c \, d)^{n/2}$.
We compute for $X=c^s d^t P_{23}$ that
\begin{equation}\nonumber
\tilde{X}=\tilde{\Omega}^\dagger \, X \, \tilde{\Omega}^\star=c^{s'} d^{t'} P_{23} \;\;\; \mbox{with} \;\;\; s'=s-2 \, f \;\; , \;\; t'=t -2 \, h
\end{equation}
and, thus, the form of $X$ remains the same. Similarly, we see that $X=a \, b \, c^s d^{2 s} P_{23}$ does not
change its form, since
\begin{equation}\nonumber
\tilde{X}= a \, b \, c^{s'} d^{2 s'} P_{23} \;\;\; \mbox{with} \;\;\; s'=s-h \; .
\end{equation}
Also $X=a^2 b \, c^{2 t} d^t P_{23}$ which gets transformed via $\tilde{\Omega}$ into
\begin{equation}\nonumber
\tilde{X}=a^2 b \, c^{2 t'} d^{t'} P_{23} \;\;\; \mbox{with} \;\;\; t'=t-f
\end{equation}
has the same form as the original $X$. Lastly, the CP transformation $X=b \, c^s d^{n-s} P_{23}$ reads, after applying $\tilde{\Omega}=c^f  d^h$,  
\begin{equation}\nonumber
\tilde{X}=b \, c^{s'} d^{n-s'} P_{23} \;\;\; \mbox{with} \;\;\; s'=s+h-f \; .
\end{equation}
Thus, we have shown that all pairs $(Z, X)$ that are mentioned in the first three lines in table \ref{tab:ZXchoices} still have the same structure in the transformed basis.

Proceeding in the same way in the case of the combination $(Z=b \, c^m d^m, X=c^s d^t P_{23})$ with the condition $t=n-2 \, m -s$
we find that this pair is transformed into
\begin{equation}\nonumber
\tilde{Z}= b \, c^{m'} d^{m'} \;\;\; , \;\;\; \tilde{X}=c^{s'} d^{t'} P_{23} \;\;\; \mbox{with} \;\;\; m'= m+f+h \;\; , \;\; s'=s-2 \, f \;\; , \;\; t'=t- 2 \, h
\end{equation}
so that the form of $(Z, X)$ as well as  of the condition are maintained, i.e. it also holds $t'=n-2 \, m'-s'$. 
Next we consider the combination $(Z=a \, b \, c^m, X=c^s d^t P_{23})$ with the condition $t=2 \, (m+s)$: $Z$ is transformed into
\begin{equation}\nonumber
\tilde{Z}=a \, b \, c^{m'} \;\;\; \mbox{with} \;\;\; m'=m+2 \, f-h
\end{equation}
via the similarity transformation $\tilde{\Omega}$. Since $\tilde{X}=c^{s'} d^{t'} P_{23}$ with $s'=s-2 f$, $t'=t- 2 h$, we also recover the form of the
condition, namely $t'=2 \, (m'+s')$. 
The combination $(Z=a^2 \, b \, d^m, X=c^s d^t P_{23})$ together with the condition $s=2 \, (m+t)$ is transformed into
\begin{equation}\nonumber
\tilde{Z}=a^2 \, b \, d^{m'} \;\;\; \mbox{with} \;\;\; m'=m+2 \, h-f \;\;\; \mbox{and} \;\;\;  \tilde{X}=c^{s'} d^{t'} P_{23} \;\;\; \mbox{with} \;\;\;  s'=s-2 \, f \; , \; t'=t-2 \, h
\end{equation}
fulfilling the constraint $s'= 2 \, (m'+t')$.
Eventually, using these results it is immediate to see that also the three remaining pairs $(Z=b \, c^m d^m, X=b \, c^s d^{n-s} P_{23})$, 
$(Z=a \, b \, c^m, X=a \, b \, c^s d^{2 s} P_{23})$
and $(Z=a^2 \, b \, d^m, X=a^2 b \, c^{2 t} d^t P_{23})$ keep their structure when the transformation $\tilde{\Omega}$ is applied.

In summary, we have shown that all combinations $(Q=a \, c^\gamma d^\delta, Z, X)$ (and  $(Q=a^2 \, c^\gamma d^\delta, Z, X)$) can be related 
via a similarity transformation to $(\tilde{Q}=a, \tilde{Z}, \tilde{X})$ (and $(\tilde{Q}=a^2, \tilde{Z}, \tilde{X})$) where $\tilde{Z}$ and $\tilde{X}$ have the 
same structure as $Z$ and $X$, respectively.  Thus, it is sufficient to consider only cases with $Q=a$ in the case of the groups
$\Delta (3 \, n^2)$ as well as $\Delta (6 \, n^2)$, $3 \nmid n$, in order to perform a comprehensive analysis of the cases in which $G_e$ is a $Z_3$ symmetry.
In the next subsection we show that also the number of pairs $(Z, X)$ that needs to be discussed can be reduced.

\mathversion{bold}
\subsection{Relations among the different choices $(Q=a, Z, X)$}
\mathversion{normal}
\label{app22}

As we will see, it is sufficient to consider the similarity transformations $\tilde{\Omega}=a$ and $\tilde{\Omega}=a^2$ as well as the possibility that
also $Y= Z X$ is a viable CP transformation in the neutrino sector that leads to the same results for the mixing in order to reduce the twelve different types of
$(Q=a, Z, X)$ to only three. We start with
\begin{equation}\nonumber
Q=a \;\;\; , \;\;\; Z=c^{n/2} \;\;\; \mbox{and} \;\;\; X=c^s d^t P_{23} \; .
\end{equation}
Taking $\tilde{\Omega}=a$ we find 
\begin{equation}\nonumber
\tilde{Z}=d^{n/2} \;\;\; \mbox{and} \;\;\; \tilde{X}=c^{s'} d^{t'} P_{23} \;\;\;\; \mbox{with} \;\;\;\; s'=n-t \;\; , \;\; t'=s-t
\end{equation}
with $s'$ and $t'$ taking all possible values between $0$ and $n-1$. If we use instead $\tilde{\Omega}=a^2$, we see
that 
\begin{equation}\nonumber
\tilde{Z}=(c \, d)^{n/2} \;\;\; \mbox{and} \;\;\; \tilde{X}=c^{s'} d^{t'} P_{23} \;\;\;\; \mbox{with} \;\;\;\; s'=-s+t \;\; , \;\; t'=n-s
\end{equation}\nonumber
with again $s'$ and $t'$ taking all possible values between $0$ and $n-1$. Obviously, $\tilde{Q}=a$ in these two (and in the following) cases.
Next we study
\begin{equation}\nonumber
Q=a \;\;\; , \;\;\;  Z=c^{n/2} \;\;\; \mbox{and} \;\;\; X=a \, b \, c^s d^{2 s} P_{23} \; .
\end{equation}
Again, we first take $\tilde{\Omega}=a$ and find 
\begin{equation}\nonumber
\tilde{Z}=d^{n/2} \;\;\; \mbox{and} \;\;\; \tilde{X}=a^2 \, b \, c^{s'} d^{t'} P_{23} \;\;\;\; \mbox{with} \;\;\;\; s'=2 \, (n-s) \;\; , \;\; t'=n-s
\end{equation}
so that $s'=2 \, t'$ as required. For $\tilde{\Omega}=a^2$, on the other hand, we get 
\begin{equation}\nonumber
\tilde{Z}=(c \, d)^{n/2} \;\;\; \mbox{and} \;\;\; \tilde{X}=b \, c^{s'} d^{t'} P_{23} \;\;\;\; \mbox{with} \;\;\;\; s'=s \;\; , \;\; t'=n-s
\end{equation}
and thus $t'=n-s'$ as needed. We, similarly, find that starting with
\begin{equation}\nonumber
Q=a \;\;\; , \;\;\;  Z=b \, c^m d^m \;\;\; \mbox{and} \;\;\; X=b \, c^s d^{n-s} P_{23} 
\end{equation}
the application of $\tilde{\Omega}=a$ leads to
\begin{equation}\nonumber
\tilde{Z}=a \, b \, c^{m'} \;\;\; \mbox{and} \;\;\; \tilde{X}=a \, b \, c^{s'} d^{t'} P_{23} \;\;\;\; \mbox{with} \;\;\;\; m'=n-m \;\; , \;\; s'=s \;\; , \;\; t'=2 \, s
\end{equation}
and thus $t'=2 \, s'$. Applying $\tilde{\Omega}=a^2$ instead gives rise to
\begin{equation}\nonumber
\tilde{Z}=a^2 b \, d^{m'} \;\;\; \mbox{and} \;\;\; \tilde{X}=a^2 b \, c^{s'} d^{t'} P_{23} \;\;\;\; \mbox{with} \;\;\;\; m'=n-m \;\; , \;\; s'=2\, (n-s) \;\; , \;\; t'=n-s
\end{equation}
and so that $s'=2 \, t'$.
Furthermore, we can relate 
\begin{equation}\nonumber
Q=a \;\;\; , \;\;\; Z=b \, c^m d^m \;\;\; \mbox{and} \;\;\; X=c^s d^{t} P_{23} \;\;\;\; \mbox{with} \;\;\;\; t=n-2 \, m-s
\end{equation}
via the transformation $\tilde{\Omega}=a$ to
\begin{equation}\nonumber
\tilde{Q}=a \;\;\; , \;\;\; \tilde{Z}= a \, b \, c^{m'}  \;\;\; \mbox{and} \;\;\; \tilde{X}=c^{s'} d^{t'} P_{23} \;\;\;\; \mbox{with} \;\;\;\; m'=n-m \;\; , \;\; s'=n-t \;\; , \;\; t'=s-t
\end{equation}
with $t'= 2 \, (m'+s')$ being fulfilled as  well as via the transformation $\tilde{\Omega}=a^2$ to
\begin{equation}\nonumber
\tilde{Q}=a \;\;\; , \;\;\; \tilde{Z}= a^2 b \, d^{m'}  \;\;\; \mbox{and} \;\;\; \tilde{X}=c^{s'} d^{t'} P_{23} \;\;\;\; \mbox{with} \;\;\;\; m'=n-m \;\; , \;\; s'=t-s \;\; , \;\; t'=n-s
\end{equation}
so that $s'= 2 \, (m'+t')$. Eventually, we notice that the case
\begin{equation}\nonumber
Q=a \;\;\; , \;\;\; Z=b \, c^m d^m \;\;\; \mbox{and} \;\;\; X=b \, c^s d^{n-s} P_{23} 
\end{equation}
can be related to another case by exploiting that $Y=Z X$ can function as CP transformation in the neutrino sector
\begin{equation}\nonumber
Y=Z X= c^{s'} d^{t'} P_{23} \;\;\;\; \mbox{with} \;\;\;\; s'=s-m \;\; , \;\; t'=n-s-m
\end{equation}
so that $t'=n-2 \, m-s'$ holds. Obviously, $Q=a$ and $Z=b \, c^m d^m$ remain untouched. This allows us to recover  the combination
\begin{equation}\nonumber
Z=b \, c^m d^m \;\;\; \mbox{and} \;\;\; X=c^s d^t P_{23} \;\;\;\; \mbox{with} \;\;\;\; t=n-2 \, m-s \; .
\end{equation}

So, starting with twelve different allowed combinations $(Q=a, Z, X)$ we end up with three, namely   
\begin{eqnarray}\nonumber
&&Q=a \;\;\; , \;\;\;  Z=c^{n/2} \;\;\; \mbox{and} \;\;\; X=c^s d^t P_{23} \; , 
\\ \nonumber
&&Q=a \;\;\; , \;\;\;  Z=c^{n/2} \;\;\; \mbox{and} \;\;\; X=a \, b \, c^s d^{2 s} P_{23} \; ,
\\ \nonumber
\mbox{and}\;\;\;&&Q=a \;\;\; , \;\;\; Z=b \, c^m d^m \;\;\; \mbox{and} \;\;\; X=b \, c^s d^{n-s} P_{23} \; ,
\end{eqnarray}
for which we  study the lepton mixing. 

\normalsize

\newpage



\begin{thebibliography}{00}

 \bibitem{nufit}
M.~C.~Gonzalez-Garcia, M.~Maltoni, J.~Salvado and T.~Schwetz,
  JHEP {\bf 1212} (2012) 123
  [arXiv:1209.3023 [hep-ph]].

\bibitem{otherglobal_1}
F.~Capozzi, G.~L.~Fogli, E.~Lisi, A.~Marrone, D.~Montanino and A.~Palazzo,
  Phys.\ Rev.\ D {\bf 89} (2014) 093018
  [arXiv:1312.2878 [hep-ph]].

\bibitem{otherglobal_2}
D.~V.~Forero, M.~Tortola and J.~W.~F.~Valle,
  Phys.\ Rev.\ D {\bf 90} (2014) 093006
  [arXiv:1405.7540 [hep-ph]].
  
  \bibitem{reviews}
    G.~Altarelli, F.~Feruglio,
  Rev.\ Mod.\ Phys.\  {\bf 82 } (2010)  2701-2729
  [arXiv:1002.0211 [hep-ph]];
  H.~Ishimori, T.~Kobayashi, H.~Ohki, Y.~Shimizu, H.~Okada, M.~Tanimoto,
  Prog.\ Theor.\ Phys.\ Suppl.\  {\bf 183 } (2010)  1-163
  [arXiv:1003.3552 [hep-th]];
 S.~F.~King and C.~Luhn,
  Rept.\ Prog.\ Phys.\  {\bf 76} (2013) 056201
  [arXiv:1301.1340 [hep-ph]].

  \bibitem{review_math}
  W.~Grimus and P.~O.~Ludl,
  J.\ Phys.\ A {\bf 45} (2012) 233001
  [arXiv:1110.6376 [hep-ph]].
  
\bibitem{GLD3D4}
 W.~Grimus and L.~Lavoura,
  JHEP {\bf 0508}, 013 (2005)
  [arXiv:hep-ph/0504153];
 W.~Grimus and L.~Lavoura,
  Phys.\ Lett.\  B {\bf 572}, 189 (2003)
  [arXiv:hep-ph/0305046].

\bibitem{A4first}
G.~Altarelli and F.~Feruglio,
  Nucl.\ Phys.\ B {\bf 741} (2006) 215
  [hep-ph/0512103];
I.~de Medeiros Varzielas, S.~F.~King and G.~G.~Ross,
  Phys.\ Lett.\ B {\bf 644} (2007) 153
  [hep-ph/0512313];
X.~-G.~He, Y.~-Y.~Keum and R.~R.~Volkas,
  JHEP {\bf 0604} (2006) 039
  [hep-ph/0601001];
  E.~Ma,
  Phys.\ Rev.\ D {\bf 70} (2004) 031901
  [hep-ph/0404199].
  
\bibitem{Lam07}
 C.~S.~Lam,
  Phys.\ Lett.\ B {\bf 656} (2007) 193
  [arXiv:0708.3665 [hep-ph]];
 C.~S.~Lam,
  Phys.\ Rev.\ Lett.\  {\bf 101} (2008) 121602
  [arXiv:0804.2622 [hep-ph]];
 C.~S.~Lam,
  Phys.\ Rev.\ D {\bf 78} (2008) 073015
  [arXiv:0809.1185 [hep-ph]].

\bibitem{BHL07}
A.~Blum, C.~Hagedorn and M.~Lindner,
  Phys.\ Rev.\ D {\bf 77} (2008) 076004
  [arXiv:0709.3450 [hep-ph]].

 \bibitem{dATFH}
 R.~de Adelhart Toorop, F.~Feruglio and C.~Hagedorn,
  Phys.\ Lett.\ B {\bf 703} (2011) 447
  [arXiv:1107.3486 [hep-ph]];
 R.~de Adelhart Toorop, F.~Feruglio and C.~Hagedorn,
  Nucl.\ Phys.\ B {\bf 858} (2012) 437
  [arXiv:1112.1340 [hep-ph]].
  
   \bibitem{residualGnuZ2}
  S.~-F.~Ge, D.~A.~Dicus and W.~W.~Repko,
  Phys.\ Lett.\ B {\bf 702} (2011) 220
  [arXiv:1104.0602 [hep-ph]];
  S.~-F.~Ge, D.~A.~Dicus and W.~W.~Repko,
  Phys.\ Rev.\ Lett.\  {\bf 108} (2012) 041801
  [arXiv:1108.0964 [hep-ph]];
  D.~Hernandez and A.~Y.~Smirnov,
  Phys.\ Rev.\ D {\bf 86} (2012) 053014
  [arXiv:1204.0445 [hep-ph]];
  D.~Hernandez and A.~Y.~Smirnov,
  Phys.\ Rev.\ D {\bf 87} (2013) 5,  053005
  [arXiv:1212.2149 [hep-ph]].
  
\bibitem{HDscan2}
M.~Holthausen and K.~S.~Lim,
  Phys.\ Rev.\ D {\bf 88} (2013) 033018
  [arXiv:1306.4356 [hep-ph]].
  
\bibitem{HMV}
C.~Hagedorn, A.~Meroni and L.~Vitale,
  J.\ Phys.\ A {\bf 47} (2014) 055201
  [arXiv:1307.5308 [hep-ph]].
  
 \bibitem{D6n2mixing}
 S.~F.~King, T.~Neder and A.~J.~Stuart,
  Phys.\ Lett.\ B {\bf 726} (2013) 312
  [arXiv:1305.3200 [hep-ph]].
  
\bibitem{D3n2mixing}
A.~S.~Joshipura and K.~M.~Patel,
  Phys.\ Lett.\ B {\bf 727} (2013) 480
  [arXiv:1306.1890 [hep-ph]].

\bibitem{HDscan}
M.~Holthausen, K.~S.~Lim and M.~Lindner,
  Phys.\ Lett.\ B {\bf 721} (2013) 61
  [arXiv:1212.2411 [hep-ph]].
  
\bibitem{Lamscan}
 C.~S.~Lam,
  Phys.\ Rev.\ D {\bf 87} (2013) 013001
  [arXiv:1208.5527 [hep-ph]].
  
\bibitem{Grimus2014}
R.~M.~Fonseca and W.~Grimus,
  JHEP {\bf 1409} (2014) 033
  [arXiv:1405.3678 [hep-ph]].
  
\bibitem{TM_pheno}
 X.~G.~He and A.~Zee,
  Phys.\ Lett.\  B {\bf 645} (2007) 427
  [arXiv:hep-ph/0607163];
W.~Grimus and L.~Lavoura,
  JHEP {\bf 0809}, 106 (2008)
  [arXiv:0809.0226 [hep-ph]];
  Y.~Lin,
  Nucl.\ Phys.\  B {\bf 824} (2010) 95
  [arXiv:0905.3534 [hep-ph]];
 W.~Grimus, L.~Lavoura and A.~Singraber,
  Phys.\ Lett.\  B {\bf 686}, 141 (2010)
  [arXiv:0911.5120 [hep-ph]].
  
  \bibitem{S4CPgeneral}
 F.~Feruglio, C.~Hagedorn and R.~Ziegler,
  JHEP {\bf 1307} (2013) 027
  [arXiv:1211.5560 [hep-ph]].
  
\bibitem{GrimusRebelo}
W.~Grimus and M.~N.~Rebelo,
  Phys.\ Rept.\  {\bf 281} (1997) 239
  [hep-ph/9506272].
  
 \bibitem{GfCPHD}
 M.~Holthausen, M.~Lindner and M.~A.~Schmidt,
  JHEP {\bf 1304} (2013) 122
  [arXiv:1211.6953 [hep-ph]].
  
 \bibitem{TB}
 P.~F.~Harrison, D.~H.~Perkins and W.~G.~Scott,
  Phys.\ Lett.\ B {\bf 530} (2002) 167
  [hep-ph/0202074];
  Z.~-z.~Xing,
  Phys.\ Lett.\ B {\bf 533} (2002) 85
  [hep-ph/0204049].
  
 \bibitem{Delta48CP} 
G.~-J.~Ding and Y.~-L.~Zhou,
  arXiv:1312.5222 [hep-ph];
 G.~-J.~Ding and Y.~-L.~Zhou,
  JHEP {\bf 1406} (2014) 023
  [arXiv:1404.0592 [hep-ph]].
  
 \bibitem{Delta96CP}
 G.~-J.~Ding and S.~F.~King,
  Phys.\ Rev.\ D {\bf 89} (2014) 093020
  [arXiv:1403.5846 [hep-ph]].
  
  \bibitem{Delta27CP}
  C.~C.~Nishi,
  Phys.\ Rev.\ D {\bf 88} (2013) 3,  033010
  [arXiv:1306.0877 [hep-ph]].
  
 \bibitem{D6n2CPZ2Z2}
 S.~F.~King and T.~Neder,
 Phys.\ Lett.\ B {\bf 736} (2014) 308
  [arXiv:1403.1758 [hep-ph]].
 
 \bibitem{ChenRatz}
M.~-C.~Chen, M.~Fallbacher, K.~T.~Mahanthappa, M.~Ratz and A.~Trautner,
  Nucl.\ Phys.\ B {\bf 883} (2014) 267
  [arXiv:1402.0507 [hep-ph]].
 
 
  \bibitem{Delta3n2}
 C.~Luhn, S.~Nasri and P.~Ramond,
  J.\ Math.\ Phys.\  {\bf 48} (2007) 073501
  [hep-th/0701188].
  
 \bibitem{Delta6n2}
 J.~A.~Escobar and C.~Luhn,
  J.\ Math.\ Phys.\  {\bf 50} (2009) 013524
  [arXiv:0809.0639 [hep-th]].
 
  \bibitem{FSind_comp}
  N.~Kawanaka and H.~Matsuyama,
  Hokkaido Math. J. {\bf 19} (3) (1990) 495;
   D.~Bump and D.~Ginzburg,
  J. Algebra {\bf 278} (1) (2004) 294; 
  C.~Ryan~Vinroot,
  J. Algebra {\bf 293} (2005) 279. 
 
  \bibitem{TB_first_column}
  C.~S.~Lam,
  Phys.\ Rev.\ D {\bf 74} (2006) 113004
  [hep-ph/0611017];
  C.~H.~Albright and W.~Rodejohann,
  Eur.\ Phys.\ J.\ C {\bf 62} (2009) 599
  [arXiv:0812.0436 [hep-ph]];
 X.~-G.~He and A.~Zee,
  Phys.\ Rev.\ D\ {\bf 84} (2011) 053004
  [arXiv:1106.4359 [hep-ph]].

\bibitem{sumruleBranco}
G.~C.~Branco, M.~N.~Rebelo, J.~I.~Silva-Marcos and D.~Wegman,
  arXiv:1405.5120 [hep-ph].

\bibitem{D150D600}
C.~S.~Lam,
  Phys.\ Rev.\ D {\bf 87} (2013) 5,  053012
  [arXiv:1301.1736 [hep-ph]].

 \bibitem{A4S4CPmodels}
 G.~J.~Ding, S.~F.~King, C.~Luhn and A.~J.~Stuart,
  JHEP {\bf 1305} (2013) 084
  [arXiv:1303.6180 [hep-ph]];
  F.~Feruglio, C.~Hagedorn and R.~Ziegler,
  Eur.\ Phys.\ J.\ C {\bf 74} (2014) 2753
  [arXiv:1303.7178 [hep-ph]];
  G.~J.~Ding, S.~F.~King and A.~J.~Stuart,
  JHEP {\bf 1312} (2013) 006
  [arXiv:1307.4212 [hep-ph]];
  C.~C.~Li and G.~J.~Ding,
  Nucl.\ Phys.\ B {\bf 881} (2014) 206
  [arXiv:1312.4401 [hep-ph]];
  C.~C.~Li and G.~J.~Ding,
  arXiv:1408.0785 [hep-ph].
 
\bibitem{pdg}
J.~Beringer {\it et al.} [Particle Data Group], Phys. \ Rev. \ D {\bf 86} (2012) 010001 and 2013 partial update for the 2014 edition.

\bibitem{jcp}
  C.~Jarlskog,
  Phys.\ Rev.\ Lett.\  {\bf 55 } (1985)  1039.
  
  \bibitem{Jenkins_Manohar_invariants}
  E.~E.~Jenkins and A.~V.~Manohar,
  Nucl.\ Phys.\ B {\bf 792} (2008) 187
  [arXiv:0706.4313 [hep-ph]].
  
  \bibitem{rephasing_invariants_original}
  G.~C.~Branco, L.~Lavoura and M.~N.~Rebelo,
  Phys.\ Lett.\ B {\bf 180} (1986) 264.
  
  \bibitem{Jenkins_Manohar_Hilbert}
  E.~E.~Jenkins and A.~V.~Manohar,
  JHEP {\bf 0910} (2009) 094
  [arXiv:0907.4763 [hep-ph]].

  \bibitem{Majorana_invariants_other}
  J.~F.~Nieves and P.~B.~Pal,
  Phys.\ Rev.\ D {\bf 36} (1987) 315;
  F.~del Aguila and M.~Zralek,
  Nucl.\ Phys.\ B {\bf 447} (1995) 211
  [hep-ph/9504228];
 F.~del Aguila, J.~A.~Aguilar-Saavedra and M.~Zralek,
  Comput.\ Phys.\ Commun.\  {\bf 100} (1997) 231
  [hep-ph/9607311];
  J.~A.~Aguilar-Saavedra and G.~C.~Branco,
  Phys.\ Rev.\ D {\bf 62} (2000) 096009
  [hep-ph/0007025];
  J.~F.~Nieves and P.~B.~Pal,
  Phys.\ Rev.\ D {\bf 64} (2001) 076005
  [hep-ph/0105305].




 \end{thebibliography}
\end{document}